%% file: jacm3.tex
\documentclass[acmjacm]{acmtrans2m}

\usepackage{latexsym}
\usepackage{amsmath}
\usepackage{amssymb}
\usepackage{epsfig}
\usepackage{epic}
\usepackage{eepic}
\usepackage{xspace}
\usepackage{rotating}
\usepackage{pstricks}

\def\permission{}
\def\firstfoot{}

\def\punto{$\hspace*{\fill}\Box$}
\newcommand{\nop}[1]{}
\newcommand{\tuple}[1]{{\langle#1\rangle}}

\newtheorem{theorem}{Theorem}[section]
\newtheorem{example}[theorem]{Example}

\newtheorem{definition}[theorem]{Definition}
\newtheorem{proposition}[theorem]{Proposition}

\newtheorem{corollary}[theorem]{Corollary}
\newtheorem{lemma}[theorem]{Lemma}
\newtheorem{remark}[theorem]{Remark}

\def\ouraxes{{\bf Ax}}

\def\following {\textit{Following}}

\def\labela {(\textit{Label}_a)_{a \in \Sigma}}
\def\child {\textit{Child}\xspace}
\def\parent {\textit{Parent}\xspace}
\def\descendant {\textit{Descendant}}
\def\descendantorself {\textit{Descendant-or-self}}
\def\nextsibling {\textit{NextSibling}}
\def\nextindocorder {\textit{NextInDocOrder}}
\def\followingsibling {\textit{Following-sibling}}

\title{Conjunctive Queries over Trees}

\author{
GEORG GOTTLOB \\
Oxford University, Oxford, United Kingdom \\
CHRISTOPH KOCH \\
Universit\"at des Saarlandes, Saarbr\"ucken, Germany \\
KLAUS U.\ SCHULZ \\
Ludwig-Maximilians-Universit\"at M\"unchen, Munich, Germany
}

\date{}

\begin{abstract}
We study  the complexity and  expressive power of  conjunctive queries
over unranked labeled trees represented using a variety of structure
relations  such  as  ``child'',  ``descendant'',  and ``following'' as
well as unary  relations for node  labels.
We  establish a framework  for characterizing structures
representing  trees for  which  conjunctive queries  can be  evaluated
efficiently. Then we completely chart the tractability frontier of the
problem and establish a dichotomy theorem
for our axis relations, i.e., we find all subset-maximal
sets of  axes for which query  evaluation is in  polynomial time and show that
for all other cases, query evaluation is NP-complete.  All
polynomial-time  results  are  obtained  immediately using  the  proof
techniques from  our framework.  Finally, we study the expressiveness
of conjunctive  queries over  trees and  show that for each
conjunctive query,
there is an equivalent acyclic  positive query (i.e., a set of acyclic
conjunctive  queries),  but that  in  general  this  query is  not  of
polynomial size.
\end{abstract}









\begin{document}

\begin{bottomstuff}
This work was partially
supported by project No.\ Z29-N04 of the Austrian Science Fund (FWF),
by a project of the German Research Foundation (DFG),
and by the REWERSE Network of Excellence of the European Union.

An   extended  abstract
\cite{GKS2004}  of   this  work  appeared  in  {\em   Proc.\  23rd  ACM
SIGMOD-SIGACT-SIGART Symposium on Principles of Database Systems (PODS
2004)}\/, Paris, France, ACM Press,  New York, USA, pp. 189 -- 200.

Contact details:
Georg Gottlob (Georg.Gottlob@comlab.ox.ac.uk),
Oxford University Computing Laboratory,
Wolfson Building,
Parks Road,
Oxford OX1 3QD,
United Kingdom.


Christoph Koch (koch@infosys.uni-sb.de),
Lehrstuhl f\"ur Informationssysteme,
Universit\"at des Saarlandes,
D-66123 Saarbr\"ucken, Germany.

Klaus U. Schulz (schulz@cis.uni-muenchen.de),
Centrum f\"ur Informations- und Sprachverarbeitung,
Ludwig-Maximilians-Universit\"{a}t M\"unchen,
D-80536 M\"unchen, Germany.
\end{bottomstuff}

\maketitle


\section{Introduction}
\label{sect:introduction}

The theory  of  conjunctive  queries  over relational structures is,
from a certain point of view, 
the  greatest success  story of the  theory of database  queries.  These
queries   correspond to the most  common queries in
database  practice,  e.g.\   SQL  select-from-where  queries  with  conditions
combined using ``and'' only. Their evaluation problem has also been considered
in different contexts and under different names, notably
as the {\em Constraint Satisfaction problem}\/ in AI
\cite{KV98,Dec2003} and the {\em H-coloring problem}\/ in graph theory
\cite{HN2004}.
Conjunctive queries are surprisingly well-behaved: Many important
properties  hold for  conjunctive  queries  but fail  for  more general  query
languages  (cf.\  \cite{CM77,AHV95,Mai83}).

\nop{
For  conjunctive queries,  the
central problems of  checking containment, query minimization, satisfiability,
and related  problems are known  to be solvable,  even under various  forms of
dependencies \cite{AHV95}.  Also, a  large part of the theoretical
work  on data integration  has focussed  on studying  the case  of conjunctive
queries  (or mappings  between  databases formulated  as conjunctive  queries,
e.g.\ \cite{LMSS95}).
Many central problems related to the theory of conjunctive queries are NP-hard
\cite{CM77}, but the contributions  from the database theory community towards
making conjunctive queries practical and the associated computational problems
feasible  have  been substantial (see e.g.\
\cite{Yan81,GLS2001,CR1997,FFG2001,GLS2002} for some work on dealing with
the complexity of conjunctive query evaluation).
}

Unranked  labeled trees  are a  clean abstraction  of HTML, XML, LDAP, and
linguistic parse trees. This
motivates the study of conjunctive queries
{\em over trees}, where the tree structures are represented using unary
node  label  relations  and  binary  relations (often referred to as
{\em axes}) such  as  {\em  Child}\/,  {\em
Descendant}\/, and {\em Following}\/.

\smallskip
\noindent
{\bf XML Queries.}
Conjunctive queries over trees are naturally related to the problem of
evaluating  queries (e.g.,  XQuery  or XSLT)  on  XML data
(cf.\  \cite{DT2003}).   However,
conjunctive queries  are a cleaner and simpler  model whose complexity
and expressiveness can be formally  studied (while XQuery and XSLT are
Turing-complete).

(Acyclic) conjunctive queries over trees are
a generalization of the  most frequently used fragment of XPath.  For
example, the  XPath query //A[B]/following::C  is equivalent to  the (acyclic)
conjunctive query
\[
Q(z) \leftarrow \textit{A}(x),\; \child(x,y),\; B(y),\;
\following(x,z),\; C(z). 
\]
While       XPath   has    been   studied    extensively   (see e.g.\
\cite{GKP2005,GKPS2005} on its  complexity, \cite{BFK2003,OMFB2002}
on its expressive power,
%
%
and \cite{Hid2003} on the satisfiability problem),
%
%
little work so far has addressed the theoretical properties of {\em cyclic}\/
conjunctive queries over trees.
Sporadic results on their complexity can be found in
\cite{MSB2001,GK2002b,GKJACM,MS2001}.

\smallskip
\noindent
{\bf Data extraction and integration.}
(Cyclic)  conjunctive queries  on  trees  have been  used  previously in  data
integration, where queries in languages such as XQuery were canonically mapped
to conjunctive  queries over  trees to  build upon the  existing work  on data
integration   with  conjunctive   queries  \cite{DT2003,DT2003b}.   Another
application is  Web information extraction using a  datalog-like language over
trees \cite{BFG2001a,GKJACM}. (Of course,  each nonrecursive datalog rule is a
conjunctive query.)

\smallskip
\noindent
{\bf Queries in computational linguistics.}
A  further  area   in  which  such  queries  are   employed  is  computational
linguistics,  where one  needs to  search in,  or check  properties  of, large
corpora  of   parsed  natural  language.    Corpora  such  as   Penn  Treebank
\cite{Treebank} are unranked trees labeled with the phrase structure of parsed
(for Treebank, financial news) text.  A query asking for prepositional phrases
following noun phrases in the same  sentence can be phrased as the conjunctive
query
\[
Q(z) \leftarrow S(x),\; \textit{Descendant}(x,y),\; \textit{NP}(y),\;
\textit{Descendant}(x,z),\; \textit{PP}(z),\; \textit{Following}(y,z). 
\]

\begin{figure}
\begin{center}
\input{ex1.tex}
\end{center}
\vspace{-5mm}
\caption{A query graph.}
\label{fig:ex1}
\end{figure}
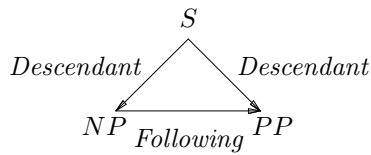

Figure~\ref{fig:ex1} shows this query in
the intuitive graphical notation that we will use throughout the article
(in which nodes correspond to variables, node labels to unary atoms, and edges
to binary atoms).

\smallskip
\noindent
{\bf Dominance constraints.}
Another
important issue  in  computational linguistics  are  conjunctions of  {\em
dominance  constraints}\/ \cite{MHF83},  which turn  out to  be  equivalent to
(Boolean)  conjunctive queries  over trees.   Dominance constraints  have been
influential  as a  means of  incompletely  specifying parse  trees of  natural
language,  in cases  where (intermediate)  results of parsing and
disambiguation  remain ambiguous.
%
%
One problem of practical importance is the
rewriting of  sets of dominance  constraints into
equivalent but  simpler sets (in particular, so-called  {\em solved forms}\/
\cite{BDNM2004}, which correspond to acyclic queries).
This implies  that studying the  expressive power of conjunctive  queries over
trees,  and  the  problem of  deciding  whether  there  is  a set  of  acyclic
conjunctive queries  equivalent to  a given conjunctive  query, is  relevant to
computational linguistics.

\smallskip
\noindent
{\bf Higher-order unification.}
The  query
evaluation problem for conjunctive queries over trees is also closely related
to the  {\em  context matching  problem}\/\footnote{To be precise,
the analogy is most direct with ranked trees.},
a variant of the well-known context-unification problem \cite{SSS1998,SSS2002}.
Some tractability frontier for the context matching problem is
outlined in \cite{SSS2001}.  However, little insight
is gained  from this for  the database context,  since the classes  studied in
\cite{SSS2001}
become  unnatural   when  formulated  as   conjunctive  queries\footnote{These
conjunctive queries  require node  inequality $\neq$ as  a binary  relation in
addition to  the tree structure relations.  If $\neq$ is  removed, the queries
become acyclic.  However, it is easy  to see that  already conjunctive queries
using  only the  inequality relation  over  a fixed  tree of  three nodes  are
NP-complete, by a reduction from Graph 3-Colorability.}.

\subsection*{Contributions}

Given the substantial number of applications  that we have hinted at above and
the nice  connection between  database theory, computational  linguistics, and
term  rewriting, it  is surprising  that conjunctive  queries over  trees have
never been  the object of  a concerted study\footnote{Of course,  as mentioned
above, there are  a number of papers that  implicitly contain relevant results
\cite{MSB2001,MS2001,Hid2003,SSS2001}. The papers \cite{HNZ1996a,HNZ1996}
address the complexity of a notion of tree homomorphisms that is uncomparable
to the one used in database theory, and the results there are orthogonal.}.

In particular, three questions seem worth studying:
\begin{enumerate}
\item
The  complexity  of  (cyclic)  conjunctive  queries on  trees  has  only  been
scratched  in  the  literature.  There  is little  understanding  of  how  the
complexity of conjunctive queries over  trees depends on the relations used to
model the tree.

\item
There is a natural connection between conjunctive queries and XPath. Since all
XPath queries  are acyclic, the  question arises whether the  acyclic positive
queries (i.e.,  unions of acyclic conjunctive queries)
are as expressive  as the full
class of conjunctive queries over trees.\footnote{This is equivalent to asking
whether for all conjunctive queries over trees there exist equivalent
{\em positive
Core XPath}\/ queries \cite{GKP2005}.}

\item
If that is  the case,   how much bigger do the
acyclic    versions    of    queries    get   than    their    cyclic
counterparts?  Except  from  being  of theoretical  interest,
first  translating queries into their acyclic
versions, if that is possible,  and then evaluating them as such may
be a practical query evaluation strategy, because there are
particularly good algorithms for evaluating such queries
\cite{Yan81,CR1997,FFG2001,GKJACM}.
\end{enumerate}

We thus  study conjunctive  queries on tree  structures represented  using the
XPath {\em axis}\/ relations {\em child}\/, {\em descendant}\/, {\em
descendant-or-self}\/, {\em following-sibling}\/, and {\em
following}\/.
Since we are free to  use these relations with any pair
of variables of  our conjunctive queries (differently from  XPath), these five
axes  render  all  others,   i.e.\  {\em  parent}\/,  {\em  ancestor}\/,  {\em
ancestor-or-self}\/,   {\em  preceding-sibling}\/,   and   {\em  preceding}\/,
redundant. Typed child axes such as {\em attribute}\/ are
redundant with the  {\em child}\/ axis and unary  relations in our framework.

For  a more elegant framework, we  study  the  axes
$\textit{Child}$,   
$\textit{Child}^*$  (=  {\em  descendant-or-self}\/),
$\textit{Child}^+$     (=      {\em     descendant}\/),
$\textit{NextSibling}$,
$\textit{NextSibling}^*$,
$\textit{NextSibling}^+$         (=         {\em follow\-ing-sibling}\/),
and $\textit{Following}$. ($\textit{NextSibling}$
and $\textit{NextSibling}^*$ are not supported in XPath but are nevertheless
considered here.) Subsequently, we denote this set of all axes considered
in this article by $\ouraxes$.

The main contributions of this article are as follows.
\begin{itemize}
\item
In \cite{GWW1992} it was shown that the {\em $H$-coloring problem}\/ (cf.\
\cite{HN2004}), and thus
Boolean conjunctive query evaluation, on directed graphs $H$ that
have the so-called {\em $\underbar{X}$-property}\/ (pronounced
``X-underbar-property'') is polynomial-time solvable.
We determine which of our axis relations have the $\underbar{X}$-property with
respect to which orders of the domain elements.
We show that the subset-ma\-xi\-mal sets of axis relations for which the
$\underbar{X}$-property yields tractable query evaluation are the three
disjoint sets
\[
\{ \child, \nextsibling, \nextsibling^*, \nextsibling^+ \},
\]\[
\{ \child^*, \child^+ \}, \quad\mbox{and}\quad \{ \following \}.
\]

\item
We prove that the conjunctive query evaluation problem for queries involving
any two axes that do not have the $\underbar{X}$-property with respect to the
same ordering of the tree nodes is NP-complete.

Thus the $\underbar{X}$-property yields a complete characterization of the
tractability frontier of the problem (under the assumption that P $\neq$ NP).

\begin{theorem}
Unless P = NP, for any $F \subseteq \ouraxes$,
the conjunctive queries over structures with unary
relations and binary relations from $F$ are in P if and only if
there is a total order $<$ such that
all binary relations in $F$ have the $\underbar{X}$-property w.r.t.\ $<$.
\end{theorem}

Moreover we have the dichotomy that for any of our tree structures, the
conjunctive query evaluation problem is either in P or NP-complete.

\begin{table}
\begin{scriptsize}
\hspace{-.5cm}
\begin{tabular}{|l||c|c|c|c|c|c|c|} 
\hline
\ & $\child$ & $\child^+$  & $\child^*$  & $\nextsibling$ & $\nextsibling^+$ & $\nextsibling^*$  & $\following$  \\
\hline
\hline
$\child$ & in P & NP-hard & NP-hard & in P & in P & in P & NP-hard \\
& (\ref{theo:child_ns_nsplus_nsstar_ptime}) &
(\ref{theo:NPhardnessChildDescendant}) &
(\ref{theo:NPhardnessChildDescendant}) &
(\ref{theo:child_ns_nsplus_nsstar_ptime}) &
(\ref{theo:child_ns_nsplus_nsstar_ptime}) &
(\ref{theo:child_ns_nsplus_nsstar_ptime}) &
(\ref{theo:NPhardnessFollowingChild}) \\
\hline
$\child^+$ & & in P & in P & NP-hard & NP-hard & NP-hard & NP-hard \\
& & (\ref{prop:descendant_dos_ptime}) &
(\ref{prop:descendant_dos_ptime}) &
(\ref{theo:child_plus_nextsibling_all}) &
(\ref{theo:child_plus_nextsibling_all}) &
(\ref{theo:child_plus_nextsibling_all}) &
(\ref{theo:NPhardnessFollowingChildTrans}) \\
\hline
$\child^*$ & & & in P & NP-hard & NP-hard & NP-hard & NP-hard \\
& & & (\ref{prop:descendant_dos_ptime}) &
(\ref{theo:child_star_nextsibling}) &
(\ref{cor:child_star_nextsibling_plus}) &
(\ref{theo:child_star_nextsibling_star}) &
(\ref{theo:NPhardnessFollowingChildTrans}) \\
\hline
$\nextsibling$ & & & & in P & in P & in P & NP-hard \\
& & & & (\ref{theo:child_ns_nsplus_nsstar_ptime}) &
(\ref{theo:child_ns_nsplus_nsstar_ptime}) &
(\ref{theo:child_ns_nsplus_nsstar_ptime}) &
(\ref{theo:NPhardnessFollowingNextsibling}) \\
\hline
$\nextsibling^+$ & & & & & in P & in P & NP-hard \\
& & & & & (\ref{theo:child_ns_nsplus_nsstar_ptime}) &
(\ref{theo:child_ns_nsplus_nsstar_ptime}) &
(\ref{theo:NPhardnessFollowingNextsibling}) \\
\hline
$\nextsibling^*$ & & & & & & in P & NP-hard \\
& & & & & & (\ref{theo:child_ns_nsplus_nsstar_ptime}) &
(\ref{theo:NPhardnessFollowingNextsibling}) \\
\hline
$\following$ & & & & & & & in P  \\
& & & & & & & (\ref{theo:following_ptime}) \\
\hline
\end{tabular} 
\end{scriptsize}
\vspace{-4mm}
\caption{Complexity results for signatures with one or two axes,
with pointers to relevant theorems.}
\label{TableSurvey}
\end{table}

Table~\ref{TableSurvey} shows the complexities of conjunctive queries
over structures containing unary relations and either one or two
axes.\footnote{It was shown in \cite{GKJACM} that conjunctive queries
over $\child$ and $\nextsibling$ are in P.
Proposition~\ref{prop:descendant_dos_ptime} is from \cite{GK2002b}.)
The other results are new.}

All NP-hardness results hold already for fixed data trees (query complexity
\cite{Var82}). The polynomial-time upper bounds are established under the
assumption that both data and query are variable (combined complexity).

\item
We study  the expressive power of conjunctive queries  on
trees.  We show
that for  each conjunctive  query over trees,  there is an  equivalent acyclic
positive query (APQ)  over the same tree relations.  The blowup in size of the
APQs produced is exponential in the worst case.

It  follows that there is
an equivalent XPath query for each conjunctive query over trees,
since each APQ can be translated into XPath (even in linear time).  

\item
Finally, we  provide a  result that  sheds some light  at the  succinctness of
(cyclic) conjunctive queries and  which demonstrates that the blow-up observed
in our translation is actually necessary.  We prove that there are conjunctive
queries over trees for which no equivalent polynomially-sized APQ exists.
\end{itemize}

The structure  of the article is as  follows.  We  start with basic  notions in
Section~\ref{sect:preliminaries}.  Section~\ref{sect:p_framework} introduces
the $\underbar{X}$-property and the associated
framework for finding classes of  conjunctive queries that can be evaluated in
polynomial time.  Section~\ref{sect:p_results} contains our
polynomial-time  complexity  results. Section~\ref{sect:np_results}  completes
our  tractability   frontier  with  the  NP-hardness   results.   In
Section~\ref{sect:expressiveness},  we  provide  our  expressiveness  results.
Finally, we present our succinctness result in Section~\ref{sect:succinctness}.


\section{Preliminaries}
\label{sect:preliminaries}

\def\dom{{\bf dom}}
\def\varsq{\textit{Var}(Q)}
\def\strucA{{\cal A}}
\def\ltbflr{<_{\textit{\footnotesize bflr}}}
\def\lebflr{\le_{\textit{\footnotesize bflr}}}

Let $\Sigma$ be  a labeling alphabet. Throughout the article, if not explicitly
stated other\-wise, we  will not assume $\Sigma$
to be fixed.  An unranked  tree is a
tree in which each node may have an unbounded number of children. We allow for
tree nodes to be labeled  with multiple labels. However, throughout the
article,
our tractability  results will support  multiple labels while  our NP-hardness
and expressiveness results will not make use of them.

We  represent  trees as  relational  structures  using  unary label  relations
$\labela$ and binary relations called axes.  For a relational structure ${\cal
A}$, let $A = |{\cal A}|$ denote the finite domain (in the case of a tree, the
nodes) and  let $||{\cal  A}||$ denote  the size of  the structure 
under any reasonable encoding scheme (see e.g.\
\cite{EF99}).  We use  the binary {\em axis}\/ relations  $\child$ (defined in
the normal way)  and $\nextsibling$ (where $\nextsibling(v,w)$ if and only
if  $w$ is the
right neighboring sibling of $v$  in the tree), their transitive and reflexive
and  transitive closures  (denoted  $\child^+$, $\nextsibling^+$,  $\child^*$,
$\nextsibling^*$),  and the  axis $\following$, defined as
%
%
\begin{equation}
\label{eq:following}
\following(x,y) = \exists z_1 \exists z_2 \; \child^*(z_1,x) \land
\nextsibling^+(z_1, z_2) \land \child^*(z_2,y).
\end{equation}
This set of axes
covers the standard XPath axes (cf.\ \cite{XPath1}) by the equivalences
$\child^+ = \descendant$,
$\child^* = \descendantorself$, and
$\nextsibling^+ = \followingsibling$.

We consider  three well-known  total orderings on  finite ordered  trees.  The
{\em  pre-order}\/ $\leq_{pre}$  corresponds  to a  depth-first  left-to-right
traversal of a  tree.  If XML-documents are represented as  trees in the usual
way, the pre-order  coincides with the {\em document order}\/.  It is given by
the  sequence   of  {\em   opening}\/  tags  of   the  XML   {\em  elements}\/
(corresponding to nodes).  The {\em post-order}\/ $\leq_{post}$ corresponds to
a bottom-up left-to-right  traversal of the tree and is  given by the sequence
of  {\em closing}\/  tags  of  elements.  Furthermore,  we  also consider  the
ordering $\leq_{bflr}$  which is given by  the sequence of opening  tags if we
traverse the tree breadth-first left-to-right.

The $k$-ary {\em conjunctive queries}\/ can be defined by positive existential
first-order formulas without disjunction and with $k$ free variables. We
will usually use the standard (datalog)
{\em rule notation}\/ for conjunctive queries (cf.\ \cite{AHV95}).

We call the 0-ary queries {\em Boolean}\/ and the unary queries {\em
  monadic}\/.
The {\em containment}\/ of queries $Q$ and $Q'$ is defined in the normal way:
Query $Q$ is said to be contained
in $Q'$ (denoted $Q \subseteq Q'$) iff, for all tree structures ${\cal A}$,
$Q'$ returns at least all tuples on ${\cal A}$ that $Q$ returns
on ${\cal A}$. (To cover Boolean queries, tuples here may be nullary.)
Two queries $Q,Q'$ are called
equivalent iff $Q \subseteq Q'$ and $Q' \subseteq Q$.

Let $Q$ be a conjunctive query and let
$\varsq$ denote the  variables appearing in $Q$.  The query  graph of $Q$ over
unary  and binary  relations is  the {\em directed}\/
multigraph $G=(V,E)$  with edge labels and multiple node labels such  that
$V=\varsq$, node $x$ is labeled $P$ iff
$Q$ contains unary atom $P(x)$,  and $E$ contains labeled directed edge $x
\stackrel{R}{\rightarrow}  y$  if and only if
$Q$  contains binary  atom  $R(x,y)$.
Figure~\ref{fig:ex1} shows an example of such a query graph.
Our notion of query graph is sometimes called {\em positive atomic
  diagram}\/  in model theory or the graph of the {\em canonical database}\/
of a query in the database theory literature.

Throughout the article, we use lower case node and variable names and upper
case label and relation names.


\section{The \underline{X}-Property}
\label{sect:p_framework}

Let $Q$ be  a conjunctive query and let $A$ denote the  finite domain, i.e.\
in case  of a tree  the set  of nodes.  A  pre-valuation for $Q$  is a
total  function $\Theta: \varsq  \rightarrow 2^A$  that assigns  to each
variable  of $Q$ a {\em nonempty}\/ subset of  $A$.
A  valuation for  $Q$ is  a total
function $\theta: \varsq \rightarrow A$.

Let ${\cal A}$ be a relational structure of unary and binary relations.
A  pre-valuation $\Theta$  is  called  {\em
arc-consistent}\/\footnote{This notion is well-known in constraint
satisfaction, cf.\ \cite{Dec2003}.}
iff  for each unary atom  $P(x)$ in $Q$  and each $v
\in \Theta(x)$,  $P(v)$ is  true (in ${\cal  A}$) and for  each binary
atom $R(x,y)$ in  $Q$, for each $v \in \Theta(x)$  there exists $w \in
\Theta(y)$ such that $R(v,w)$ is true and for each $w \in \Theta(y)$ there
exists $v \in \Theta(x)$ such that $R(v,w)$ is true.

\begin{proposition}[(Folklore)]
\label{prop:compute_preval}
There is an algorithm which checks in time
$O(||{\cal A}|| \cdot
|Q|)$ whether an arc-consistent pre-valuation of $Q$
on $\strucA$ exists, and if it does, returns one.
\end{proposition}

\def\rem{\mathit{Remove}}

\begin{proof}
We phrase the problem of computing $\Theta$ by
deciding, for each $x, v$, whether $v \not\in
\Theta(x)$ as an instance ${\cal P}$ of propositional Horn-SAT. The
propositional predicates are the atoms $\rem(x,v)$ (where $x \in
\mathit{Vars}(Q)$, $v \in A$ are constants), and the Horn clauses are
\begin{tabbing}
\hspace{8mm} \=
   $\{ \rem(x, v) \leftarrow .
   \mid P(x) \in Q,\; v \in A,\; \neg P^\strucA(v) \}$ $\cup$ \\
\> $\{ \rem(x, v) \leftarrow \bigwedge \{ \rem(y,w) \mid R^\strucA(v, w) \}.
   \mid R(x,y) \in Q,\; v \in A \}$ $\cup$ \\
\> $\{ \rem(y, w) \leftarrow \bigwedge \{ \rem(x,v) \mid R^\strucA(v, w) \}.
   \mid R(x,y) \in Q,\; w \in A \}$
\end{tabbing}
Let $\rem$ be the binary relation defined by ${\cal P}$ and let
\[
T = (\mathit{Vars}(Q) \times A) - \rem
\]
be the complement of that relation.
If there is a variable $x$ such that for no node $v$, $(x,v) \in T$,
no arc-consistent pre-valuation of $Q$ on $\strucA$
exists and $Q$ is not satisfied. Otherwise,
the pre-valuation defined by
\[
\Theta(x) \mapsto \{ v \mid (x,v) \in T \},
\]
for each $x$, is obviously arc-consistent and
contains all arc-consistent pre-valuations of $Q$ and $\strucA$.

Program ${\cal P}$ can be computed and solved (e.g.\ using Minoux'
algorithm
\cite{Min88}), and the solution complemented, in time linear in the
size of the program, which is $O(||{\cal A}|| \cdot |Q|)$.
\end{proof}

Actually, this algorithm computes the unique subset-ma\-xi\-mal
arc-consistent pre-valuation of $Q$ on $\strucA$.

A valuation $\theta$ is called  {\em consistent}\/ if it satisfies the
query. In this case, for a Boolean query, we also say that the
  structure is a {\em model}\/ of the query and the valuation a {\em
    satisfaction}\/.  Obviously, a valuation is
consistent if and only if the pre-valuation $\Theta$ defined
by $\Theta(x) \mapsto \{ \theta(x)  \}$ is arc-consistent.  Let $<$ be
a total  order on $A  = |{\cal A}|$  and $\Theta$ be  a pre-valuation.
Then the valuation $\theta$ with  $\theta(x) \mapsto v$ iff $v$ is the
smallest node  in $\Theta(x)$ w.r.t.\  $<$ is called the  {\em minimum
valuation}\/ w.r.t.\ $<$ in $\Theta$.

\begin{definition}
\label{def:foolproof}
\em
Let ${\cal  A}$ be  a relational structure,  $R$ a binary  relation in
${\cal A}$, and $<$ a total order on $A = |{\cal A}|$.
Then, $R$ is said to have the {\em $\underbar{X}$-property w.r.t.\ $<$}\/ iff
for all $n_0,n_1,n_2,n_3 \in A$ such that $n_0 < n_1$ and $n_2 < n_3$,
\[
R(n_1,n_2) \land  R(n_0,n_3) \Rightarrow R(n_0,n_2).
\]
\end{definition}

\begin{figure}
\begin{center}
\input{xunderbar.tex}
\end{center}
\caption{The $\underbar{X}$-property. Graph (a) and its illustration by
arcs between two bars (b).
For crossing arcs $R(u, v)$ and $R(u', v')$, say $u < u'$ and
$v' < v$, there must be an arc $R(u, v')$.}
\label{fig:xunderbar}
\end{figure}
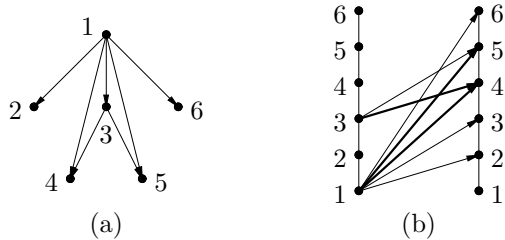

Figure~\ref{fig:xunderbar} illustrates why the property is called
$\underbar{X}$ (read as ``X-underbar'').
Let us consider two vertical bars both representing the order
$<$ bottom-up (i.e., with the smallest value at the bottom). Let each edge
$(u,v)$ in $R$ be represented by an arc from node $u$ on the left bar to node
$v$ on the right bar. Then, whenever there are two crossing arcs $(u, v)$ and
$(u',v')$ in this diagram, then there must be an arc $(\mbox{min}(u,u'),
\mbox{min}(v,v'))$, the ``underbar'', in the diagram as well.

\begin{remark}
\em
The $\underbar{X}$-property\footnote{In \cite{GKS2004}, this property was
called {\em hemichordality}\/.}
was introduced in \cite{GWW1992}, where it was shown that
the $H$-coloring problem (or equivalently the conjunctive query evaluation
problem) on graphs $H$ with the $\underbar{X}$-property is
polynomial-time solvable (see also \cite{HN2004}).
In the remainder of this section, we rephrase this result
as a tool for efficiently evaluating conjunctive queries.
\end{remark}

Let ${\cal A}$ be a  structure of unary and binary relations
and let $<$ be a total order on $|{\cal A}|$. 
Structure $\strucA$ is said to have the $\underbar{X}$-property w.r.t.\ $<$
if all binary relations $R$ in ${\cal A}$ have the $\underbar{X}$-property
w.r.t.\ $<$.

\begin{lemma}
\label{lem:ArcConsistency}
Let ${\cal A}$ be a structure with the $\underbar{X}$-property
w.r.t.\ $<$ and let $\Theta$ be an arc-consistent pre-valuation on ${\cal A}$
for a given conjunctive query over the relations of ${\cal A}$.
Then, the minimum valuation in $\Theta$ w.r.t.\ $<$ is consistent.
\end{lemma}

\begin{proof}
Let $\theta$ denote the minimum  valuation in $\Theta$ w.r.t.\ $<$.
To prove $\theta$ consistent, we only need to
show the  following: If $R(x,y)$ is  any binary atom of  $Q$ with variables
$x, y$ then $R(x,y)$ holds  under assignment $\theta$, i.e.\
$R(\theta(x), \theta(y))$ is true in ${\cal A}$.

Let $\theta(x) = n_0$ and $\theta(y)  = n_2$.
Since $\Theta$ is arc-consistent
there exists a node $n_1\in \Theta(x)$ such that $R(n_1,n_2)$ and a node
$n_3\in \Theta(y)$ such that $R(n_0,n_3)$. If $n_0=n_1$ or $n_2=n_3$
then $R(\theta(x), \theta(y)) = R(n_0,n_2)$ is true and we are done.
Otherwise, since $\theta$ is a minimum valuation
we have $n_0 < n_1$ (because $n_0 = \theta(x) = \min \Theta(x)$,
$n_1 \in \Theta(x)$, and $n_0 \neq n_1$) and $n_2 < n_3$
(because $n_2 = \theta(y) = \min \Theta(y)$,
$n_3 \in \Theta(y)$, and $n_2 \neq n_3$).
Then it follows from Definition~\ref{def:foolproof} that $R(n_0,n_2)$.
\end{proof}

Clearly, if no arc-consistent pre-valuation of $Q$ on $\strucA$ exists,
there is no consistent valuation for $Q$ on $\strucA$.

\nop{

Note that the reversal of Lemma~\ref{lem:ArcConsistency} is also true.
That is, the $\underbar{X}$-property
characterizes precisely those structures for
which the minimum valuations of arc-consistent pre-valuations are globally
consistent.

\begin{proposition}
Let ${\cal A}$ be a structure and let $<$ be a total order on $|{\cal A}|$.
If for every conjunctive query $Q$ over the relations of ${\cal A}$
and every  arc-consistent pre-valuation $\Theta$ of $Q$ on ${\cal A}$, 
the minimum valuation in $\Theta$ w.r.t.\ $<$ is consistent,
then ${\cal A}$ has the $\underbar{X}$-property w.r.t.\ $<$.
\end{proposition}

\begin{proof}
Die Praemisse von Condition~1 charakterisiert genau ein
arc-konsistentes Netz f\"ur die Anfrage $R(x,y)$ wo $x$ Domain $\{n_0,n_1\}$ und $y$ Domain $\{n_2,n_3\}$ hat. Die zwei R-Bedingungen sichern Arc-konsistenz.
Wenn wir nun schliessen koennen, dass die Minimum-Evaluierung Loesung ist,
muss also $R(n_0,n_2)$ gelten. Bedingung 2 geht analog.
\end{proof}
} 

\begin{theorem}
\label{theo:efficient_eval}
Given a structure $\strucA$ with the $\underbar{X}$-property w.r.t.\ $<$
and a Boolean conjunctive query $Q$ over ${\cal A}$,
$Q$ can be evaluated on $\strucA$ in time $O(||\strucA|| \cdot |Q|)$.
\end{theorem}

\begin{proof}
By   Lemma~\ref{lem:ArcConsistency}, all we need to do
to check  whether a Boolean
query $Q$  is satisfied is to try to compute the
subset-maximal arc-consistent pre-valuation $\Theta$  with respect to $Q$.
By Proposition~\ref{prop:compute_preval}, this can be
done in time $O(||\strucA|| \cdot |Q|)$.
If it  exists, $Q$ returns true; otherwise, $Q$ returns false.
\end{proof}

If follows that checking whether a given tuple
$\tuple{a_1, \dots, a_k}$
is in the result of a $k$-ary conjunctive query on
structures with the $\underbar{X}$-property w.r.t.\ some order
can be decided in time $O(||\strucA|| \cdot |Q|)$ as well.
All we need to do is to add (new) singleton unary relations
$X_1 = \{a_1\}, \dots, X_k = \{a_k\}$ to $\strucA$ and to
rewrite the query
$
Q(x_1, \dots, x_k) \leftarrow \Phi(x_1, \dots, x_k)
$
into the Boolean query
$
Q \leftarrow \Phi(x_1, \dots, x_k) \land X_1(x_1) \land \dots \land X_k(x_k).
$
A $k$-ary conjunctive query $Q$ over ${\cal A}$  with $A = |\strucA|$
can thus be evaluated on $\strucA$ in time
$O(|A|^k \cdot ||\strucA|| \cdot |Q|)$.

For relations that are subsets of the given total order
$\leq$ (the reflexive closure of $<$),
a slightly stronger condition for the $\underbar{X}$-property w.r.t.\ $<$ can
be given.

\begin{lemma}
\label{lem:ArcConsistency2}
Let ${\cal A}$ be a structure,
$<$ a total  order on $A = |{\cal A}|$, and $R$ a  binary relation of
${\cal A}$ such that $R\subseteq \leq$.
Then,
$R$ has the $\underbar{X}$-property w.r.t.\ $<$ iff
for all $n_0, n_1, n_2, n_3 \in A$ such that $n_0 < n_1 \leq n_2 < n_3$,
\[
R(n_1,n_2) \land  R(n_0,n_3)
\Rightarrow R(n_0,n_2).
\]
\end{lemma}

\begin{proof}
Obviously, if $R$ has the
$\underbar{X}$-property w.r.t.\ $<$, then  the condition of
Definition~\ref{def:foolproof} implies that the condition of our
lemma holds.
Conversely, since for all $n_1, n_2$, $R(n_1, n_2) \Rightarrow n_1 \le n_2$,
by our lemma, for all $n_0, n_1, n_2, n_3\in A$ such that
$n_0 < n_1$ and $n_2 < n_3$,
$R(n_1, n_2) \land R(n_0, n_3) \Rightarrow R(n_0, n_2)$.
\end{proof}

A symmetric version of Lemma 3.5 holds for relations $R\subseteq \geq$.

\begin{lemma}
\label{lem:ArcConsistency2inv}
\em
Let ${\cal A}$ be a structure, $<$ a total order on
$A=\vert {\cal A}\vert$, and $R$ a binary relation of ${\cal A}$
such that $R\subseteq \geq$. Then, $R$ has the
$\underbar{X}$-property w.r.t.\ $<$ iff for all
$n_0, n_1, n_2, n_3\in A$ such that $n_0 < n_1 \leq n_2 < n_3$,
\[
R(n_2,n_1) \land R(n_3,n_0) \Rightarrow R(n_2,n_0).
\]
\end{lemma}

\begin{proof}
Let $R' = R^{-1}$. By Lemma~\ref{lem:ArcConsistency2},
$R'$ has the $\underbar{X}$-property w.r.t.\ $<$ precisely if
for all $n_0, n_1, n_2, n_3 \in A$,
$n_0 < n_1 \leq n_2 < n_3 \land R'(n_1,n_2) \land  R'(n_0,n_3)
\Rightarrow R'(n_0,n_2)$.
Thus, $R$ has the $\underbar{X}$-property w.r.t.\ $<$
iff the condition of our lemma holds.
\end{proof}


\section{Polynomial-Time Results}
\label{sect:p_results}

\def\ltpost{{<_{\textit{\footnotesize post}}}}
\def\leqpost{{\leq_{\textit{\footnotesize post}}}}
\def\leqbflr{{\leq_{\textit{\footnotesize bflr}}}}
\def\ltpre{{<_{\textit{\footnotesize pre}}}}
\def\leqpre{{\leq_{\textit{\footnotesize pre}}}}

The results of Section~\ref{sect:p_framework}
provide us with a simple technique for proving polynomial-time complexity
results for
conjunctive queries over trees.
Indeed, there is a wealth of inclusions of axis relations in the total orders
introduced in Section~\ref{sect:preliminaries}:
\begin{enumerate}
\item
all the axes in $\ouraxes$ are subsets of the pre-order $\leqpre$,

\item
$\child^{-1}$, $(\child^+)^{-1}$, $(\child^*)^{-1}$,
$\following$, $\nextsibling$, $\nextsibling^+$, and
$\nextsibling^*$ are subsets of the
post-order $\leqpost$, and

\item
$\child$, $\child^+$, $\child^*$,
$\nextsibling$, $\nextsibling^+$, and
$\nextsibling^*$ are subsets of the order
$\leqbflr$.
\end{enumerate}

Using Lemma~\ref{lem:ArcConsistency2},
it is straightforward to show that

\begin{theorem}
\label{theo:x_underbar_results}
The axes
\begin{enumerate}
\item
$\child^+$ and $\child^*$ have the $\underbar{X}$-property w.r.t.\ $\ltpre$,
\item
$\following$ has the $\underbar{X}$-property w.r.t.\ $\ltpost$, and
\item
$\child$, $\nextsibling$, $\nextsibling^*$, and $\nextsibling^+$ have the
$\underbar{X}$-property w.r.t.\ $\ltbflr$.
\end{enumerate}
\end{theorem}

\begin{proof}
All proof arguments use Lemma~\ref{lem:ArcConsistency2}.

We first show  that  $\child^*$ has the
$\underbar{X}$-property w.r.t.\ $\ltpre$.
(The  proof for
$\child^+$ is similar.) Consider the nodes $n_0, \dots, n_3$ such that
$n_0 \,\ltpre\, n_1 \,\leqpre\, n_2 \,\ltpre\, n_3$, $\child^*(n_0,n_3)$,
and   $\child^*(n_1,n_2)$.   It   is  simple   to   see  that
$\leqpre$  is   the  disjoint  union   of  $\child^*$  and
$\following$. Therefore, either $\child^*(n_0,n_1)$, which implies
$\child^*(n_0,n_2)$,  or  $\following(n_0,n_1)$.  The  latter
case would  yield $n_3  \ltpre n_1$, a  contradiction.

Next, we show that $\following$ has the $\underbar{X}$-property w.r.t.\
$\ltpost$.
Assume that
\[
n_0  \,\ltpost\,  n_1  \,\leqpost\,  n_2  \,\ltpost\,  n_3
\]
and $\following(n_1,n_2)$, $\following(n_0,n_3)$. Clearly,
the relation
$\leqpost$  is  the  disjoint  union of  $\following$  and  the
inverse of $\child^*$. Since $n_0 \ltpost n_1$ is true,
either
$\child^*(n_1,n_0)$
or         $\following(n_0,n_1)$ must hold.
In   both   cases  it   follows   that
$\following(n_0,n_2)$.
Thus, $\following$  has the $\underbar{X}$-property w.r.t.\ $\ltpost$.

The fact that \child has the $\underbar{X}$-property w.r.t.\ $\ltbflr$
follows vacuously from the
characterization of Lemma~\ref{lem:ArcConsistency2}:
Assume that $n_0 \ltbflr n_1 \lebflr n_2 \ltbflr n_3$ and that
$\child(n_1, n_2)$
(thus $n_0 \ltbflr n_1 \ltbflr n_2 \ltbflr n_3$) and
$\child(n_0,n_3)$.
Because of $\child(n_0,n_3)$, the node $n_1$ is at most one level
below $n_0$ in the tree.
There are two cases,
(1) $\following(n_0,n_1)$ and $n_0$, $n_1$ are on the same level in the
tree or
(2) $\following(n_1,n_3)$ and $n_1$, $n_3$ are on the same level in the tree.
In case (1), since $n_3$ is a child of $n_0$ and $n_2$ is a child of $n_1$,
$n_3 \ltbflr n_2$, contradiction. 
In case (2), since $n_2$ is a child of $n_1$, $n_2$ is one level below $n_3$ in
the tree and thus $n_3 \ltbflr n_2$, contradiction.

It is  easy to verify
that   $\nextsibling$,   $\nextsibling^*$, and $\nextsibling^+$
have the $\underbar{X}$-property w.r.t.\ $\ltbflr$ using
Lemma~\ref{lem:ArcConsistency2}.
\end{proof}

Now, it follows immediately from Theorem~\ref{theo:efficient_eval} that

\begin{corollary}[\cite{GK2002b}]
\label{prop:descendant_dos_ptime}
Conjunctive queries over
$$
\tau_1 := \langle \labela, \child^+, \child^* \rangle
$$
are in polynomial time w.r.t.\ combined complexity.
\end{corollary}

\begin{corollary}
\label{theo:following_ptime}
Conjunctive queries over the signature
$$
\tau_2 := \langle \labela, \following \rangle
$$ 
are in polynomial time w.r.t.\ combined complexity.
\end{corollary}

\begin{corollary}
\label{theo:child_ns_nsplus_nsstar_ptime}
Conjunctive queries over the signature
\[
\tau_3 := \langle \labela, \child, \nextsibling,
\nextsibling^*, \nextsibling^+ \rangle
\]
are in polynomial time w.r.t.\ combined complexity.
\end{corollary}

\begin{figure}
\begin{center}
\epsfig{width=8.5cm, file=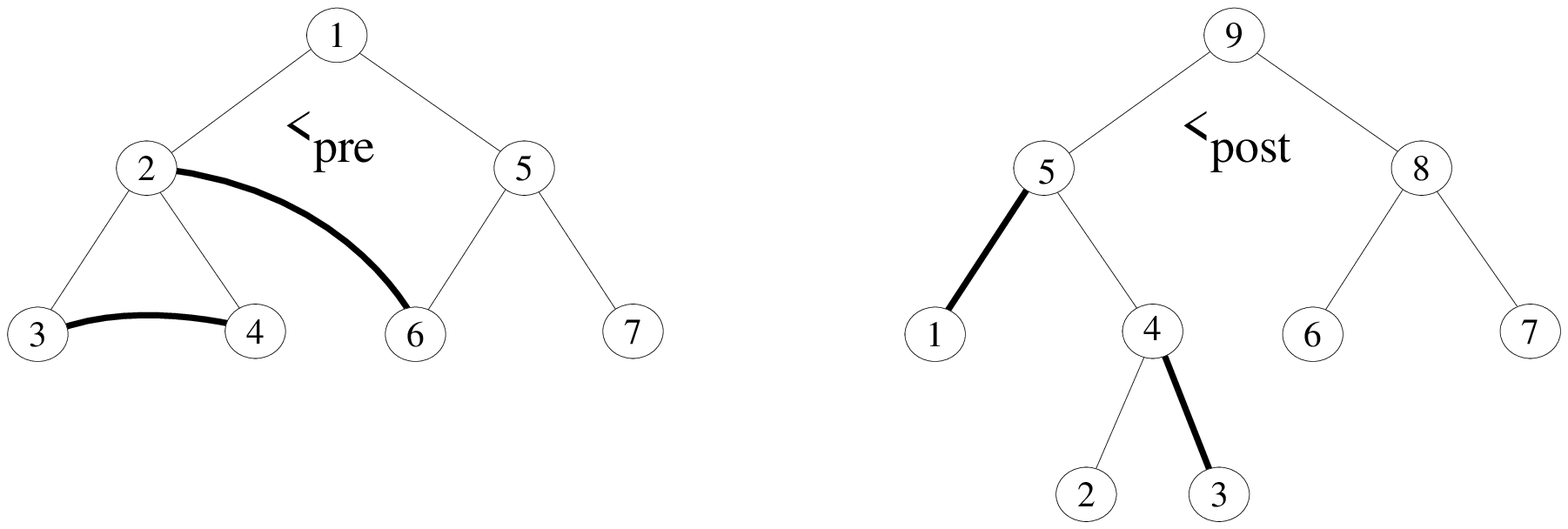} \\
$\mbox{(a)} \hspace{4.3cm} \mbox{(b)}$
\end{center}
\vspace{-2mm}
\caption{(a) $\following$ does not have the
$\underbar{X}$-property w.r.t.\ $\ltpre$; (b) $\descendant^{-1}$ and
$\descendantorself^{-1}$ do not have the $\underbar{X}$-property w.r.t.\
$\ltpost$.}
\label{fig:counterexamples1}
\end{figure}

\begin{example}
\em
The remaining inclusions between axis relations and total orders
introduced at the beginning of this section do not extend to
the $\underbar{X}$-property. 
For example, Figure~\ref{fig:counterexamples1}~(a) illustrates that
$\following$ does not satisfy Lemma~\ref{lem:ArcConsistency2}  w.r.t.\
pre-order $\ltpre$. (While $2 \ltpre 3 \ltpre 4 \ltpre 6$,
$\following(2,6)$ and $\following(3,4)$ hold,
$\following(2,4)$ does not hold.)

Figure~\ref{fig:counterexamples1}~(b) shows that
$\descendant^{-1}$  and    $\descendantorself^{-1}$  do not  satisfy
the condition of Lemma~\ref{lem:ArcConsistency2inv} w.r.t.\
post-order $\ltpost$. (While $1 \ltpost 3 \ltpost 4 \ltpost 5$,
$\descendant^{-1}(1,5)$ and
$\descendant^{-1}(3,4)$ hold, $\descendant^{-1}(1,4)$ does not hold.)
%
%
\end{example}

For total order $<$, let
$
\textit{Succ}_< :=
\{ \tuple{x,y} \mid x < y \; \land \nexists z\; x < z < y \}.
$
It is trivial to verify that $\textit{Succ}_<$, $<$, and $\leq$
have the $\underbar{X}$-property w.r.t.\ $<$.
Thus, we may for instance add the relations $<_{pre}$ (document order) and
$\textit{Succ}_{<_{pre}}$
(``next node in document order'') to $\tau_1$,
while retaining polynomial-time combined complexity.



\section{NP-Hardness Results}
\label{sect:np_results}

In this section,  we study the complexity of  the conjunctive query evaluation
problem for  the remaining sets of axis  relations.
For  all cases for  which our  techniques based  on the
$\underbar{X}$-property  do not yield a polynomial-time  complexity result, we
are able  to prove NP-hardness. All NP-hardness results hold already  for
query complexity, i.e., in a setting where the data tree, and thus in
particular the labeling alphabet, is fixed and only the query is assumed
variable.

All  reductions are  from {\em  one-in-three 3SAT}\/,  which is  the following
NP-complete problem: Given a set $U$ of variables, a collection
${\mathbf C}$  of clauses over $U$ such  that each clause $C  \in {\mathbf C}$
has $|C| =  3$, is there a truth  assignment for $U$ such that  each clause in
${\mathbf C}$ has  exactly one true literal?  1-in-3  3SAT remains NP-complete
if all clauses contain only positive literals \cite{Sch1978}.

Below, we will use shortcuts of the form $\chi^k(x,y)$, where
$\chi$ is an axis, in queries
to denote chains of $k$ $\chi$-atoms leading from variable
$x$ to $y$. For example, $\child^2(x,y)$ is a shortcut for
$\child(x,z), \child(z,y)$, where $z$ is a new variable.

The first theorem strengthens a known result for combined complexity
\cite{MSB2001} to query complexity.

\begin{theorem}
\label{theo:NPhardnessChildDescendant}
Conjunctive queries over the signatures
\begin{eqnarray*}
\tau_4 &:=& \langle \labela, \child, \child^+ \rangle \\
\tau_5 &:=& \langle \labela, \child, \child^* \rangle
\end{eqnarray*}
are NP-complete w.r.t.\ query complexity. 
\end{theorem}

\begin{proof}
Here, as in all other proofs of this section, we only need to show NP-hardness.
Let $C_1,\ldots,C_m$ be a 1-in-3 3SAT instance with positive 
literals only. We assume that $C_i$ is an ordered sequence of three
positive literals. 
We may assume without loss of generality that no clause contains a particular
literal more than once.
We reduce this instance to one of the Boolean conjunctive query evaluation
problem for $\tau_4$ ($\tau_5$).

The fixed data tree over alphabet $\{X,Y,L_1,L_2,L_3\}$
is shown in Figure~\ref{fig:child_childs}.

For the query, we introduce variables $x_i, y_i$ for
$1\leq i\leq m$ and in addition a variable $z_{k,l,i,j}$
whenever the $k$-th literal of $C_i$ coincides with
the $l$-th literal of $C_j$ ($1\leq i\leq m$, $1\leq j\leq m$,
$i\neq j$, $1\leq k,l\leq 3$).

The Boolean query consists of the following atoms:
\begin{itemize}
\item
for $1\leq i\leq m$,
\[
X(x_i), \; Y(y_i), \; \child^3(x_i, y_i),
\]

\item
for each variable $z_{k,l,i,j}$,
\[
L_k(z_{k,l,i,j}), \child^\circ(y_i,z_{k,l,i,j}),
\mbox{\sl Child}^{8+k-l}(x_j,z_{k,l,i,j})
\]
where $\circ$ is ``$+$'' on signature $\tau_4$ and ``$\ast$'' on $\tau_5$.
\end{itemize}

\begin{figure}
\begin{center}
\input{child_childs_data}
\end{center}
\label{fig:QueryComplexityNeu}
\vspace{-5mm}
\caption{Data tree of the proof of
Theorem~\ref{theo:NPhardnessChildDescendant}.}
\label{fig:child_childs}
\end{figure}
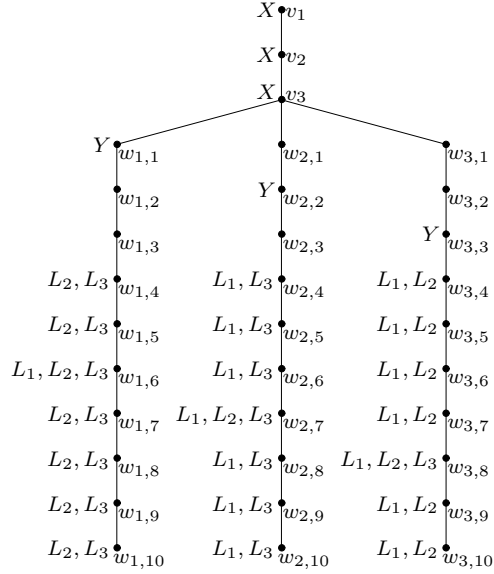

``$\Rightarrow$''.
To prove correctness of the reduction, we first show that
given any solution mapping
$
\sigma : \{1,\ldots,m\}\rightarrow \{1,2,3\}
$
of $C_1,\ldots,C_m$ (i.e., $\sigma(i) = k'$ iff $\sigma$ selects the $k'$-th
literal from $C_i$) we can define a satisfaction $\theta$ of the query.
We first define a valuation $\theta$ of our query and then show
that all query atoms are satisfied. We set
\begin{itemize}
\item
   $\theta(x_i) := v_{\sigma(i)}$ for $1 \leq i\leq m$,
\item
  $\theta(y_i) := w_{\sigma(i),\sigma(i)}$ for $1\leq i\leq m$, and
\item
for each variable $z_{k,l,i,j}$,
  $\theta(z_{k,l,i,j}) := w_{\sigma(i), 5+k-l+\sigma(j)}$.
%
\end{itemize}

We now prove that $\theta$ is  a satisfaction of the query.
Our choice of $\theta$
implies that the variables $x_i$ and $y_i$ are mapped to nodes
with labels $X$ and $Y$, respectively. Furthermore, $\theta(y_i) =
w_{\sigma(i),\sigma(i)}$ can be reached from $\theta(x_i) = v_{\sigma(i)}$
with three child-steps. For any variable of the form $z_{k,l,i,j}$,
$\theta(z_{k,l,i,j}) = w_{\sigma(i),5+k-l+\sigma(j)}$ is always
a $\child^\circ$ of $w_{\sigma(i),\sigma(i)}$.
If $\sigma(i) \neq k$, then
$\theta(z_{k,l,i,j})= w_{\sigma(i),5+k-l+\sigma(j)}$
has label $L_k$ because $4 \leq 5+k-l+\sigma(j) \leq 10$ and
the nodes $w_{\sigma(i),4}, \dots, w_{\sigma(i), 10}$ all have (at least)
the two labels $L_{k'}$ for which $\sigma(i) \neq k'$.
If $\sigma(i)=k$, then $\sigma(j) = l$. By going $8+k-l$ steps downward from
$v_{\sigma(j)}$, passing through $w_{k,k}$, we reach node $w_{k,5+k}$, which
has label $L_k$. Since
$\theta(z_{k,l,i,j}) = w_{\sigma(i),5+k-l+\sigma(j)} = w_{k,5+k}$,
the query atoms $\child^{8+k-l}(x_j,z_{k,l,i,j})$ are satisfied.
Therefore, $\theta$ is indeed a satisfaction of our query.

``$\Leftarrow$''.
To finish the proof we show that from any satisfaction $\theta$
of the query we obtain a corresponding solution for the 1-in-3 3SAT
instance $C_1,\ldots,C_m$.
If $\theta(x_i) = v_k$, we interpret
this as the $k$-th literal of clause $C_i$ being chosen to be true.
Obviously, under any valuation of the query, we select
precisely one literal from each clause $C_i$.
We have to verify that if a literal $L$ occurs
in two clauses $C_i$ and $C_j$ and we select $L$ in $C_i$, we also
select $L$ in $C_j$.
Let $L$ be the $k$-th literal of $C_i$ and let $\theta(x_i) = v_k$
(i.e., $L$ is selected in $C_i$).
Then
$\theta(z_{k,l,i,j})= w_{k,5+k}$
because that is the only node below $\theta(y_i) = w_{k,k}$
that has label $L_k$.
The query contains the atom
$\child^{8+k-l}(x_j, z_{k,l,i,j})$
for variable $z_{k,l,i,j}$.
From node $w_{k,5+k}$, by $8+k-l$ upward steps we arrive at node $v_l$.
Hence $\theta(x_j) = v_l$, and we select $L$ from clause $C_j$.

Some nodes in the data tree carry multiple labels.
However, since the $\child$  axis is available in
both $\tau_4$ and $\tau_5$, multiple labels can be
eliminated by pushing them down to new children in the data tree and
modifying the queries accordingly.
\end{proof}

\begin{theorem}
\label{theo:NPhardnessFollowingChild}
Conjunctive queries over the signature
\[
\tau_6 := \langle \labela, \child, \following \rangle
\]
are NP-complete w.r.t.\ query complexity. 
\end{theorem}

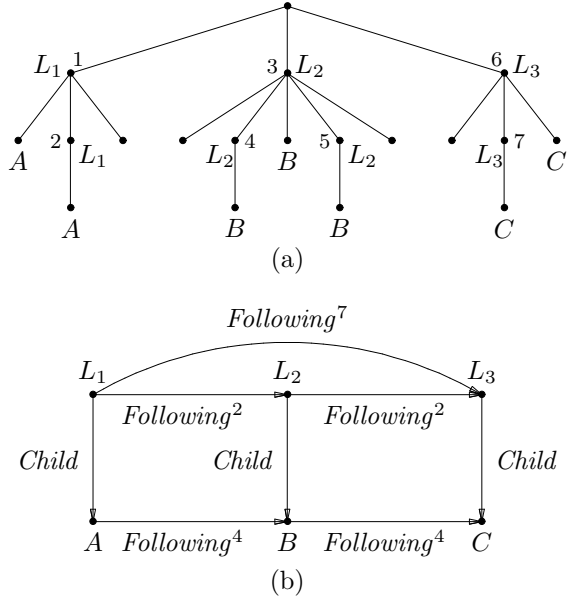
\begin{figure}
\begin{center}
\input{child_following.tex}
\end{center}
\vspace{-4mm}
\caption{Clause gadget of proof of
Theorem~\ref{theo:NPhardnessFollowingChild}.}
\label{fig:followingchildNP_chr}
\end{figure}

\begin{proof}
Figure~\ref{fig:followingchildNP_chr} shows (a) the fragment
of a data tree and (b) a query over the labeling alphabet
$\Sigma = \{ A, B, C, L_1, L_2, L_3 \}$.

Observe that the labels
$L_1$, $L_2$, and $L_3$ occur only once each in
Figure~\ref{fig:followingchildNP_chr}~(b). We will refer to the nodes
(= query variables)
labeled $L_1$, $L_2$, and $L_3$ by $v_1$, $v_2$, and $v_3$, respectively.
For the following discussion, we have annotated some of the nodes of the data
tree with numbers (1--7).
Below, node $1$ (resp.  $3$, $6$) is called  the {\em topmost
position} of variable $v_1$ (resp.  $v_2$, $v_3$). We start with two
simple observations.
\begin{enumerate}
\item
{\em In any satisfaction $\theta$ of the query on the data tree,
at most one of the
variables $v_1$,  $v_2$, and $v_3$  is mapped to its  topmost position
under  $\theta$.} In  fact, assume,  e.g., that  $\theta(v_1)=1$. From
node  1, node  3 (resp.  6) cannot  be reached  by a  sequence  of $2$
(resp. $7$)  $\following$-steps. Hence we  have $\theta(v_2)\neq
3$ and $\theta(v_3)\neq 6$.

\item
{\em In any solution $\theta$ of the problem, at least one of the
variables $v_1$,  $v_2$, and $v_3$  is mapped to its  topmost position
under  $\theta$.}   In  fact,  assume  that  $\theta(v_1)   =  2$  and
$\theta(v_2) \neq 3$. The  atoms in the query (in particular, on
the variables corresponding to nodes on the bottom of the query graph) require
that  $\theta(v_2)  \neq 4$.  Hence  $\theta(v_2)  =  5$ is  the  only
remaining  possibility.  But now  the query requires
that $\theta(v_3) \neq 7$. Hence $\theta(v_3) =6$.
\end{enumerate}

Thus, precisely the three partial assignments 
\begin{eqnarray*}
&(a)& \theta(v_1) := 1,\; \theta(v_2) := 4,\; \theta(v_3) := 7 \\
&(b)& \theta(v_1) := 2,\; \theta(v_2) := 3,\; \theta(v_3) := 7 \\ 
&(c)& \theta(v_1) := 2,\; \theta(v_2) := 5,\; \theta(v_3) := 6
\end{eqnarray*} 
can be extended to a satisfaction of the query.
Precisely one of the variables $v_1$, $v_2$, and $v_3$ is mapped to 
its topmost position under each of the above assignments. Conversely, for each
variable there is a satisfying assignment in which it takes its topmost
position.

Given a clause  $C$, an {\em ordered}\/ list of three positive literals, we
interpret a satisfaction $\theta$ in which  variable $v_k$ is mapped  to
its  topmost position as  the selection
of the $k$-th literal from $C$ to be true.
The encoding described above thus assures that 
exactly  one variable  of clause $C$  is
selected and becomes true.

Now consider a 1-in-3 3SAT problem instance over positive literals with clauses
$C_1, \ldots,  C_m$.
We encode such an instance as a conjunctive query over $\tau_6$ and a fixed
data tree over labeling alphabet
$\Sigma = \{ A,B,C, L_1, L_2, L_3 \}$. This tree
consists of two copies of the tree of
Figure~\ref{fig:followingchildNP_chr}~(a) under a common root, i.e.,
\begin{center}
\input{child_following_qc.tex}
\end{center}
where $T$ denotes the tree of Figure~\ref{fig:followingchildNP_chr}~(a).

\begin{table}
\begin{center}
\begin{tabular}{rr|ccc}
&$k \backslash l$ & 1 & 2 & 3 \\
\hline
& 1 & 10 & 13 & 18 \\
& 2 & 5 & 8 & 13 \\
& 3 & 2 & 5 & 10
\end{tabular}
\end{center}
\vspace{-4mm}
\caption{The function $\textit{NAND}(k,l)$.}
\label{tab:distances}
\end{table}

The query is obtained as follows.
Each clause $C_i$ is represented using two copies of the query
gadget of Figure~\ref{fig:followingchildNP_chr}~(b) (a ``left'' copy
$Q_i$ and a ``right'' copy $Q_i'$).
We wire the two sets of subqueries $Q_1, \dots, Q_m, Q_1', \dots, Q_m'$ as
follows.

Consider first the integer function $\textit{NAND}(k, l)$
defined by Table~\ref{tab:distances}.
We can enforce that two variables, $x$ and $y$, labeled $L_k$ and $L_l$ in
their respective subqueries, cannot both match the topmost node labeled
$L_k$ resp.\ $L_l$ in the left,  respective right, part of the data tree
by adding an atom of the form $\following^{\textit{NAND}(k,l)}(x,y)$
to the query.

For each pair of clauses $C_i$, $C_j$,
variable $x$  such that $Q_i$ (resp., $Q_i'$)
contains the unary atom $L_k(x)$,
and variable $y$
such that $Q_j'$ (resp., $Q_j$) contains the unary atom $L_l(y)$, if
\begin{itemize}
\item
the $k$-th literal of $C_i$ occurs also in $C_j$ and

\item
the $k$-th literal of $C_i$ and
the $l$-th literal of $C_j$ are different,
\end{itemize}
then we add an atom $\following^{\textit{NAND}(k,l)}(x, y)$ to the query.

These query atoms make sure that  if a  literal is chosen  to be true  in one
clause, it must be  selected to be true in all other  clauses as well.  In the
case that $i=j$, the idea is to make sure that both copies of the query gadget
of each  clause, $Q_i$ and $Q_i'$,  make the same choice  of selected literal.
The case that $i \neq j$  models the interaction between distinct clauses. Thus
our query assures that  each literal is assigned the same  truth value in all
clauses.

Using two copies of the query gadget for each clause and two copies of the
tree gadget of Figure~\ref{fig:followingchildNP_chr}~(a) in the data tree is
necessary, as we cannot use $\following^k$-atoms to make sure that two
variables are not both assigned their topmost positions in the data tree
(corresponding to ``true'') if the data tree consists just of the tree of
Figure~\ref{fig:followingchildNP_chr}~(a) and these two topmost positions in
the data tree coincide.

This concludes the construction, which can be easily implemented to run in
logarithmic space.
It is not difficult to verify that the fixed data tree satisfies the query
precisely if the 1-in-3 3SAT instance is satisfiable.
\end{proof}

\begin{theorem}
\label{theo:NPhardnessFollowingChildTrans}
Conjunctive queries over the signatures 
\begin{eqnarray*}
\tau_7 &:=& \langle \labela, \child^+, \following \rangle, \\
\tau_8 &:=& \langle \labela, \child^*, \following \rangle
\end{eqnarray*}
are NP-complete w.r.t.\ query complexity. 
\end{theorem}

\begin{proof}
The same encoding as in the  previous proof can be used, with the only
difference  that  $\child^*$  resp.\  $\child^+$ is  used  instead  of
$\child$ in  the query.   In fact, if  the topmost position  for $v_1$
(resp.\ $v_2$,  $v_3$) is chosen,  there are two possible  matches for
``A'' (resp.\, three for ``B'' and  two for ``C''). This has no impact
on the constraints across clauses  or the constraints that at most one
variable of each  clause is assigned to its  topmost position. To make
sure that {\em at least}\/ one variable of each clause is assigned its
topmost  position, the constraints  of the  query assure  that either
``A'',  ``B'', or ``C''  are assigned  to the  correspondingly labeled
node at  depth two  in the  subtree of the  clause (rather  than depth
three).
\end{proof}

Since $\following$ can be defined by a conjunctive query over
$\child^*$ and $\nextsibling^+$ (see Equation~(\ref{eq:following}) in
Section~\ref{sect:preliminaries}),

\begin{corollary}
\label{cor:child_star_nextsibling_plus}
Conjunctive queries over the signature
\[
\tau_9 := \langle \labela, \child^*, \nextsibling^+ \rangle
\]
are NP-complete w.r.t.\ query complexity. 
\end{corollary}

\begin{theorem}
\label{theo:child_star_nextsibling}
Conjunctive queries over the signature
\[
\tau_{10} := \langle \labela, \child^*, \nextsibling \rangle
\]
are NP-complete w.r.t.\ query complexity. 
\end{theorem}

\begin{proof}
If we replace $\following$ by
\[
\following'(x,y) := \exists z_1 \exists z_2 \; \child^*(z_1,x) \land
\nextsibling(z_1, z_2) \land \child^*(z_2,y),
\]
we can reuse the construction of the proof of
Theorem~\ref{theo:NPhardnessFollowingChild} (in the modified form
of the proof of Theorem~\ref{theo:NPhardnessFollowingChildTrans}).
\end{proof}

\begin{theorem}
\label{theo:child_star_nextsibling_star}
Conjunctive queries over the signature
\[
\tau_{11} := \langle \labela, \child^*, \nextsibling^* \rangle
\]
are NP-complete w.r.t.\ query complexity. 
\end{theorem}

\begin{figure}
\begin{center}
\input{auxiliary_nodes.tex}
\end{center}
\vspace{-4mm}
\caption{Data tree of proof of Theorem~\ref{theo:child_star_nextsibling_star}.}
\label{fig:auxiliary_nodes}
\end{figure}

\begin{proof}
The proof basically uses the same argument as
Corollary~\ref{cor:child_star_nextsibling_plus}. However, to
deal with $\nextsibling^*$ rather than $\nextsibling^+$,
we need a way to ensure that $\nextsibling^*$ moves at least one step
to the right. We thus replace each occurrence of $\following$ in the
construction of  the proof of
Theorem~\ref{theo:NPhardnessFollowingChild} by
\begin{multline*}
\following'(x,y) := \exists z_1 \exists z_2 \exists z_3\;
\child^*(z_1,x) \land
\nextsibling^*(z_1, z_2) \, \land \\
H(z_2) \land \nextsibling^*(z_2, z_3) \land \child^*(z_3,y).
\end{multline*}

The modified data tree is as shown in
Figure~\ref{fig:auxiliary_nodes}. It uses specially labeled auxiliary
nodes inserted between each pair of adjacent siblings in the data tree
of the proof of Theorem~\ref{theo:NPhardnessFollowingChild}.
\end{proof}

\begin{theorem}
\label{theo:child_plus_nextsibling_all}
Conjunctive queries over the signatures 
\begin{eqnarray*}
\tau_{12} &:=& \langle \labela, \child^+, \nextsibling   \rangle,\\
\tau_{13} &:=& \langle \labela, \child^+, \nextsibling^+ \rangle, \\
\tau_{14} &:=& \langle \labela, \child^+, \nextsibling^* \rangle
\end{eqnarray*}
are NP-complete w.r.t.\ query complexity. 
\end{theorem}

\begin{proof}
The proofs are analogous to the proofs for the respective signatures
with $\child^*$ rather than $\child^+$, except that we modify the
respective data trees as follows: Each edge $\tuple{u,w}$ is replaced
by two edges $\tuple{u,v}, \tuple{v,w}$, where $v$ is a new node.
Now, to make a $\following$-step between two nodes corresponding to original
tree nodes, we can use the relation
\[
\following''(x,y) := \exists z_1 \exists z_2 \exists z_3\;
\child^+(z_1,x) \land
\nextsibling^\alpha(z_2, z_3) \land \child^+(z_3,y).
\]
where $\alpha$ is ``$1$'' for $\tau_{12}$, ``$+$'' for $\tau_{13}$,
and ``$*$'' for $\tau_{14}$.
\end{proof}

\begin{theorem}
\label{theo:NPhardnessFollowingNextsibling}
Conjunctive queries over the signatures 
\begin{eqnarray*}
\tau_{15} &:=& \langle \labela, \following, \nextsibling   \rangle,\\
\tau_{16} &:=& \langle \labela, \following, \nextsibling^+ \rangle,\\
\tau_{17} &:=& \langle \labela, \following, \nextsibling^* \rangle
\end{eqnarray*}
are NP-complete w.r.t.\ query complexity. 
\end{theorem}

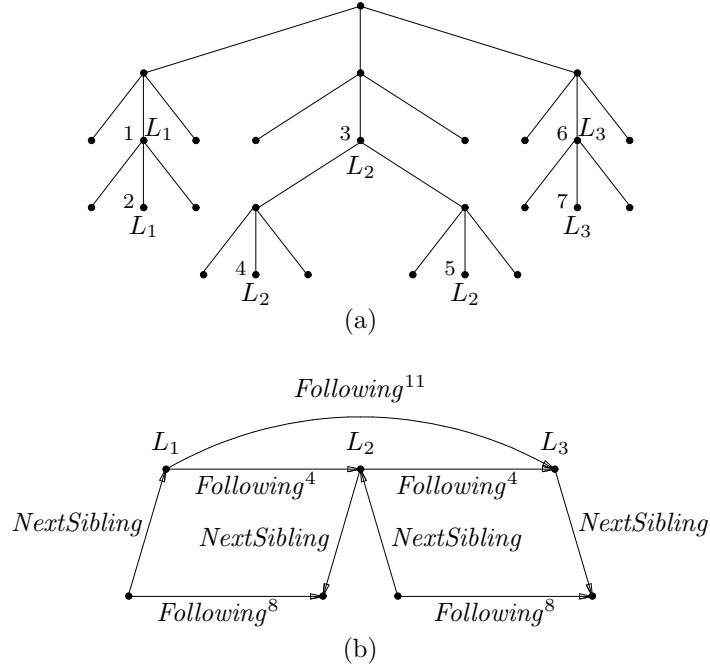
\begin{figure}
\begin{center}
\input{following_nextsibling.tex}
\end{center}
\vspace{-6mm}
\caption{Encoding the selection of exactly one of the positive
literals of a clause as a conjunctive query over signature $\tau_{15}$.}
\label{fig:followingnextsiblingNP}
\end{figure}

\begin{proof}
We first look at signature $\tau_{15}$. Consider the data tree shown
in Figure~\ref{fig:followingnextsiblingNP}~(a) and the query of
Figure~\ref{fig:followingnextsiblingNP}~(b).

As in the proof of Theorem~\ref{theo:NPhardnessFollowingChild},
there is again one variable per label $L_1$ ($L_2$, $L_3$), which we call
$v_1$ ($v_2$, $v_3$).
Again, at most one variable $v_1$, $v_2$, and $v_3$
can be mapped to  its topmost position.
The  query shown in Figure~\ref{fig:followingnextsiblingNP}~(a) requires that
precisely the partial assignments 
\begin{eqnarray*}
&& \theta(v_1) := 1,\; \theta(v_2) := 4,\; \theta(v_3) := 7\\
&& \theta(v_1) := 2,\; \theta(v_2) := 3,\; \theta(v_3) := 7\\ 
&& \theta(v_1) := 2,\; \theta(v_2) := 5,\; \theta(v_3) := 6
\end{eqnarray*} 
can be extended to solutions of the query.

This provides us with an encoding for the selection of exactly one literal 
from a given clause with three positive literals.
The full reduction from 1-in-3 3SAT over positive literals
can be obtained analogously to the proof of
Theorem~\ref{theo:NPhardnessFollowingChild}.

The same reduction can be used to prove the corresponding result for the 
signatures $\tau_{16}$ and $\tau_{17}$.
\end{proof}

\nop{
\begin{remark}
\em
Note that our NP-hardness resuls -- which characterize
maximal polynomial signatures -- also exhibit that
relations from one polynomial signature cannot be defined as conjunctive
queries over a different polynomial signature, unless $P = NP$.
For example, $\following$ cannot be defined as a conjunctive query over
$\child^*$, $\child^+$, $\nextindocorder$, $\nextindocorder^+$, or
$\nextindocorder^*$.
\end{remark}
} 


\section{Expressiveness}
\label{sect:expressiveness}

In this section, we study the expressive power of conjunctive queries
over trees. The main result is that for each conjunctive query over
trees, an equivalent  {\em acyclic positive query}\/ (APQ)  can be
found. However, these APQs are in general exponentially larger.
As we show in Section~\ref{sect:succinctness}, this is necessarily so.

We introduce a number of technical notions.
In Section~\ref{sect:preliminaries}, query graphs were introduced
as directed (multi)graphs. Below, we will deal with two kinds of
cycles in query graphs; {\em directed cycles}\/, the standard notion of cycles
in directed graphs, and the more general
{\em undirected cycles}\/, which are cycles
in the undirected {\em shadows}\/ of query graphs.\footnote{The shadow of
a directed graph is obtained by replacing each directed edge from node $u$ to
node $v$ by an undirected edge between $u$ and $v$.}
The standard notion of conjunctive
query acyclicity in the case that relations are at most binary
refers to the absence of undirected cycles from the shadow of the query graph.

Let $F \subseteq \ouraxes$ be a set of axes. We  denote  by  CQ[$F$]  the
conjunctive  queries over  signature $\tuple{\labela,  F}$.
By PQ[$F$] we denote  the positive (first-order) queries (written  as finite
unions of conjunctive queries)  over $F$. We  denote the acyclic positive
queries -- that is, unions of {\em acyclic}\/ conjunctive queries --
over $F$ by APQ[$F$].

\begin{remark}
\em
Given a  set of XPath axes $F$, let $F^{-1}$  denote their inverses (e.g.,
$\parent$  for  $\child$;  see  \cite{XPath1} for the names of the inverse
XPath axes).   
It is easy to
show that for any set $F$ of XPath axes,
positive Core XPath[$F \cup F^{-1}$],
the positive, navigation-only fragment of the XPath language
\cite{GKPS2005},
captures the unary APQ[$F$]
on trees in which each node has (at most) one label.
No proof of this is presented here because
a formal definition of XPath is tedious and the result follows 
immediately from such a definition. (Positive Core
XPath queries are acyclic and support logical disjunction.)
\end{remark}

\begin{figure*}
\vspace{-1cm}
\hspace{-2.8cm}
\input{cq_apq.tex}
\vspace{-6mm}
\caption{Translation of a conjunctive query into an APQ.}
\label{fig:cq_apq}
\end{figure*}
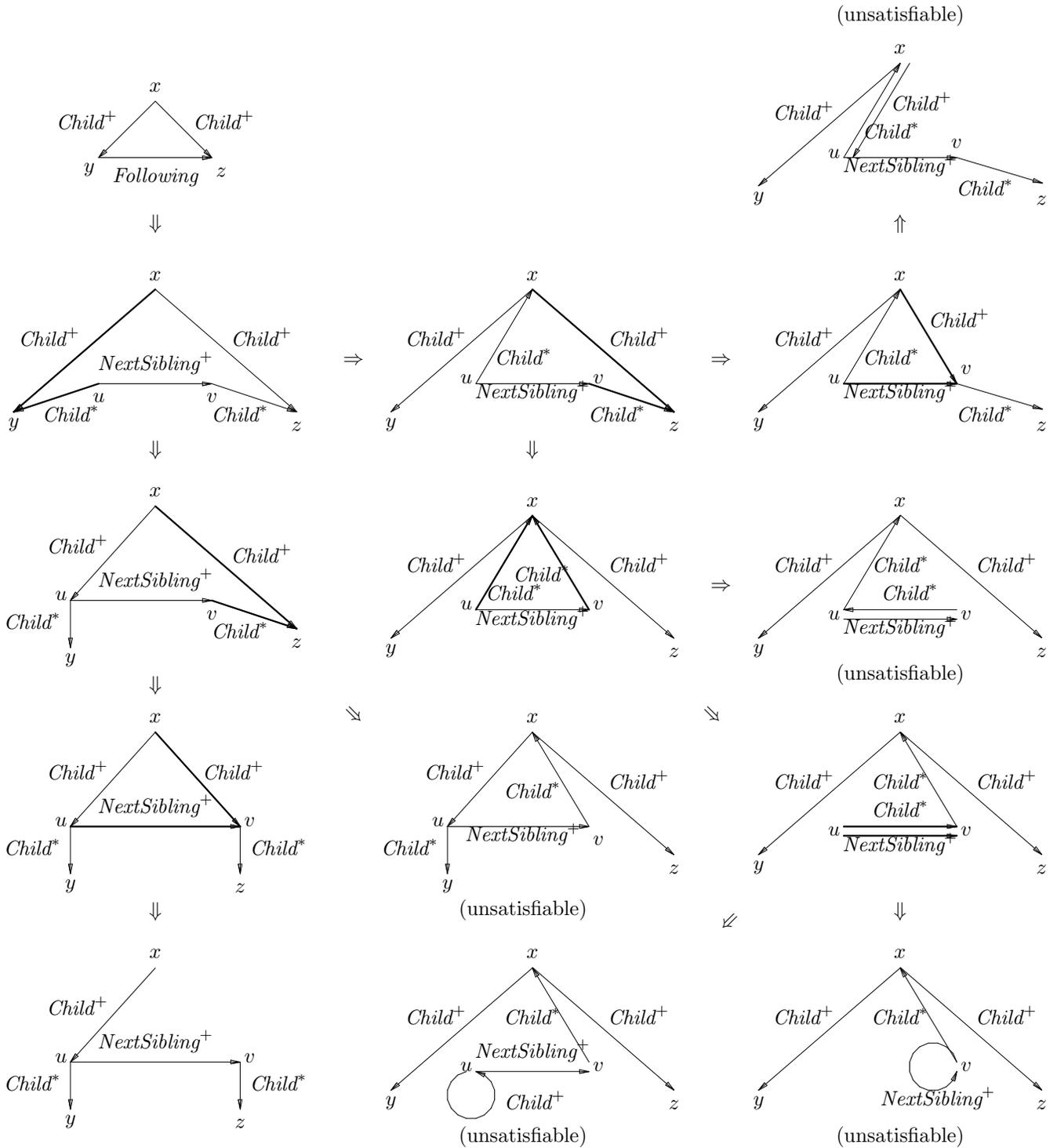

Before we can get to the main result of this section,
Theorem~\ref{theo:CQ_in_APQ},
we need to define the notion of
a {\em join lifter}\/, for which we will subsequently give an intuition
and an example. After providing two lemmata, we will be able to prove
Theorem~\ref{theo:CQ_in_APQ}. The proof of the main result employs a
rewrite system whose workings are illustrated in a detailed example
in Figure~\ref{fig:cq_apq} (Example~\ref{ex:cq_apq}).
The reader may find it helpful to start with
that example before reading on sequentially from here.

\begin{definition}
\label{def:join_lifter}
\em
Let $F$ be a set of binary relations.
A positive quantifier-free formula $\psi_{R,S}(x,y,z)$
in Disjunctive Normal Form (DNF)
is called a {\em join lifter}\/ over $F$ for binary relations $R$ and $S$ if
\begin{enumerate}
\item
each conjunction of $\psi_{R,S}(x,y,z)$ is of one of the following five forms:
\begin{enumerate}
\item
$P(x,y) \land P'(y,z)$

\item
$P(y,x) \land P'(x,z)$

\item
$P(x,z) \land y = z$

\item
$P(y,z) \land x = z$

\item
$P(x,z) \land x = y$
\end{enumerate}
where $P, P' \in F$ and

\item
for all trees $\strucA$ and nodes $a, b, c$,
\[
(\strucA, a, b, c) \vDash \phi_{R, S}[a, b, c]
\;\Leftrightarrow\;
(\strucA, a, b, c) \vDash \psi_{R, S}[a, b, c].
\]
where $\phi_{R,S}(x, y, z) = R(x, z) \land S(y, z)$.

(Subsequently, we will write this as
$\psi_{R,S} \equiv \phi_{R,S}$.)
\end{enumerate}
\end{definition}

A join lifter $\psi_{R,S}$ can be used to rewrite a
conjunctive query $Q$ that contains atoms $R(x,z), S(y,z)$ -- the role of
such pairs of atoms will be clarified below, in the proof of
Lemma~\ref{lem:CQ_in_APQ}-- into a union of
conjunctive queries (one conjunctive query for each conjunction $C$ of the
DNF formula
$\psi_{R,S}$, by replacing $R(x,z), S(y,z)$ by $C$) such that none of
the conjunctive queries obtained is larger than $Q$.
In fact, each of conjunctive queries obtained is either
shortened (because equality atoms $v=w$ in conjunctions of form (c), (d) or (e)
can be eliminated after
substituting variable $v$ by $w$ everywhere in the query)
or the join on $z$ is intuitively lifted ``up'' in the query graph using
a conjunction of form (a) or (b).

\begin{example}
\em
The formula
\[
\psi_{\child, \nextsibling}(x,y,z) = \child(x,y) \land \nextsibling(y,z)
\]
is a join lifter for $\child$ and $\nextsibling$ because it satisfies
the syntactic requirement (1) -- the formula is a single conjunction of form
(a) --  and the equivalence (2)
\[
\psi_{\child, \nextsibling}(x,y,z) \equiv 
\phi_{\child, \nextsibling}(x,y,z) = \child(x,z) \land \nextsibling(y,z).
\]
of Definition~\ref{def:join_lifter}.
Conjunctions of form (a) such as this one lift the join occurring in
$\phi_{\child, \nextsibling}$ one level up in the
query graph -- here from variable $z$ in $\phi_{\child, \nextsibling}$
to variable $y$ in $\psi_{\child, \nextsibling}$ when rewriting $\phi_{\child,
  \nextsibling}$ by $\psi_{\child, \nextsibling}$.
\end{example}

Moving joins {\em upward}\/
is only meaningful in queries whose query graphs do not have directed cycles.
As demonstrated by the following lemma, such cycles can always be eliminated.

\begin{lemma}
\label{lem:directed_cycles}
Let $Q$ be a $CQ[\ouraxes]$ that contains a directed cycle
\[
R_1(x_1, x_2), R_2(x_2, x_3), \dots, R_{k-1}(x_{k-1}, x_k), R_k(x_k, x_1).
\]
If $R_1, \dots, R_k \in \{ \child^*, \nextsibling^* \}$, then
$Q$ is equivalent to the query obtained by adding $x_1 = x_2 = \dots = x_k$
to the body of $Q$.
Otherwise, $Q$ is unsatisfiable.
\end{lemma}

\begin{proof}
The graph of the
relation $\child \cup \nextsibling \cup \following$ is acyclic. Therefore,
a query with a cycle
can only be satisfied if all variables in the
cycle are mapped to the same node. If the cycle contains
an irreflexive axis (any axis besides $\child^*$ and $\nextsibling^*$),
the query is unsatisfiable.
\end{proof}

\begin{lemma}
\label{lem:CQ_in_APQ}
Let $F \subseteq \ouraxes$ be a set of axes and let there be join lifters
$\psi_{R,S}$ over $F$ for each pair $(R,S)$ of relations in $F$.
Then, each $CQ[F]$ can be rewritten into an equivalent $APQ[F]$ in
singly exponential time.
\end{lemma}

\begin{proof}
Given a conjunctive query $Q_0$,
we execute the following algorithm.
Let ${\cal Q}$ be a set of conjunctive queries,
initially $\{ Q_0 \}$.
Repeat the following until the query graphs of all queries in ${\cal Q}$ are
forests.
\begin{enumerate}
\item
Choose any conjunctive query $Q$ from ${\cal Q}$
whose query graph is not a forest.

\item
If $Q$ contains a directed cycle in which a predicate other than
$\nextsibling^*$ or $\child^*$
appears, $Q$ is unsatisfiable (by Lemma~\ref{lem:directed_cycles})
and is removed from ${\cal Q}$.

\item
For each directed cycle in $Q$ that consists exclusively of $\child^*$ and
$\nextsibling^*$ atoms, we identify the variables occurring in it.
(That is, if $x_1, \dots, x_n$ are precisely all the variables of the
cycle, we replace each occurrence of any of these variables in the body
{\em or head} of $Q$ by $x_1$.)
Atoms of the form $\child^*(x_1,x_1)$ or $\nextsibling^*(x_1,x_1)$ are removed.
%

In order to assure safety, we add an atom $\mathit{Node}(x_1)$
if $x_1$ now does not occur in any remaining atom.
(The predicate $\mathit{Node}$ matches any node and can be defined as
$R(x_1, x_1')$, where $R$ is a predicate of the directed cycle just
eliminated -- either $\child^*$ or $\nextsibling^*$ -- and $x_1'$ is
a new variable.)

By Lemma~\ref{lem:directed_cycles}, the outcome of this transformation
is equivalent to the input query.


\item
Now there are no directed cycles left in the query graph, but undirected
cycles may remain.
If $Q$ contains undirected cycles,
we choose a variable $z$ that is in an undirected cycle such that
there is no directed path in the query graph leading from $z$ to
another variable that is in an undirected cycle as well.
(Such a choice is possible because there are no directed cycles in the
query graph.)
The cycle contains two atoms $R(x, z), S(y, z)$.

Now, we use join lifter $\psi_{R,S}$ to replace these two atoms.
Let $\psi_{R,S}$ be the DNF
$\psi_{R,S}^{(1)} \lor \dots \lor \psi_{R,S}^{(k)}$ such that the
$\psi_{R,S}^{(i)}$
are conjunctions of atoms. We create copies $Q_1, \dots, Q_k$ of $Q$ and
replace $R(x, z), S(y, z)$ in each $Q_i$ by  $\psi_{R,S}^{(i)}$.
If $\psi_{R,S}^{(i)}$
contains an equality atom $v = w$,
we replace each occurrence of variable $w$ in $Q_i$
by $v$ and remove the equality atom.
Finally, we replace $Q$ in ${\cal Q}$ by $Q_1, \dots, Q_k$.
\end{enumerate}

First we show that this algorithm indeed terminates.
The elimination of directed cycles -- steps (2) and (3) --
is straightforward,
but we need to consider in more detail how the algorithm deals with undirected
cycles. The idea here is to eliminate undirected cycles from the bottom to
the top (with respect to the direction of edges in the query graph.)
This is done by rewriting bottom atoms $R(x,z), S(y,z)$ of undirected
cycles using the join lifters $\psi_{R,S}$.
While $R(x,z), S(y,z)$ are two binary atoms that involve $z$,
each conjunction in join lifter $\psi_{R,S}$ contains only one binary atom
over $z$ apart from a possible equality atom.
Therefore, each rewrite step either removes $z$ from at least one cycle
or identifies $z$ with either $x$ or $y$ via an equality atom
(which, for our purposes, means to remove $z$ entirely, and thus also from
any cycle it appears in).

Let $|V|$ be the number of variables and $|E|$ be the number of
binary atoms in $Q_0$.
The number of atoms in a conjunctive query never increases by the
rewrite steps (each conjunction of the formulae $\psi_{R,S}$ is of length two).
For a given bottommost variable $z$ of the query graph that is in an
undirected cycle, there can be at most $|E|$ incoming edges
(i.e., binary atoms) for $z$.
After at most $|E|-1$ appropriate iterations of our algorithm, there is
only one incoming edge for $z$ or $z$ has been eliminated.
Consequently, after no more than $|V| \cdot |E|$ iterations of our algorithm
on a conjunctive query (in each of which a join lifter can be applied),
the conjunctive query is necessarily acyclic.

In each such loop, a single query may be replaced by at most $k$ others,
where $k$ is the maximum number of conjunctions occurring in a
join lifter -- a constant (no greater than three in this article).
Thus, we make no more than $k^{|V| \cdot |E|}$ iterations in total
until all conjunctive queries in ${\cal Q}$ are acyclic, i.e.\ their
query graphs are forests.
This is the termination condition of our algorithm.

Thus, ${\cal Q}$ cannot contain more than $k^{|V| \cdot |E|}$
conjunctive queries, all of size $\le |Q_0|$.
Since the cycle detection and transformation procedures in (2) to (4)
can be easily implemented
to run in polynomial time each, the overall running time of our algorithm
is singly exponential.

The query computed by the algorithm is equivalent to $Q_0$. This follows by
induction from the fact that  the steps (2) to (4) each produce equivalent
rewritings. (The individual arguments are provided with steps (2) to (4).)
Thus, on termination,
${\cal Q}$ is a union of acyclic conjunctive queries -- an APQ --
equivalent to $Q_0$.

Note that step (4) can introduce new directed cycles into a query; therefore,
it may be necessary to repeat steps (2) and (3) after an application of step
(4), as done by our algorithm.
\end{proof}

Note that the rewriting technique of the previous algorithm is {\em
  nondeterministic}\/ (by the choice of next query to rewrite in step (1)),
but we do not prove confluence of our rewrite system since it is not essential
to our main theorem, stated next.


\begin{theorem}
\label{theo:CQ_in_APQ}
(1) For $F \subseteq \{ \child, \child^*, \child^+ \}$,
$
CQ[F] \subseteq APQ[F].
$
\\
(2) For $F \subseteq \{ \child, \child^+,
\nextsibling, \nextsibling^*, \nextsibling^+ \}$,
\[
CQ[F] \subseteq APQ[F].
\]
(3) For
$F \subseteq \{ \child, \child^*, \child^+,
\nextsibling, \nextsibling^*, \nextsibling^+ \}$,
\[
CQ[F] \subseteq APQ[F \cup \{ \child^+ \}].
\]
\end{theorem}

\begin{proof}
Consider the DNF formulae
\[
\psi_{R, S}(x, y, z) =
\left\{
\begin{array}{lll}
R(x, z) \land x = y
& \dots &
\mbox{$R = S \in \{ \child, \nextsibling \}$,} \\[1 ex]
(R(x, z) \land R(y, x)) \;\lor
& \dots &
   \mbox{$R = S \in \{ \child^*, \nextsibling^* \}$} \\
\; (R(x, y) \land R(y, z))
&& \\[1 ex]
(R(x, z) \land R(y, x)) \;\lor
& \dots &
   \mbox{$R = S \in \{ \child^+, \nextsibling^+ \}$} \\
\; (R(x, y) \land R(y, z)) \;\lor
&& \\
\; (R(x, z) \land x = y)
&& \\[1 ex]
(R(x, z) \land y = z) \;\lor
& \dots &
   \mbox{$R \in \{ \child, \nextsibling \}$, $S = R^*$} \\
\; (R(x, z) \land S(y, x))
&& \\[1 ex]
(R(x, z) \land x = y) \;\lor
& \dots &
   \mbox{$R \in \{ \child, \nextsibling \}$, $S = R^+$} \\
\; (R(x, z) \land S(y, x))
&& \\[1 ex]
(R(x, z) \land y = z) \;\lor
& \dots &
   \mbox{$R = \chi^+$, $S = \chi^*$} \\
\; (R(x,z) \land S(y, x)) \;\lor
&&
   \mbox{where $\chi \in \{ \child, \nextsibling \}$} \\
\; (R(y,z) \land S(x, y))
&& \\[1 ex]
R(x, z) \land S(y, x)
& \dots &
R \in \{ N, N^*, N^+ \}, N = \nextsibling, \\
&& S \in \{ \child, \child^+ \} \\[1 ex]
(R(x, z) \land y = z) \;\lor
& \dots &
R \in \{ N, N^*, N^+ \}, N = \nextsibling, \\
\; (R(x, z) \land \child^+(y, x))
&& S = \child^* \\[1 ex]
\psi_{S,R}(y, x, z) & \dots & \mbox{otherwise} 
\end{array}
\right.
\]
which are defined for all
\[
R, S \in \{ \child, \child^*, \child^+, \nextsibling, \nextsibling^*,
\nextsibling^+ \}.
\]

The $\psi_{R,S}$ are join lifters for each $R,S$.
The syntactic properties
of join lifters  of Definition~\ref{def:join_lifter}
can be easily verified by inspection. Moreover, indeed for all $R$, $S$,
$\psi_{R,S}(x,y,z) \equiv \phi_{R,S}(x,y,z) = R(x,z) \land S(y,z)$.
The arguments required to show this are very simple and are omitted.
(For example, $\phi_{\child, \child} \equiv \child(x,z) \land x=y$ because
each node in a tree can have only at most one parent.)

Thus, the $\psi_{R,S}$ are indeed join lifters. Now observe that
$\psi_{R,\child^*}$ for $R \in \{ \nextsibling, \nextsibling^+,
\nextsibling^* \}$ uses the $\child^+$ axis, but all other $\psi_{R,S}$ only
use the relations $R$ and $S$ (plus equality).
From Lemma~\ref{lem:CQ_in_APQ}, it follows that for $F$ such that
$\child^* \not\in F$ or 
$\nextsibling, \nextsibling^+, \nextsibling^* \not\in F$,
each $CQ[F]$ can be translated into an equivalent $APQ[F]$ (parts 1 and 2 of
our theorem) and otherwise,
each $CQ[F]$ can be translated into an equivalent $APQ[F \cup \{ \child^+ \}]$
(part 3).
\end{proof}

In all three cases of Theorem~\ref{theo:CQ_in_APQ},
the conjunctive queries can be rewritten into equivalent APQs
in singly exponential time.

Similar techniques to those  of the previous two proofs were used in 
\cite{OMFB2002} to eliminate backward axes from XPath expressions
and in \cite{Schw2000} to rewrite  first-order queries over
trees given by certain regular path relations.
%
%
The special cases of Theorem~\ref{theo:CQ_in_APQ} that
$CQ[\{ \child \}] \subseteq APQ[\{ \child \}]$ and
$CQ[\{ \child, \child^* \}] \subseteq APQ[\{ \child, \child^* \}]$
are implicit in \cite{BFK2003}.
%

\begin{example}
\em
Consider the query
$Q_0(x,y) \leftarrow \child^*(x,y) \land \nextsibling^*(x,y)$.
Since $\psi_{\nextsibling^*, \child^*}(x,x,y) =
(\nextsibling^*(x,y) \land x = y) \lor (\nextsibling^*(x,y) \land
\child^+(x,x))$,
we set ${\cal Q} = \{ Q, Q' \}$ with
$Q(x,x) \leftarrow \nextsibling^*(x,x)$,
which is further simplified to $Q(x,x) \leftarrow \mathit{Node}(x)$,
and
$Q'(x,y) \leftarrow \nextsibling^*(x,y) \land \child^+(x,x))$.
$Q'$ is unsatisfiable due to the directed cycle defined by its second atom
and is removed from ${\cal Q}$.
We obtain the APQ $\{ Q \}$ which is equivalent to $Q_0$.
\end{example}

\begin{example}
\label{ex:cq_apq}
\em
Figure~\ref{fig:cq_apq} illustrates the query rewriting algorithm of the proof
of Lemma~\ref{lem:CQ_in_APQ} using the join lifters of the proof of
Theorem~\ref{theo:CQ_in_APQ}
by means of an example.
The example query $Q$ is that from the introduction, but since
Theorem~\ref{theo:CQ_in_APQ} does not handle the $\following$ axis,
we first rewrite it using $\child^*$ and $\nextsibling^+$. 
All conjunctive queries that we obtain are unsatisfiable, except for one,
shown at the bottom left corner of Figure~\ref{fig:cq_apq}.
Thus, for $Q$ there exists an equivalent acyclic {\em conjunctive}\/ query.

Note that in Figure~\ref{fig:cq_apq} we make an exception from the
conventions followed throughout this article by labeling the
nodes of the query graphs with the variable names in order to allow for the
variables to be tracked through the rewrite steps more easily.
\punto
\end{example}

We complement Theorem~\ref{theo:CQ_in_APQ} by two further translation theorems.

\begin{theorem}
\label{theo:cq_apq_following1}
If $Q$ is a $CQ[F]$ such that
\[
F \subseteq \{ \child, \nextsibling, \nextsibling^*, \nextsibling^+,
\following \},
\]
then $Q$ can be rewritten into an equivalent
$APQ[F \cup \{ \nextsibling^+ \}]$ in singly exponential time.
\end{theorem}

\begin{proof}
We extend $\psi_{R,S}$ from the proof of Theorem~\ref{theo:CQ_in_APQ}
by join lifter formulae for $S = \following$ and $R \in F$:
\begin{eqnarray*}
\psi_{\nextsibling, \following}(x,y,z) &:=&
(\nextsibling(x,z) \land x=y) \;\lor \\ 
&&(\nextsibling(x,z) \land \following(y,x)) 
\\
\psi_{\nextsibling^+, \following}(x,y,z) &:=&
(\nextsibling^+(x,z) \land x=y) \;\lor \\
&& (\nextsibling^+(x,z) \land \following(y,x)) \;\lor \\
&& (\nextsibling^+(x,y) \land \nextsibling^+(y,z)) 
\\
\psi_{\nextsibling^*, \following}(x,y,z) &:=&
(\nextsibling^*(x,z) \land \following(y,x)) \;\lor \\
&& (\nextsibling^*(x,y) \land \nextsibling^+(y,z)) 
\\
\psi_{\child, \following}(x,y,z) &:=&
(\child(x,z) \land x=y) \;\lor \\
&& (\child(x,z) \land \following(y,x)) \;\lor \\
&& (\child(x,y) \land \nextsibling^+(y,z))
\end{eqnarray*}
\begin{eqnarray*}
\psi_{\following, \following}(x,y,z) &:=&
(\following(x,z) \land x=y) \;\lor \\
&& (\following(x,z) \land \following(y,x)) \;\lor \\
&& (\following(x,y) \land \following(y,z))
\end{eqnarray*}
%
%
Each $\psi_{R,S}$ is defined using only relations $R, S, =$ and
$\nextsibling^+$.
Now the theorem follows immediately from Lemma~\ref{lem:CQ_in_APQ}.
\end{proof}

\begin{theorem}
\label{theo:cq_apq_following2}
If $Q$ is a $CQ[F]$ such that $F \subseteq \ouraxes$,
then $Q$ can be rewritten into an equivalent
$APQ[F \cup \{ \child^+, \nextsibling^+ \}]$
in singly exponential time.
\end{theorem}

\begin{proof}
Given query $Q$,
we first rewrite all occurrences of $\following$ using $\child^*$ and
$\nextsibling^+$ using Equation~(\ref{eq:following}) from
Section~\ref{sect:preliminaries}.
In order to be economical with axes, we rewrite all $n$ occurrences of
$\child^*$ using $\child^+$. We define an APQ consisting of $2^n$ copies
of $Q$ such that in the $m$-th copy of $Q$, the $k$-th $\child^*(x,y)$ atom is
replaced by $\child^+(x,y)$ if the $k$-th bit of $m$ represented in binary is,
say, $1$ and by $x=y$ otherwise (that is, all occurrences of variable $y$ in
the query are replaced by $x$). Clearly, since $\child^*(x,y) \Leftrightarrow
\child^+(x,y) \lor x=y$, the APQ obtained in this way is equivalent to $Q$.
Then we apply the algorithm of the proof of Lemma~\ref{lem:CQ_in_APQ}
using the join lifters as in the proof of Theorem~\ref{theo:CQ_in_APQ}~(3)
to each of the $2^n$ modified
conjunctive queries and compute the union of the APQs
obtained. Of course, the overall transformation can be again effected in
exponential time.
\end{proof}

It follows that the acyclic positive queries capture the positive queries
over trees.

\begin{corollary}
$PQ[\ouraxes] = APQ[\ouraxes]$.
\end{corollary}

\begin{remark}
\em
Since $\child^+$ and $\nextsibling^+$
are XPath axes (``descendant'' and ``following-sibling''),
it follows from Theorem~\ref{theo:cq_apq_following2}
that each unary conjunctive query over XPath axes
can also be formulated as an XPath query.
This is in contrast to full first-order logic (i.e., with negation) on trees,
which is known to be stronger than acyclic first-order logic on trees resp.\
Core XPath \cite{Marx2005}.
\end{remark}

Obviously, the $CQ[F]$ are not closed under union. On trees
of one node only, conjunctive queries are equivalent to ones which do
not use binary atoms. It is easy to see that the query
$\{ x \mid A(x) \lor B(x) \}$
has no conjunctive counterpart.

\begin{proposition}
For any $F \subseteq \ouraxes$, $CQ[F] \neq APQ[F]$.
\end{proposition}

There are signatures with axes for which all conjunctive queries can be
rewritten into APQ's in polynomial time.\footnote{As shown in the
next section, there are also signatures for
which this is not possible.}

\begin{proposition}[\cite{GKJACM}]
Any $CQ[\{ \child, \nextsibling \}]$
can be rewritten into an equivalent acyclic
$CQ[\{ \child, \nextsibling, \nextsibling^* \}]$
in linear time.
\end{proposition}

\begin{remark}
\em
It is  easy to  verify by inspecting  the proof in  \cite{GKJACM} that
rewriting each CQ[$\child, \nextsibling$]
into  an equivalent a\-cy\-clic CQ[$\child, \nextsibling$] in linear
time is also
possible.   (The   proof  there also   deals   with   relations   such   as
$\textit{FirstChild}$. If  these are not  present, $\nextsibling^*$ is
not required.)
\end{remark}


\section{Succinctness}
\label{sect:succinctness}

The translations from conjunctive queries into APQs of
the Theorems~\ref{theo:CQ_in_APQ}, \ref{theo:cq_apq_following1}
and \ref{theo:cq_apq_following2}
run in exponential time and can produce APQs of exponential size.
In this section, we show that this situation cannot be improved upon:
there are conjunctive queries over trees that cannot be
polynomially translated into equivalent APQs.

By the size $|Q|$ of a Boolean conjunctive query $Q$,
we denote the number of atoms in its body. The size of an APQ is given by the
sum of the sizes of the constituent conjunctive queries.

Let $D_n$ denote the $n$-diamond Boolean conjunctive query
\begin{multline*}
D_n \leftarrow Y_1(y_1) \land
\bigwedge_{i=1}^n
\big(
\child^+(y_i, x_i) \land
X_i(x_i) \land
\child^+(x_i, y_{i+1}) \land \\
\child^+(y_i, x_i') \land
X_i'(x_i') \land
\child^+(x_i', y_{i+1}) \land
Y_{i+1}(y_{i+1})
\big).
\end{multline*}
A graphical representation of $D_n$ is provided in
Figure~\ref{fig:diamond}~(a).

The following is the main result of this section:

\begin{theorem}
\label{theo:dn}
There is no family $({\cal Q}_n)_{n \ge 1}$
of queries in $APQ[\ouraxes]$
such that each ${\cal Q}_n$ is of size polynomial in $n$ and is
equivalent to $D_n$.
\end{theorem}

Before we can show this, we have to provide a few definitions.

We use the acronyms ABCQ for {\em acyclic Boolean conjunctive queries}\/ and
DABCQ ({\em directed ABCQ}\/)
for Boolean conjunctive queries whose
query graphs are acyclic. That is, the query graph of a DABCQ is a directed
acyclic graph, while the query graph of an ABCQ is a forest
(because conjunctive query acyclicity is defined with respect to the
undirected shadows of query graphs).
By Lemma~\ref{lem:directed_cycles}, an equivalent $DABCQ[F]$ exists
(and can be computed efficiently) for each Boolean
$CQ[F]$ that is satisfiable, for any $F \subseteq \ouraxes$.
The queries $D_n$ are $DABCQ[\{ \child^+ \}]$.

For a DABCQ $Q$,
let $\Pi_Q \subseteq \textit{Var}(Q)^*$ denote the set of {\em
  variable-paths}\/ in the query graph of $Q$ 
from variables that have in-degree zero to variables that have
out-degree  zero. For example, if $Q$ is the left- and bottommost query
of Figure~\ref{fig:cq_apq}, then $\Pi(Q) = \{ xuy, xuvz \}$.
We say that a label $L$ {\em occurs}\/ in variable-path $\pi \in \Pi_Q$ iff
there is a variable $x$ in $\pi$ for which $Q$ contains a unary atom $L(x)$.

By a {\em path-structure}\/, we denote a tree structure in which the graph
of the $\child$-relation is a path.
Given a variable-path $x_1.\cdots.x_k$, the associated {\em label-path}\/ is 
the path-structure of $k$ nodes
in which the $i$-th node is labeled $L$ iff $Q$ contains
atom $L(x_i)$. Observe that some nodes of this structure
may be unlabeled, and some may have several labels.
Given a set $P$ of variable-paths, let $LP(P)$ denote the
corresponding label-paths.

We say that a path-structure is {\em $k$-scattered}\/  if
(*) it consists of at least $k$ nodes,
(*) each node has at most one label,
(*) no two nodes have the same label, and
(*) if node $v$ has a label and node $v'$ ($v \neq v'$) either is
the topmost node, the bottommost node, or has a label, then
the distance between $v$ and $v'$ is at least $k$.

\def\slps{\mathit{PS}(n, p(n))}

\nop{ 
\begin{remark}
The motivation for introducing $k$-scattered path structures can be explained
as follows: the nodes of a path structure can be interpreted as natural numbers
$1,2,\ldots,m$. An atom $\mbox{\sl Child}(x,y)$ then reads as $x=y+1$, an
atom $\mbox{\sl Child}^\ast(x,y)$ (resp. $\mbox{\sl Child}^+(x,y)$)
as $x\leq y$ (resp. $x < y$). When looking at satisfiability in
$k$-scattered structures where $k$ is sufficiently large,
we may sometimes refine $\mbox{\sl Child}^\ast$-atoms (i.e.,
$\leq$-constraints) to equality constraints and atoms $\mbox{\sl
  Child}^+(x,y)$ (i.e., $\leq$-constraints) by atoms $\mbox{\sl Child}(x,y)$
without affecting satisfiability (see proof of Lemma 7.6 below). This helps to
transfer queries to a special form.
\end{remark}
} 

In order to prove our theorem, we need two technical lemmata.
The first states, {\em essentially}\/,
that on sufficiently scattered
path structures, each ABCQ is equivalent
to an ABCQ that only uses the axes $\child^+$ and $\child^*$.
This is somewhat reminiscent of results
on the locality of first-order queries (cf.\ e.g.\ \cite{Lib2004}).

\begin{lemma}
\label{lem:scattered_cleanup}
Let $Q$ be an $ABCQ[\ouraxes]$ that is true on at least one
$|Q|$-scattered path-structure.
Then, there is an $ABCQ[\{ \child^+, \child^* \}]$ $Q'$ such that
$Q' \subseteq Q$, $|Q'| \le |Q|$, and
$Q'$ is true on all $|Q|$-scattered path-structures on which $Q$ is true.
\end{lemma}

The second lemma states that two $DABCQ[\{ \child^*, \child^+ \}]$
$Q$ and $Q'$ with $LP(\Pi_Q) \neq LP(\Pi_{Q'})$ differ in the sets of
path structures on which they are true.

\begin{lemma}
\label{lem:excluded_path}
Let $Q$ and $Q'$ be two $DABCQ[\{ \child^*, \child^+ \}]$ and $\Gamma$ be
a set of labels.
If there is a label-path in $LP(\Pi_{Q'})$ in which all labels from $\Gamma$
occur, but there is no label-path in $LP(\Pi_Q)$
in which all labels from $\Gamma$ occur, then there is a
path-structure ${\cal M}$ on which $Q$ is true but $Q'$ is not.
\end{lemma}

Now we can prove our theorem. The proofs of the two lemmata follow at the
end of the section.

\begin{figure}
\begin{center}
\input{diamond.tex}
\end{center}
\vspace{-6mm}
\caption{Query $D_n$ (a) and path structures $\slps$ (b).}
\label{fig:diamond}
\end{figure}
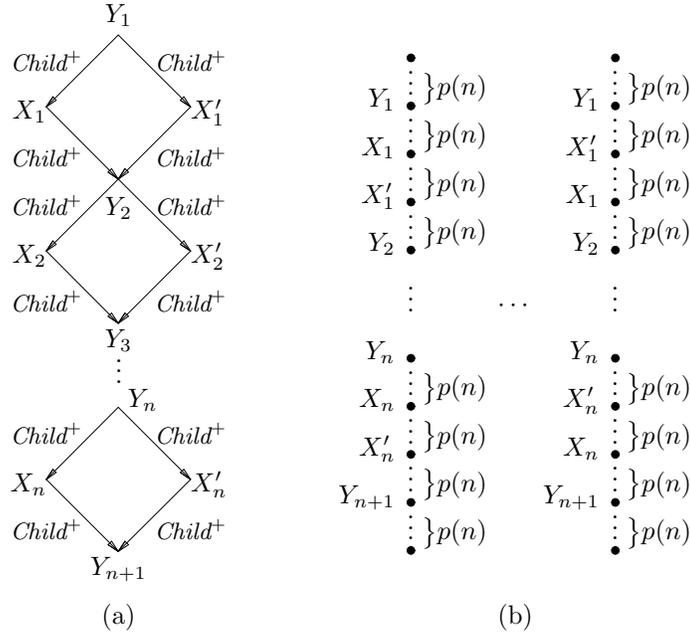

\begin{proof}[of Theorem~\ref{theo:dn}]
By contradiction.
Assume there  is a  (Boolean)  APQ
${\cal Q}$, that is, a finite union of ABCQs,
which is equivalent to $D_n$, and that ${\cal Q}$ is of size bounded by
polynomial $p(n)$.
Let $s$ be a path of $p(n)$ unlabeled nodes.
The regular expression
\begin{multline*}
s.Y_1.s.(X_1.s.X_1' \mid X_1'.s.X_1).s.Y_2.s.(X_2.s.X_2' \mid
X_2'.s.X_2).s.Y_3.s.\cdots. \\
s.Y_n.s.(X_n.s.X_n' \mid X_n'.s.X_n).s.Y_{n+1}.s
\end{multline*}
defines a set of $p(n)$-scattered path-structures over alphabet
\[
\Sigma = \{ X_1,  \dots, X_n, X_1', \dots, X_n', Y_1, \dots, Y_{n+1} \},
\]
as ske\-tched in Figure~\ref{fig:diamond}~(b).
We refer to the set of these structures as $\slps$.
It is easy to see that $D_n$ is true on each of the structures in
$\slps$.

There are $2^n$ structures in $\slps$ and $D_n$ is true on all of them,
but there are no more than $p(n)$ ABCQs in ${\cal Q}$.
Therefore, there is an ABCQ $Q \in {\cal Q}$ which is true
for at least
$2^{n - \mathrm{log}\, p(n)}$
structures in $\slps$ and is not true on any structure on which $D_n$
is not true.

As the path structures in $\slps$ are ($p(n) \ge |Q|$)-scattered,
by Lemma~\ref{lem:scattered_cleanup}, there is an
$ABCQ[\{ \child^+, \child^*\}]$ $Q'$ with $|Q'| \le |Q|$, $Q' \subseteq Q$,
and which  is true on all structures in $\slps$ on which $Q$ is true.

In each path-structure ${\cal A}$
of $\slps$, for any $1 \le j \le n$, precisely
one node $v$ is labeled $X_j$ and precisely one {\em different}\/
node $w$ is labeled $X_j'$.
Thus if query $Q'$ contains unary atoms $X_j(x_j)$ and $X_j'(x_j')$,
a mapping $\theta$ can only be a satisfaction of $Q'$ on
${\cal A}$ if $\theta(x_j) = v$ and $\theta(x_j') = w$.
But if there is a variable-path in $\Pi_{Q'}$
with $x_j$ above $x_j'$ (respectively, $x_j'$ above $x_j$) and $w$ above $v$
(respectively, $v$ above $w$) in ${\cal A}$, no satisfaction of $Q'$
on the structure can exist.
Let $i_1, \dots, i_m$ be {\em precisely}\/ those pairwise distinct
indexes for which, for $1 \le j \le m$, $Q'$
does not contain a variable-path containing two
variables $x_{i_j}$ and $x'_{i_j}$ such that $X_{i_j}(x_{i_j})$ and
$X'_{i_j}(x'_{i_j})$ are unary atoms of $Q'$.
Then $Q'$ is true on at most $2^m$ path-structures of $\slps$.

We assumed that $Q$ is true on at least $2^{n - \mathrm{log}\, p(n)}$
structures of $\slps$ and showed that $Q'$ is true on the same.
But then $m \ge n - \mathrm{log}\; p(n)$.

Since the (undirected) query graph of $Q'$ is a forest,
the number of paths in $\Pi_{Q'}$ is not greater than
the square of  the number  of its  variables.
As $|Q'| \le p(n)$, $|\Pi_{Q'}| \le p(n)^2$.

Now, if $n > 3 \cdot \mathrm{log}\; p(n)$, then
$m > 2 \cdot \mathrm{log}\; p(n)$ and there are more choices
\[
\Gamma = \{ E_1 \in \{ X_{i_1}, X'_{i_1} \}, \dots,
E_m \in \{ X_{i_m}, X'_{i_m} \} \}
\]
than there are paths in $\Pi_{Q'}$.
Assume there are  two {\em distinct}\/ such choices  $\Gamma, \Gamma'$
and a variable-path $\pi \in \Pi_{Q'}$ such that
all labels of $\Gamma \cup \Gamma'$ occur in $\pi$.
Then there is an index $i_j$
such that $X_{i_j}, X'_{i_j} \in \Gamma \cup \Gamma'$. This is in
contradiction to the assumptions we made about the
indexes $i_1, \dots, i_m$.
Thus there must be (at least) one such choice $\Gamma$
such that no single path in $\Pi_{Q'}$ exists in which all the labels of
$\Gamma$ occur.
Since there is a path in $D_n$ which contains all the labels of
$\Gamma$, by Lemma~\ref{lem:excluded_path}, there is a
model ${\cal M}$ of $Q'$ which is not a model of $D_n$. Since
$Q' \subseteq Q$, ${\cal M}$ is also a model of $Q$.

This is in
contradiction with our assumption that $Q \subseteq D_n$.
Consequently, for $n > 3 \cdot \mathrm{log}\; p(n)$,
there cannot be an ABCQ of size bounded by polynomial $p(n)$
that is contained in $D_n$ and is true on exponentially
many structures of $\slps$. It follows that for sufficiently large $n$
there cannot be an APQ
equivalent to $D_n$ that is of polynomial size.
\end{proof}

Now it remains to prove the two technical lemmata we used in the proof of our
succinctness result.


\subsection*{Proof of Lemma~\ref{lem:scattered_cleanup}}

We say that a Boolean query $Q'$ is a {\em faithful simplification}\/ of a
Boolean query $Q$ {\em w.r.t.\ a class of structures}\/ ${\bf A}$ if
$|Q'| \le |Q|$, $Q' \subseteq Q$, and $Q'$ is true on
structures of ${\bf A}$ on which $Q$ is true.

Below, by $G_Q$, we refer to the directed graph obtained
from the query graph of $Q$ by removing all
edges besides the $\child$ edges.
We will in particular consider the connected components of this graph,
subsequently called the $G_Q$-components.
We say that $C_0$ is a {\em parent component}\/ of connected
component $C$ of such a graph
iff there is a variable $x$ in $C_0$ and a variable
$y$ in $C$ such that
there is an atom $\child^+(x,y)$ or $\child^*(x,y)$ in $Q$.
(Of course, $C \neq C_0$ because the query graph of $Q$ is a forest.)
The ancestors of a component are obtained by upward reachability through the
parent relation on $G_Q$-components.

Lemma~\ref{lem:scattered_cleanup} immediately follows from the following
four lemmata.

\begin{lemma}
\label{lem:chase}
Let $Q$ be an $ABCQ[\ouraxes]$ that is
true on at least one path structure.
Then there is an $ABCQ[\{ \child, \child^*, \child^+ \}]$ $Q'$
that is a faithful simplification of $Q$ w.r.t.\ the
path-structures and in which each $G_{Q'}$-component is a path.
\end{lemma}

\begin{proof}
Query $Q$ cannot contain a $\nextsibling$, $\nextsibling^+$, or
$\following$-atom,
because if it does, $Q$ is false on all path-structures, contradicting
our assumption that $Q$ is true on at least one path-structure.

Let $Q'$ be the query obtained from $Q$ by iteratively applying the following
three rules until a fixpoint is reached.
\begin{itemize}
\item
if $Q'$ contains atom
$\nextsibling^*(x,y)$, remove it and substitute all occurrences of
variable $y$ in $Q'$ by $x$;

\item
if there are atoms $\child(x,z), \child(y,z)$ in $Q'$,
remove $\child(y,z)$ and
substitute every occurrence of $y$ in $Q'$ by $x$;

\item
if there are atoms $\child(x,y), \child(x,z)$ in $Q'$,
remove $\child(x,z)$ and
substitute every occurrence of $z$ in $Q'$ by $y$.
\end{itemize}
It is easy to verify that $Q'$ is a faithful simplification of $Q$ w.r.t.\
the path structures. Moreover, none of the rewrite rules can introduce a
cycle into $Q'$, thus it is an $ABCQ[\{ \child, \child^*, \child^+ \}]$.
There are neither atoms $\child(x,y)$ and $\child(x,z)$ with $y \neq z$ nor
atoms  $\child(x,z)$ and $\child(y,z)$ with $x \neq y$ in $Q'$, so
each $G_{Q'}$-component is a path. 
\end{proof}

Since each $G_Q$-component is a path, we may give the $k$ variables inside
$G_Q$-component $C$ the names $x^C_1, \dots, x^C_k$.
We use $|C|$ to denote the number of variables $k$ in $G_Q$-component $C$.
We will think of the node names of a path structure as an initial segment of
the integers, thus $v > w$ if and only if $v$ is below $w$ in the path.

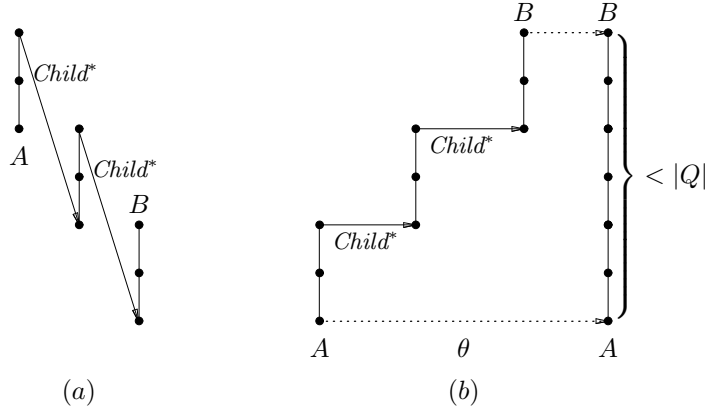
\begin{figure}
\begin{center}
\input{stretch.tex}
\end{center}
\caption{A query (a) and
the query embedded into a path structure with $B$ as high
above $A$ as possible (b). The unlabeled edges are child-edges.}
\label{fig:distance}
\end{figure}

\begin{lemma}
\label{lem:scat_satisfaction_prop}
Let $Q$ be an $ABCQ[\{ \child, \child^*, \child^+ \}]$
in which each $G_Q$-component is a path and that is
true on at least one $|Q|$-scattered path structure ${\cal A}$.
Then,
\begin{enumerate}
\item[(a)]
any $G_Q$-component contains at most one label atom and

\item[(b)]
if $C_1, \dots, C_m$ is a path of $G_Q$-components and $Q$ contains unary
atoms $L(x^{C_1}_k)$ and $L'(x^{C_m}_l)$ with $L \neq L'$,
then the node labeled $L$ in ${\cal A}$ is above the node labeled $L'$.

\item[(c)]
if $C_1, \dots, C_m$ is a path of $G_Q$-components and $Q$ contains unary
atoms $L(x^{C_1}_{k_1})$ and $L(x^{C_m}_{k_m})$,
then for no $1 \le j \le m$ and $L' \neq L$ there can be a unary atom
$L'(x^{C_j}_{k_j})$ in $Q$.
\end{enumerate}
\end{lemma}

\begin{proof}
(a) Assume that there is a $G_Q$-component $C$ with two label atoms,
\[
\child(x^C_1, x^C_2), \dots, \child(x^C_{|C|-1}, x^C_{|C|}),
L(x^C_k), L'(x^C_l)
\]
with either $L \neq L'$ or $k \neq l$, a $|Q|$-scattered path-structure
${\cal A}$, and a satisfaction $\theta$ of $Q$ on ${\cal A}$.
Since $|C| \le |Q| - 2$, $|\theta(x^C_k) - \theta(x^C_l)| < |Q|-2$.
However, ${\cal A}$ is a
$|Q|$-scattered path-structure and thus cannot contain two labels on a
subpath of length $|Q|$. Contradiction.

(b) Let $\theta$ be a satisfaction of $Q$ on ${\cal A}$.
Assume that $\theta(x^{C_1}_k) > \theta(x^{C_m}_l)$, i.e., the node labeled
$L$ is below the node labeled $L'$ in ${\cal A}$.
Then, for each $1 \le i < m$,
there is an atom $R(x^{C_i}_{j_i}, x^{C_{i+1}}_{j_{i+1}})$ in $Q$
with $R$ either $\child^+$ or $\child^*$.
So $\theta(x^{C_i}_{j_i}) \le \theta(x^{C_{i+1}}_{j_{i+1}})$ and consequently
$\theta(x^{C_i}_1) \le \theta(x^{C_{i+1}}_{|C_{i+1}|}) =
\theta(x^{C_{i+1}}_1) + |C_{i+1}| - 1$.
But then $\theta(x^{C_1}_k) - \theta(x^{C_m}_l)
\le \Sigma_{i=1}^m (|C_i| - 1) < |Q|$. (See Figure~\ref{fig:distance} for
an illustration.)
This is in contradiction with our assumption that ${\cal A}$ is a
$|Q|$-scattered path structure and thus
$|\theta(x^{C_1}_k) - \theta(x^{C_m}_l)| \ge |Q|$.

(c) follows immediately from (b) and the fact that in a $|Q|$-scattered
path structure each label occurs at at most one node.
\end{proof}

Below, we will call the
$G_Q$-components of an $ABCQ[\{ \child, \child^*, \child^+ \}]$ $Q$
{\em successor-repellent}\/ if
for any two atoms $R(x,y), R'(x',y')$ in $Q$ with
$x = x', y \neq y'$ or $x \neq x', y = y'$,
neither $R = \child$ nor $R' = \child$.
%
%
%
The naming of this term is due to the following fact:
Let $Q$ be a successor-repellent $ABCQ[\{ \child, \child^*, \child^+ \}]$.
Then for any two components $C,C'$ such that $C'$ is a successor of
$C$ and for any satisfaction $\theta$ of $Q$ (on a path structure),
$\theta(x^C_{|C|}) \le \theta(x^{C'}_1)$. 


\begin{lemma}
Let $Q$ be an $ABCQ[\{ \child, \child^*, \child^+ \}]$ that is
true on at least one $|Q|$-scattered path structure and in which each
$G_Q$-component is a path.
Then there is an $ABCQ[\{ \child, \child^*, \child^+ \}]$ $Q'$ that
is a faithful simplification of $Q$ w.r.t.\ the $|Q|$-scattered
path-structures and whose $G_{Q'}$-components are successor-repellent.
\end{lemma}

\begin{proof}
We construct the query $Q'$ as follows.
Initially, let $Q' := Q$.
As often as possible,
for any path of $G_Q$-components $C_1, \dots, C_m$ such that
there are atoms
\[
R_1(x^{C_1}_{j_1}, x^{C_2}_{j_1'}), \dots
R_{m-1}(x^{C_{m-1}}_{j_{m-1}}, x^{C_m}_{j_{m-1}'}),
L(x^{C_1}_k), L(x^{C_m}_l)
\]
in $Q'$ with $R_1, \dots R_{m-1} \in \{ \child^+, \child^* \}$, but there is no
label atom over components $C_2, \dots, C_{m-1}$,
replace all occurrences of variable $x^{C_m}_l$ in $Q'$ by
$x^{C_1}_k$ and
delete atom $R_{m-1}(x^{C_{m-1}}_{j_{m-1}}, x^{C_m}_{j_{m-1}'})$.
Moreover, for each $1 \le i < m - 1$,
if $R_i = \child^*$, remove the atom
$\child^*(x^{C_i}_{j_i}, x^{C_{i+1}}_{j_i'})$
and substitute $x^{C_{i+1}}_{j_i'}$ by $x^{C_i}_{j_i}$
and if $R_i = \child^+$, replace $R_i$ by $\child$.
Note that this query is an ABCQ.
Then apply the algorithm
of Lemma~\ref{lem:chase} to turn the $G_{Q'}$-components into paths.
(See Figure~\ref{fig:compmerge} for an example of this construction.)
To conclude with our construction, we replace
each atom $R(x^{C_i}_k,x^{C_j}_l)$ of $Q'$,
where $R$ is either $\child^+$ or $\child^*$,
by $R(x^{C_i}_{|C_i|}, x^{C_j}_1)$.

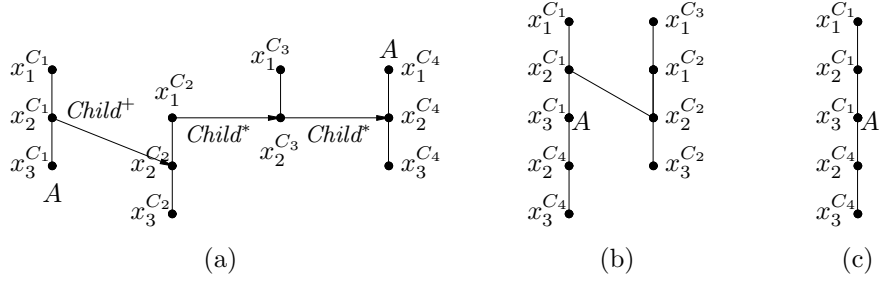
\begin{figure}
\hspace{-1cm}
\input{compmerge.tex}
\caption{A path of components with two occurrences of the same label (a),
the same query after atom replacement and variable substitution (b),
and (c) the query after applying the algorithm of Lemma~\ref{lem:chase}.
The unlabeled edges are child-edges.}
\label{fig:compmerge}
\end{figure}


Clearly, $Q'$ is a successor-repellent ABCQ with $|Q'| \le |Q|$. 
Since $Q$ is true on at least one $|Q|$-scattered path structure,
it follows from Lemma~\ref{lem:scat_satisfaction_prop}~(a) and (c)
that there are no two $G_{Q'}$-components $C_1, C_m$
such that both are labeled $L$ and $C_1$ is an ancestor of $C_m$.

It is also easy to verify that $Q' \subseteq Q$:
Given an arbitrary tree structure ${\cal A}$ and a satisfaction $\theta'$ for
$Q'$ on ${\cal A}$,
we can construct a satisfaction $\theta$ of $Q$ as
$x \mapsto \theta'(y)$ if the construction of $Q'$ from $Q$
substituted $x$ by $y$ and $x \mapsto \theta'(x)$ otherwise.
That $\theta$ is a satisfaction is obvious for all atoms of $Q$ apart from
atom $R_{m-1}(x^{C_{m-1}}_{j_{m-1}}, x^{C_m}_{j_{m-1}'})$, which we simply
deleted. But it is not hard to convince oneself that if
$R_{m-1}(\theta(x^{C_{m-1}}_{j_{m-1}}), \theta(x^{C_m}_{j_{m-1}'}))$ does not
hold, then it must be true that for every satisfaction $\theta_0$ of $Q$ on
${\cal A}$,
$
\child^*(\theta(x^{C_{m-1}}_{j_{m-1}}), \theta_0(x^{C_{m-1}}_{j_{m-1}}))
$ and therefore
$\child^+(\theta(x^{C_m}_{j_{m-1}'}), \theta_0(x^{C_m}_{j_{m-1}'}))$.
But then,
$\theta_0(x^{C_1}_k) \neq \theta_0(x^{C_m}_l)$
and ${\cal A}$ must have at
least two distinct nodes labeled $L$.
This is in contradiction with our assumption that
$Q$ is true on at least one $|Q|$-scattered path structure.

Moreover, $Q'$ is true on all $|Q|$-scattered path-structures
on which $Q$ is true. To show this,
let $Q$ be true on some $|Q|$-scattered path-structure ${\cal A}$ with
satisfaction $\theta$.
We construct a satisfaction $\theta'$ for $Q'$ on ${\cal A}$ using the
following algorithm.
\begin{tabbing}
1 \hspace{4mm} \=
      for each $G_{Q'}$-component $C_j$ do \\
\> // process components according to some topological ordering w.r.t.\ \\
\> // $\child^+, \child^*$: if there is an atom $\child^+(x^{C_i}_k,
      x^{C_j}_l)$ or $\child^+(x^{C_i}_k, x^{C_j}_l)$ \\
\> // in $Q'$, then $\theta'(x^{C_i}_1), \dots, \theta'(x^{C_i}_{|C_i|})$
   has been computed before. \\
2  \> begin \\
3  \> \hspace{2mm} \=
         if $Q'$ contains a label atom $L(x^{C_j}_l)$ then \\
4  \> \> \hspace{2mm} \=
      $\theta'(x^{C_j}_1) := \theta(x^{C_j}_l) - l + 1$; \\
   \> \> \> // $\theta(x^{C_j}_l)$ is the unique node of the path structure
               which has
               label $L$; \\
5  \> \> else $\theta'(x^{C_j}_{1}) := 1 +
              \mbox{max}\{ \theta'(x^{C_i}_{|C_i|}) \mid
              \mbox{$C_i$ is a parent $G_{Q'}$-component of $C_j$} \}$; \\
   \> \> \> // let max($\emptyset$) = 0 \\
\\
6  \> \> for the remaining $1 < k \le |C_j|$ do \\
7  \> \> \>   $\theta'(x^{C_j}_k) := \theta'(x^{C_j}_1) + k - 1$; \\
8  \> end;
\end{tabbing}

Clearly, this algorithm defines $\theta'$ for all variables of $Q'$.
Since for any $x$, $\theta'(x)$ cannot be greater than max$\{ v \mid
\mbox{path-structure node $v$ has a label} \} + |Q'|$, $\theta'$ maps into the
($|Q|$-scattered) path structure.

Lines 6-7 assure that all the $\child$-atoms of $Q'$ are true.
Line 4 assures that the label-atoms are true: otherwise, $\theta$ could not
be a satisfaction of $Q$.
For a component $C_j$ without a label-atom, line 5 assures that
all atoms of the form $R(x^{C_i}_{|C_i|}, x^{C_j}_1)$,
for $R$ either $\child^+$
or $\child^*$, are satisfied because $\theta'(x^{C_j}_1) =
\theta'(x^{C_i}_{|C_i|}) + 1$.

Finally, lines 3-4 handle the case that component $C_j$
contains a label atom $L(x^{C_j}_l)$.
By Lemma~\ref{lem:scat_satisfaction_prop}~(a)
the choice of label atom for the component in line 3 is deterministic. 
What has to be shown is that
\[
    \theta'(x^{C_j}_1) \ge 1 + \mbox{max}\{ \theta'(x^{C_i}_{|C_i|}) \mid
              \mbox{$C_i$ is a parent $G_{Q'}$-component of $C_j$} \}.
\]
It is easy to verify by induction that
\[
\mbox{max}\{ \theta'(x^{C_i}_{|C_i|}) \mid
 \mbox{$C_i$ is a parent $G_{Q'}$-component of $C_j$} \} <
v + |Q| - |C_j|,
\]
where $v$ is the bottommost among the nodes of the path structure
carrying labels $L_0$ that appear in the ancestor components of $C_j$.
Thus, if all these labels $L_0$ occur above $\theta(x^{C_j}_l)$, we are done.
(In a $|Q|$-scattered path structure,
$|\theta(x^{C_j}_l) - v| \ge |Q|$.)

We know that by our construction,
label $L$ does not occur in any of the ancestor-components of $C_j$.
But then, if all labels that occur
in ancestor-components of $C_j$ differ from $L$,
by Lemma~\ref{lem:scat_satisfaction_prop}~(b) the path-structure node $v$
must be above $\theta'(x^{C_j}_1)$, otherwise $\theta$ would not be a
satisfaction of $Q$.
\end{proof}

\begin{lemma}
Let $Q$ be an $ABCQ[\{ \child, \child^*, \child^+ \}]$ such that
the components of $G_Q$ are successor-repellent and
each $G_Q$-component contains at most one label-atom.
Then, the query $Q'$ obtained
from $Q$ by replacing each occurrence of predicate $\child$ by $\child^+$
is equivalent to $Q$.
\end{lemma}

\begin{proof}
Since $\child \subseteq \child^+$, it is obvious that $Q \subseteq Q'$.
For the other direction, let $\theta'$ be any satisfaction of $Q'$.
We define a valuation $\theta$ for $Q$ from $\theta'$.
For every $G_Q$-component $C$, let
$\theta(x^C_k) := \theta'(x^C_l) + k - l$
if there is a label-atom over variable $x^C_l$  -- as shown above,
there is at most one such variable per component -- or
$\theta(x^C_k) := \theta'(x^C_1) + k - 1$ if component $C$ does not contain
a label-atom.
It is now easy to verify that $\theta$ is indeed
a satisfaction for $Q$: The label- and $\child$-atoms of $Q$ are satisfied by
definition.
Since $\theta(x^{C_i}_{|C_i|}) \le \theta'(x^{C_i}_{|C_i|})$ and
$\theta(x^{C_j}_1) \ge \theta'(x^{C_j}_1)$,
$R(\theta'(x^{C_i}_{|C_i|}), \theta'(x^{C_j}_1))$,
where $R$ is either $\child^*$ or $\child^+$, implies
$R(\theta(x^{C_i}_{|C_i|}), \theta(x^{C_j}_1))$.
Thus, $Q' \subseteq Q$ and consequently $Q' \equiv Q$.
\end{proof}

\subsection*{Proof of Lemma~\ref{lem:excluded_path}}

\begin{proof}[of Lemma~\ref{lem:excluded_path}]
We define a number of restrictions of the set $\Pi_Q$ of variable-paths in
$Q$. For labels $X$, let $\Pi_Q|_X$ denote  the set
of variable-paths in $\Pi_Q$  which contain a
variable with label $X$.
Let $\Pi_Q|_{\neg X} = \Pi_Q - \Pi_Q|_X$ and
$\Pi_Q|_{\phi  \land \psi}  = \Pi_Q|_\phi \cap  \Pi_Q|_\psi$.
For variables $x$, let $\Pi_Q|_x$ denote the set of all variable-paths
in $\Pi_Q$ in which $x$ occurs.
Let  $LC(\psi)$ denote the label-paths in
$LP(\Pi_Q|_\psi)$ concatenated in any (say, lexicographic) order.

Let $\Gamma = \{ E_1, \dots, E_m\}$.
There is a variable-path $x_1.\cdots.x_k \in \Pi_{Q'}$ and
query $Q'$ contains atoms $E_1(x_{i_1}), \dots, E_m(x_{i_m})$ such that,
w.l.o.g., $1 \le i_1 \le \dots \le i_m \le k$.
By assumption, there is no such variable-path in $\Pi_Q$.

\paragraph*{Construction of path-structure ${\cal M}$}
We define ${\cal M}$ as the path structure
\begin{multline*}
LC(\neg E_1).LC(E_1 \land \neg E_2).
LC(E_1 \land E_2 \land \neg E_3).\cdots.
LC(E_1 \land \dots \land E_{m-1} \land \neg E_m)
\end{multline*}
Since $\Pi_Q|_{E_1 \land \dots \land E_m}$ is empty, ${\cal M}$
is a concatenation of {\em all} paths in $LP(\Pi_Q)$.

\paragraph*{${\cal M}$ is  a model of  $Q$}
We show that $Q$ is true on any concatenation of the label-paths of
$LP(\Pi_Q)$.
Consider the partial function $\theta$
from variables of $Q$ to nodes in ${\cal M}$ defined as
\begin{eqnarray*}
\theta(x) = v &\Leftrightarrow&
\mbox{$v$ is the topmost node in ${\cal M}$ such that
for all $\pi.x.\pi' \in \Pi_Q|_x$}, \\
&& \mbox{$\pi.x$ can be matched in the path from the root of ${\cal M}$
to $v$.}
\end{eqnarray*}

We say that a variable-path $\pi \in \Pi_Q$ can be matched in a subpath
$\pi'$ of a path structure iff
each of the variables $x$ in $\pi$
can be mapped to a node $\alpha(x)$ in $\pi'$ such that if
$L(x)$ is an atom in $Q$, $\alpha(x)$ carries label $L$, and if $x$ occurs
before $y$ in $\pi$, $\alpha(x)$ occurs before $\alpha(y)$ in $\pi'$.

As ${\cal M}$ is a concatenation of all  paths in $LP(\Pi_Q)$,
for each $x$, the label-paths of all prefixes of paths in $\Pi_Q(x)$
occur in ${\cal M}$.
Thus $\theta$ is  defined for all variables in $Q$. 

The valuation $\theta$  is also consistent.
By definition, $\theta$ satisfies all unary (``label'') atoms.
Consider a binary atom $\child^+(x,y)$ or $\child^*(x,y)$.
(Thus there is a path $\pi.x.y.\pi' \in \Pi_Q$.)
Assume that  $\theta(x) = v$ and $\theta(y) =  w$.
By definition,
$v$ is the topmost node such that all variable-paths with a prefix $\pi_0.x$
can be  matched in the  subpath of ${\cal M}$  from the root  to $v$.
For  each such $\pi_0.x$,  $\pi_0.x.y$  must  match  the  path  from
the  root  of  ${\cal M}$  to $w$. Thus, $w$ must be below $v$ in ${\cal M}$.

\paragraph*{${\cal M}$ is not a model  of $Q'$}
Assume there is a satisfaction  $\theta$ of $Q'$ on ${\cal M}$.
\begin{itemize}
\item[(1)]
By definition, $\theta(x_{i_1})$ cannot be a node in $LC(\neg E_1)$.

\item[($j \rightarrow j+1$)]
Induction step:
Assume that $\theta(x_{i_j})$ cannot be a node in the prefix
$LC(\neg E_1).\cdots.LC(E_1 \land  \dots \land E_{j-1} \land \neg E_j)$
of ${\cal M}$.
For $\theta$ to be a satisfaction, $\theta(x_{i_{j+1}})$ must
either be a descendant of $\theta(x_{i_j})$ or
$\theta(x_{i_{j+1}}) = \theta(x_{i_j})$.
By the induction hypothesis, $\theta(x_{i_{j+1}})$ cannot be in
$LC(\neg E_1).\cdots.LC(E_1 \land  \dots \land E_{j-1} \land \neg E_j)$.
But by definition $\theta(x_{i_{j+1}})$  cannot be a node in
$LC(E_1 \land  \dots \land E_j \land \neg E_{j+1})$ either.
It follows that $\theta(x_{i_{j+1}})$ cannot be a node in
$LC(\neg E_1).\cdots.LC(E_1 \land  \dots \land E_j \land \neg E_{j+1})$.
\end{itemize}
So $\theta(x_{i_m})$ must remain undefined.
Contradiction with our assumption that $\theta$ is a satisfaction of $Q'$
on ${\cal M}$.
\end{proof}

We illustrate the construction by an example.

\begin{figure}
\begin{center}
\input{diamond_example.tex}
\end{center}
\vspace{-6mm}
\caption{Example of the path structure construction of the proof of
Lemma~\ref{lem:excluded_path}.}
\label{fig:diamond_example}
\end{figure}
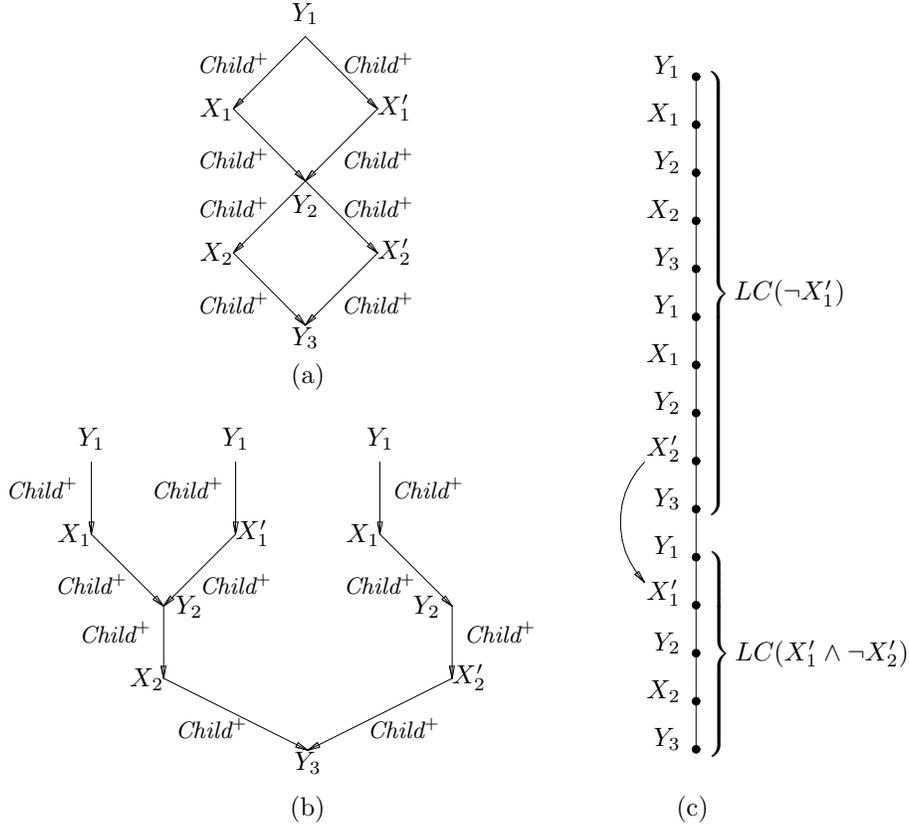

\begin{example}
\em
Consider the 2-diamond query $D_2$ shown in
Figure~\ref{fig:diamond_example}~(a) and the
ABCQ    $Q$   of
Figure~\ref{fig:diamond_example}~(b).  In $Q$
there is no path that contains both $E_1 = X_1'$ and $E_2 = X_2'$,
while $D_2$ contains such a path.
The path-structure
${\cal M} = LC(\neg X_1').LC(X_1' \land \neg X_2')$ constructed as
described above is shown in Figure~\ref{fig:diamond_example}~(c).
It consists of a concatenation of the two paths $Y_1.X_1.Y_2.X_2.Y_3$ and
$Y_1.X_1.Y_2.X_2'.Y_3$  -- which do not contain $X_1'$
(and which  we can  add to ${\cal M}$ in  any order)
--  with the path $Y_1.X_1'.Y_2.X_2.Y_3$, which contains
$X_1'$ but not $X_2'$ and which is therefore appended to ${\cal M}$ after
the other two paths. It is easy to see that indeed $Q$ is true on
${\cal M}$. However, $D_2$ is false on ${\cal M}$.
(The  unique occurrence  of  $X_1'$
in  ${\cal M}$  is a descendant of the unique occurrence of $X_2'$.)
This witnesses that $Q \not\subseteq D_2$.
\punto
\end{example}


\nop{
\section*{Acknowledgments}

We thank Moshe Vardi for valuable suggestions that helped to improve
this article.
} 

\bibliographystyle{acmtrans}
\bibliography{bibtex}



\end{document}

%% file: ex1.tex
\setlength{\unitlength}{0.00083333in}
\begingroup\makeatletter\ifx\SetFigFont\undefined%
\gdef\SetFigFont#1#2#3#4#5{%
  \reset@font\fontsize{#1}{#2pt}%
  \fontfamily{#3}\fontseries{#4}\fontshape{#5}%
  \selectfont}%
\fi\endgroup%
{\renewcommand{\dashlinestretch}{30}
\begin{picture}(1050,918)(0,-10)
\drawline(525,720)(75,270)
\blacken\drawline(117.426,333.640)(75.000,270.000)(138.640,312.426)(117.426,333.640)
\drawline(525,720)(975,270)
\blacken\drawline(911.360,312.426)(975.000,270.000)(932.574,333.640)(911.360,312.426)
\drawline(75,270)(975,270)
\blacken\drawline(900.000,255.000)(975.000,270.000)(900.000,285.000)(900.000,255.000)
\put(525,795){\makebox(0,0)[b]{\smash{{{\SetFigFont{10}{12.0}{\rmdefault}{\mddefault}{\updefault}$S$}}}}}
\put(225,495){\makebox(0,0)[rb]{\smash{{{\SetFigFont{10}{12.0}{\rmdefault}{\mddefault}{\updefault}$\descendant$}}}}}
\put(825,495){\makebox(0,0)[lb]{\smash{{{\SetFigFont{10}{12.0}{\rmdefault}{\mddefault}{\updefault}$\descendant$}}}}}
\put(1050,120){\makebox(0,0)[b]{\smash{{{\SetFigFont{10}{12.0}{\rmdefault}{\mddefault}{\updefault}$PP$}}}}}
\put(0,120){\makebox(0,0)[b]{\smash{{{\SetFigFont{10}{12.0}{\rmdefault}{\mddefault}{\updefault}$NP$}}}}}
\put(525,45){\makebox(0,0)[b]{\smash{{{\SetFigFont{10}{12.0}{\rmdefault}{\mddefault}{\updefault}$\following$}}}}}
\end{picture}
}

%% file: xunderbar.tex
\setlength{\unitlength}{0.00083333in}
\begingroup\makeatletter\ifx\SetFigFont\undefined%
\gdef\SetFigFont#1#2#3#4#5{%
  \reset@font\fontsize{#1}{#2pt}%
  \fontfamily{#3}\fontseries{#4}\fontshape{#5}%
  \selectfont}%
\fi\endgroup%
{\renewcommand{\dashlinestretch}{30}
\begin{picture}(3078,1530)(0,-10)
\put(602,1320){\blacken\ellipse{48}{48}}
\put(602,1320){\ellipse{48}{48}}
\put(602,870){\blacken\ellipse{48}{48}}
\put(602,870){\ellipse{48}{48}}
\put(152,870){\blacken\ellipse{48}{48}}
\put(152,870){\ellipse{48}{48}}
\put(1052,870){\blacken\ellipse{48}{48}}
\put(1052,870){\ellipse{48}{48}}
\put(377,420){\blacken\ellipse{48}{48}}
\put(377,420){\ellipse{48}{48}}
\put(2177,345){\blacken\ellipse{48}{48}}
\put(2177,345){\ellipse{48}{48}}
\put(2177,570){\blacken\ellipse{48}{48}}
\put(2177,570){\ellipse{48}{48}}
\put(2177,1020){\blacken\ellipse{48}{48}}
\put(2177,1020){\ellipse{48}{48}}
\put(2177,1245){\blacken\ellipse{48}{48}}
\put(2177,1245){\ellipse{48}{48}}
\put(2177,1470){\blacken\ellipse{48}{48}}
\put(2177,1470){\ellipse{48}{48}}
\put(2927,345){\blacken\ellipse{48}{48}}
\put(2927,345){\ellipse{48}{48}}
\put(2927,570){\blacken\ellipse{48}{48}}
\put(2927,570){\ellipse{48}{48}}
\put(2927,795){\blacken\ellipse{48}{48}}
\put(2927,795){\ellipse{48}{48}}
\put(2927,1020){\blacken\ellipse{48}{48}}
\put(2927,1020){\ellipse{48}{48}}
\put(2927,1245){\blacken\ellipse{48}{48}}
\put(2927,1245){\ellipse{48}{48}}
\put(2927,1470){\blacken\ellipse{48}{48}}
\put(2927,1470){\ellipse{48}{48}}
\put(2177,795){\blacken\ellipse{48}{48}}
\put(2177,795){\ellipse{48}{48}}
\put(827,420){\blacken\ellipse{48}{48}}
\put(827,420){\ellipse{48}{48}}
\path(2177,1470)(2177,345)
\path(2177,345)(2927,795)
\blacken\path(2870.405,743.550)(2927.000,795.000)(2854.971,769.275)(2870.405,743.550)
\path(2177,345)(2927,570)
\blacken\path(2859.473,534.082)(2927.000,570.000)(2850.853,562.816)(2859.473,534.082)
\path(2927,1470)(2927,345)
\path(602,1320)(602,870)
\blacken\path(587.000,945.000)(602.000,870.000)(617.000,945.000)(587.000,945.000)
\blacken\path(194.426,933.640)(152.000,870.000)(215.640,912.426)(194.426,933.640)
\path(152,870)(602,1320)
\path(602,1320)(1052,870)
\blacken\path(988.360,912.426)(1052.000,870.000)(1009.574,933.640)(988.360,912.426)
\thicklines
\path(2177,345)(2927,1020)
\blacken\thinlines
\path(2881.287,958.678)(2927.000,1020.000)(2861.218,980.977)(2881.287,958.678)
\thicklines
\path(2177,345)(2927,1245)
\blacken\thinlines
\path(2890.509,1177.781)(2927.000,1245.000)(2867.463,1196.986)(2890.509,1177.781)
\thicklines
\path(2177,795)(2927,1020)
\blacken\thinlines
\path(2859.473,984.082)(2927.000,1020.000)(2850.853,1012.816)(2859.473,984.082)
\path(2177,345)(2927,1470)
\blacken\path(2897.878,1399.276)(2927.000,1470.000)(2872.917,1415.917)(2897.878,1399.276)
\path(2177,795)(2927,1245)
\blacken\path(2870.405,1193.550)(2927.000,1245.000)(2854.971,1219.275)(2870.405,1193.550)
\path(602,870)(827,420)
\blacken\path(780.043,480.374)(827.000,420.000)(806.875,493.790)(780.043,480.374)
\path(602,870)(377,420)
\blacken\path(397.125,493.790)(377.000,420.000)(423.957,480.374)(397.125,493.790)
\path(602,1320)(827,420)
\blacken\path(794.258,489.123)(827.000,420.000)(823.362,496.399)(794.258,489.123)
\path(602,1320)(377,420)
\blacken\path(380.638,496.399)(377.000,420.000)(409.742,489.123)(380.638,496.399)
\put(2102,270){\makebox(0,0)[rb]{{\SetFigFont{10}{12.0}{\rmdefault}{\mddefault}{\updefault}1}}}
\put(2102,495){\makebox(0,0)[rb]{{\SetFigFont{10}{12.0}{\rmdefault}{\mddefault}{\updefault}2}}}
\put(2102,720){\makebox(0,0)[rb]{{\SetFigFont{10}{12.0}{\rmdefault}{\mddefault}{\updefault}3}}}
\put(2102,945){\makebox(0,0)[rb]{{\SetFigFont{10}{12.0}{\rmdefault}{\mddefault}{\updefault}4}}}
\put(2102,1170){\makebox(0,0)[rb]{{\SetFigFont{10}{12.0}{\rmdefault}{\mddefault}{\updefault}5}}}
\put(2102,1395){\makebox(0,0)[rb]{{\SetFigFont{10}{12.0}{\rmdefault}{\mddefault}{\updefault}6}}}
\put(3002,270){\makebox(0,0)[lb]{{\SetFigFont{10}{12.0}{\rmdefault}{\mddefault}{\updefault}1}}}
\put(3002,495){\makebox(0,0)[lb]{{\SetFigFont{10}{12.0}{\rmdefault}{\mddefault}{\updefault}2}}}
\put(3002,720){\makebox(0,0)[lb]{{\SetFigFont{10}{12.0}{\rmdefault}{\mddefault}{\updefault}3}}}
\put(3002,1170){\makebox(0,0)[lb]{{\SetFigFont{10}{12.0}{\rmdefault}{\mddefault}{\updefault}5}}}
\put(3002,1395){\makebox(0,0)[lb]{{\SetFigFont{10}{12.0}{\rmdefault}{\mddefault}{\updefault}6}}}
\put(3002,945){\makebox(0,0)[lb]{{\SetFigFont{10}{12.0}{\rmdefault}{\mddefault}{\updefault}4}}}
\put(527,1320){\makebox(0,0)[rb]{{\SetFigFont{10}{12.0}{\rmdefault}{\mddefault}{\updefault}1}}}
\put(77,795){\makebox(0,0)[rb]{{\SetFigFont{10}{12.0}{\rmdefault}{\mddefault}{\updefault}2}}}
\put(302,345){\makebox(0,0)[rb]{{\SetFigFont{10}{12.0}{\rmdefault}{\mddefault}{\updefault}4}}}
\put(902,345){\makebox(0,0)[lb]{{\SetFigFont{10}{12.0}{\rmdefault}{\mddefault}{\updefault}5}}}
\put(1127,795){\makebox(0,0)[lb]{{\SetFigFont{10}{12.0}{\rmdefault}{\mddefault}{\updefault}6}}}
\put(602,645){\makebox(0,0)[b]{{\SetFigFont{10}{12.0}{\rmdefault}{\mddefault}{\updefault}3}}}
\put(2552,45){\makebox(0,0)[b]{{\SetFigFont{10}{12.0}{\rmdefault}{\mddefault}{\updefault}(b)}}}
\put(602,45){\makebox(0,0)[b]{{\SetFigFont{10}{12.0}{\rmdefault}{\mddefault}{\updefault}(a)}}}
\end{picture}
}

%% file: child_childs_data.tex
\setlength{\unitlength}{0.00087489in}
\begingroup\makeatletter\ifx\SetFigFont\undefined%
\gdef\SetFigFont#1#2#3#4#5{%
  \reset@font\fontsize{#1}{#2pt}%
  \fontfamily{#3}\fontseries{#4}\fontshape{#5}%
  \selectfont}%
\fi\endgroup%
{\renewcommand{\dashlinestretch}{30}
\begin{picture}(2205,3415)(0,-10)
\put(2025,886){\makebox(0,0)[rb]{\smash{{{\SetFigFont{8}{9.6}{\rmdefault}{\mddefault}{\updefault}$L_1, L_2$}}}}}
\put(90,2281){\circle*{40}}
\put(1080,2551){\circle*{40}}
\put(1080,2281){\circle*{40}}
\put(1080,2821){\circle*{40}}
\put(1080,3091){\circle*{40}}
\put(1080,3361){\circle*{40}}
\put(2070,2551){\circle*{40}}
\put(2070,2281){\circle*{40}}
\put(90,931){\circle*{40}}
\put(90,661){\circle*{40}}
\put(90,391){\circle*{40}}
\put(90,121){\circle*{40}}
\put(1080,931){\circle*{40}}
\put(1080,661){\circle*{40}}
\put(1080,391){\circle*{40}}
\put(1080,121){\circle*{40}}
\put(2070,931){\circle*{40}}
\put(2070,661){\circle*{40}}
\put(2070,121){\circle*{40}}
\put(2070,391){\circle*{40}}
\put(90,2011){\circle*{40}}
\put(90,1741){\circle*{40}}
\put(90,1471){\circle*{40}}
\put(90,1201){\circle*{40}}
\put(1080,2011){\circle*{40}}
\put(1080,1741){\circle*{40}}
\put(1080,1471){\circle*{40}}
\put(1080,1201){\circle*{40}}
\put(2070,2011){\circle*{40}}
\put(2070,1741){\circle*{40}}
\put(2070,1201){\circle*{40}}
\put(2070,1471){\circle*{40}}
\drawline(90,2551)(90,2281)
\drawline(90,2281)(90,2011)
\drawline(1080,2551)(1080,2281)
\drawline(1080,2281)(1080,2011)
\drawline(1080,3091)(1080,2821)
\drawline(1080,3361)(1080,3091)
\drawline(2070,2551)(2070,2281)
\drawline(2070,2281)(2070,2011)
\drawline(1080,2821)(90,2551)
\drawline(1080,2821)(2070,2551)
\drawline(1080,2821)(1080,2551)
\drawline(90,1201)(90,931)
\drawline(1080,1201)(1080,931)
\drawline(2070,1201)(2070,931)
\drawline(90,931)(90,661)
\drawline(90,661)(90,391)
\drawline(90,391)(90,121)
\drawline(1080,931)(1080,661)
\drawline(1080,661)(1080,391)
\drawline(1080,391)(1080,121)
\drawline(2070,931)(2070,661)
\drawline(2070,661)(2070,391)
\drawline(2070,391)(2070,121)
\drawline(90,2011)(90,1741)
\drawline(90,1741)(90,1471)
\drawline(90,1471)(90,1201)
\drawline(1080,2011)(1080,1741)
\drawline(1080,1741)(1080,1471)
\drawline(1080,1471)(1080,1201)
\drawline(2070,2011)(2070,1741)
\drawline(2070,1741)(2070,1471)
\drawline(2070,1471)(2070,1201)
\put(0,2506){\makebox(0,0)[b]{\smash{{{\SetFigFont{8}{9.6}{\rmdefault}{\mddefault}{\updefault}$Y$}}}}}
\put(990,3316){\makebox(0,0)[b]{\smash{{{\SetFigFont{8}{9.6}{\rmdefault}{\mddefault}{\updefault}$X$}}}}}
\put(1170,2821){\makebox(0,0)[b]{\smash{{{\SetFigFont{8}{9.6}{\rmdefault}{\mddefault}{\updefault}$v_3$}}}}}
\put(1170,3046){\makebox(0,0)[b]{\smash{{{\SetFigFont{8}{9.6}{\rmdefault}{\mddefault}{\updefault}$v_2$}}}}}
\put(1170,3316){\makebox(0,0)[b]{\smash{{{\SetFigFont{8}{9.6}{\rmdefault}{\mddefault}{\updefault}$v_1$}}}}}
\put(990,3046){\makebox(0,0)[b]{\smash{{{\SetFigFont{8}{9.6}{\rmdefault}{\mddefault}{\updefault}$X$}}}}}
\put(990,2821){\makebox(0,0)[b]{\smash{{{\SetFigFont{8}{9.6}{\rmdefault}{\mddefault}{\updefault}$X$}}}}}
\put(990,2236){\makebox(0,0)[b]{\smash{{{\SetFigFont{8}{9.6}{\rmdefault}{\mddefault}{\updefault}$Y$}}}}}
\put(225,2191){\makebox(0,0)[b]{\smash{{{\SetFigFont{8}{9.6}{\rmdefault}{\mddefault}{\updefault}$w_{1,2}$}}}}}
\put(1215,2191){\makebox(0,0)[b]{\smash{{{\SetFigFont{8}{9.6}{\rmdefault}{\mddefault}{\updefault}$w_{2,2}$}}}}}
\put(2205,2191){\makebox(0,0)[b]{\smash{{{\SetFigFont{8}{9.6}{\rmdefault}{\mddefault}{\updefault}$w_{3,2}$}}}}}
\put(225,2461){\makebox(0,0)[b]{\smash{{{\SetFigFont{8}{9.6}{\rmdefault}{\mddefault}{\updefault}$w_{1,1}$}}}}}
\put(1215,2461){\makebox(0,0)[b]{\smash{{{\SetFigFont{8}{9.6}{\rmdefault}{\mddefault}{\updefault}$w_{2,1}$}}}}}
\put(2205,2461){\makebox(0,0)[b]{\smash{{{\SetFigFont{8}{9.6}{\rmdefault}{\mddefault}{\updefault}$w_{3,1}$}}}}}
\put(1035,616){\makebox(0,0)[rb]{\smash{{{\SetFigFont{8}{9.6}{\rmdefault}{\mddefault}{\updefault}$L_1, L_3$}}}}}
\put(45,346){\makebox(0,0)[rb]{\smash{{{\SetFigFont{8}{9.6}{\rmdefault}{\mddefault}{\updefault}$L_2, L_3$}}}}}
\put(2025,346){\makebox(0,0)[rb]{\smash{{{\SetFigFont{8}{9.6}{\rmdefault}{\mddefault}{\updefault}$L_1, L_2$}}}}}
\put(1035,76){\makebox(0,0)[rb]{\smash{{{\SetFigFont{8}{9.6}{\rmdefault}{\mddefault}{\updefault}$L_1, L_3$}}}}}
\put(225,841){\makebox(0,0)[b]{\smash{{{\SetFigFont{8}{9.6}{\rmdefault}{\mddefault}{\updefault}$w_{1,7}$}}}}}
\put(225,571){\makebox(0,0)[b]{\smash{{{\SetFigFont{8}{9.6}{\rmdefault}{\mddefault}{\updefault}$w_{1,8}$}}}}}
\put(225,301){\makebox(0,0)[b]{\smash{{{\SetFigFont{8}{9.6}{\rmdefault}{\mddefault}{\updefault}$w_{1,9}$}}}}}
\put(225,31){\makebox(0,0)[b]{\smash{{{\SetFigFont{8}{9.6}{\rmdefault}{\mddefault}{\updefault}$w_{1,10}$}}}}}
\put(1215,31){\makebox(0,0)[b]{\smash{{{\SetFigFont{8}{9.6}{\rmdefault}{\mddefault}{\updefault}$w_{2,10}$}}}}}
\put(1215,571){\makebox(0,0)[b]{\smash{{{\SetFigFont{8}{9.6}{\rmdefault}{\mddefault}{\updefault}$w_{2,8}$}}}}}
\put(1215,841){\makebox(0,0)[b]{\smash{{{\SetFigFont{8}{9.6}{\rmdefault}{\mddefault}{\updefault}$w_{2,7}$}}}}}
\put(2205,841){\makebox(0,0)[b]{\smash{{{\SetFigFont{8}{9.6}{\rmdefault}{\mddefault}{\updefault}$w_{3,7}$}}}}}
\put(2205,571){\makebox(0,0)[b]{\smash{{{\SetFigFont{8}{9.6}{\rmdefault}{\mddefault}{\updefault}$w_{3,8}$}}}}}
\put(2205,301){\makebox(0,0)[b]{\smash{{{\SetFigFont{8}{9.6}{\rmdefault}{\mddefault}{\updefault}$w_{3,9}$}}}}}
\put(2205,31){\makebox(0,0)[b]{\smash{{{\SetFigFont{8}{9.6}{\rmdefault}{\mddefault}{\updefault}$w_{3,10}$}}}}}
\put(45,76){\makebox(0,0)[rb]{\smash{{{\SetFigFont{8}{9.6}{\rmdefault}{\mddefault}{\updefault}$L_2, L_3$}}}}}
\put(1980,1966){\makebox(0,0)[b]{\smash{{{\SetFigFont{8}{9.6}{\rmdefault}{\mddefault}{\updefault}$Y$}}}}}
\put(1035,1696){\makebox(0,0)[rb]{\smash{{{\SetFigFont{8}{9.6}{\rmdefault}{\mddefault}{\updefault}$L_1, L_3$}}}}}
\put(2025,1696){\makebox(0,0)[rb]{\smash{{{\SetFigFont{8}{9.6}{\rmdefault}{\mddefault}{\updefault}$L_1, L_2$}}}}}
\put(45,1426){\makebox(0,0)[rb]{\smash{{{\SetFigFont{8}{9.6}{\rmdefault}{\mddefault}{\updefault}$L_2, L_3$}}}}}
\put(2025,1426){\makebox(0,0)[rb]{\smash{{{\SetFigFont{8}{9.6}{\rmdefault}{\mddefault}{\updefault}$L_1, L_2$}}}}}
\put(1035,1156){\makebox(0,0)[rb]{\smash{{{\SetFigFont{8}{9.6}{\rmdefault}{\mddefault}{\updefault}$L_1, L_3$}}}}}
\put(225,1921){\makebox(0,0)[b]{\smash{{{\SetFigFont{8}{9.6}{\rmdefault}{\mddefault}{\updefault}$w_{1,3}$}}}}}
\put(225,1651){\makebox(0,0)[b]{\smash{{{\SetFigFont{8}{9.6}{\rmdefault}{\mddefault}{\updefault}$w_{1,4}$}}}}}
\put(225,1381){\makebox(0,0)[b]{\smash{{{\SetFigFont{8}{9.6}{\rmdefault}{\mddefault}{\updefault}$w_{1,5}$}}}}}
\put(225,1111){\makebox(0,0)[b]{\smash{{{\SetFigFont{8}{9.6}{\rmdefault}{\mddefault}{\updefault}$w_{1,6}$}}}}}
\put(1215,1111){\makebox(0,0)[b]{\smash{{{\SetFigFont{8}{9.6}{\rmdefault}{\mddefault}{\updefault}$w_{2,6}$}}}}}
\put(1215,1381){\makebox(0,0)[b]{\smash{{{\SetFigFont{8}{9.6}{\rmdefault}{\mddefault}{\updefault}$w_{2,5}$}}}}}
\put(1215,1651){\makebox(0,0)[b]{\smash{{{\SetFigFont{8}{9.6}{\rmdefault}{\mddefault}{\updefault}$w_{2,4}$}}}}}
\put(1215,1921){\makebox(0,0)[b]{\smash{{{\SetFigFont{8}{9.6}{\rmdefault}{\mddefault}{\updefault}$w_{2,3}$}}}}}
\put(2205,1921){\makebox(0,0)[b]{\smash{{{\SetFigFont{8}{9.6}{\rmdefault}{\mddefault}{\updefault}$w_{3,3}$}}}}}
\put(2205,1651){\makebox(0,0)[b]{\smash{{{\SetFigFont{8}{9.6}{\rmdefault}{\mddefault}{\updefault}$w_{3,4}$}}}}}
\put(2205,1381){\makebox(0,0)[b]{\smash{{{\SetFigFont{8}{9.6}{\rmdefault}{\mddefault}{\updefault}$w_{3,5}$}}}}}
\put(2205,1111){\makebox(0,0)[b]{\smash{{{\SetFigFont{8}{9.6}{\rmdefault}{\mddefault}{\updefault}$w_{3,6}$}}}}}
\put(45,886){\makebox(0,0)[rb]{\smash{{{\SetFigFont{8}{9.6}{\rmdefault}{\mddefault}{\updefault}$L_2, L_3$}}}}}
\put(45,616){\makebox(0,0)[rb]{\smash{{{\SetFigFont{8}{9.6}{\rmdefault}{\mddefault}{\updefault}$L_2, L_3$}}}}}
\put(45,1696){\makebox(0,0)[rb]{\smash{{{\SetFigFont{8}{9.6}{\rmdefault}{\mddefault}{\updefault}$L_2, L_3$}}}}}
\put(45,1156){\makebox(0,0)[rb]{\smash{{{\SetFigFont{8}{9.6}{\rmdefault}{\mddefault}{\updefault}$L_1, L_2, L_3$}}}}}
\put(1035,886){\makebox(0,0)[rb]{\smash{{{\SetFigFont{8}{9.6}{\rmdefault}{\mddefault}{\updefault}$L_1, L_2, L_3$}}}}}
\put(1035,1426){\makebox(0,0)[rb]{\smash{{{\SetFigFont{8}{9.6}{\rmdefault}{\mddefault}{\updefault}$L_1, L_3$}}}}}
\put(2025,76){\makebox(0,0)[rb]{\smash{{{\SetFigFont{8}{9.6}{\rmdefault}{\mddefault}{\updefault}$L_1, L_2$}}}}}
\put(2025,1156){\makebox(0,0)[rb]{\smash{{{\SetFigFont{8}{9.6}{\rmdefault}{\mddefault}{\updefault}$L_1, L_2$}}}}}
\put(1035,346){\makebox(0,0)[rb]{\smash{{{\SetFigFont{8}{9.6}{\rmdefault}{\mddefault}{\updefault}$L_1, L_3$}}}}}
\put(2025,616){\makebox(0,0)[rb]{\smash{{{\SetFigFont{8}{9.6}{\rmdefault}{\mddefault}{\updefault}$L_1, L_2, L_3$}}}}}
\put(1215,301){\makebox(0,0)[b]{\smash{{{\SetFigFont{8}{9.6}{\rmdefault}{\mddefault}{\updefault}$w_{2,9}$}}}}}
\put(90,2551){\circle*{40}}
\end{picture}
}

%% file: child_following.tex
\setlength{\unitlength}{0.00087489in}
\begingroup\makeatletter\ifx\SetFigFont\undefined%
\gdef\SetFigFont#1#2#3#4#5{%
  \reset@font\fontsize{#1}{#2pt}%
  \fontfamily{#3}\fontseries{#4}\fontshape{#5}%
  \selectfont}%
\fi\endgroup%
{\renewcommand{\dashlinestretch}{30}
\begin{picture}(3296,3592)(0,-10)
\put(3043,2695){\makebox(0,0)[b]{\smash{{{\SetFigFont{8}{9.6}{\rmdefault}{\mddefault}{\updefault}7}}}}}
\put(1648,2740){\circle*{40}}
\put(1648,3145){\circle*{40}}
\put(1648,3550){\circle*{40}}
\put(1333,2740){\circle*{40}}
\put(1018,2740){\circle*{40}}
\put(1963,2740){\circle*{40}}
\put(1963,2335){\circle*{40}}
\put(1333,2335){\circle*{40}}
\put(2278,2740){\circle*{40}}
\put(343,3145){\circle*{40}}
\put(343,2740){\circle*{40}}
\put(343,2335){\circle*{40}}
\put(2953,3145){\circle*{40}}
\put(2953,2740){\circle*{40}}
\put(2953,2335){\circle*{40}}
\put(28,2740){\circle*{40}}
\put(658,2740){\circle*{40}}
\put(2638,2740){\circle*{40}}
\put(3268,2740){\circle*{40}}
\put(1648,1210){\circle*{40}}
\put(478,1210){\circle*{40}}
\put(2818,1210){\circle*{40}}
\put(1648,445){\circle*{40}}
\put(2818,445){\circle*{40}}
\put(478,445){\circle*{40}}
\drawline(1648,3550)(1648,3145)
\drawline(1648,3550)(343,3145)
\drawline(1648,3550)(2953,3145)
\drawline(1648,3145)(1648,2740)
\drawline(1648,3145)(1018,2740)
\drawline(1648,3145)(1333,2740)
\drawline(1648,3145)(2278,2740)
\drawline(1648,3145)(1963,2740)
\drawline(1333,2740)(1333,2335)
\drawline(1963,2740)(1963,2335)
\drawline(343,3145)(343,2740)
\drawline(343,3145)(658,2740)
\drawline(343,2740)(343,2335)
\drawline(343,3145)(28,2740)
\drawline(2953,3145)(2638,2740)
\drawline(2953,3145)(2953,2740)
\drawline(2953,3145)(3268,2740)
\drawline(2953,2740)(2953,2335)
\drawline(1648,1210)(1648,445)
\blacken\drawline(1633.000,520.000)(1648.000,445.000)(1663.000,520.000)(1633.000,520.000)
\drawline(478,1210)(1648,1210)
\blacken\drawline(1573.000,1195.000)(1648.000,1210.000)(1573.000,1225.000)(1573.000,1195.000)
\drawline(478,1210)(478,445)
\blacken\drawline(463.000,520.000)(478.000,445.000)(493.000,520.000)(463.000,520.000)
\drawline(1648,1210)(2818,1210)
\blacken\drawline(2743.000,1195.000)(2818.000,1210.000)(2743.000,1225.000)(2743.000,1195.000)
\drawline(2818,1210)(2818,445)
\blacken\drawline(2803.000,520.000)(2818.000,445.000)(2833.000,520.000)(2803.000,520.000)
\drawline(478,445)(1648,445)
\blacken\drawline(1573.000,430.000)(1648.000,445.000)(1573.000,460.000)(1573.000,430.000)
\drawline(1648,445)(2818,445)
\blacken\drawline(2743.000,430.000)(2818.000,445.000)(2743.000,460.000)(2743.000,430.000)
\put(1648,1975){\makebox(0,0)[b]{\smash{{{\SetFigFont{10}{12.0}{\rmdefault}{\mddefault}{\updefault}(a)}}}}}
\put(1648,2560){\makebox(0,0)[b]{\smash{{{\SetFigFont{10}{12.0}{\rmdefault}{\mddefault}{\updefault}$B$}}}}}
\put(1963,2155){\makebox(0,0)[b]{\smash{{{\SetFigFont{10}{12.0}{\rmdefault}{\mddefault}{\updefault}$B$}}}}}
\put(1333,2155){\makebox(0,0)[b]{\smash{{{\SetFigFont{10}{12.0}{\rmdefault}{\mddefault}{\updefault}$B$}}}}}
\put(1783,3145){\makebox(0,0)[b]{\smash{{{\SetFigFont{10}{12.0}{\rmdefault}{\mddefault}{\updefault}$L_2$}}}}}
\put(3268,2560){\makebox(0,0)[b]{\smash{{{\SetFigFont{10}{12.0}{\rmdefault}{\mddefault}{\updefault}$C$}}}}}
\put(3088,3145){\makebox(0,0)[b]{\smash{{{\SetFigFont{10}{12.0}{\rmdefault}{\mddefault}{\updefault}$L_3$}}}}}
\put(208,3145){\makebox(0,0)[b]{\smash{{{\SetFigFont{10}{12.0}{\rmdefault}{\mddefault}{\updefault}$L_1$}}}}}
\put(28,2560){\makebox(0,0)[b]{\smash{{{\SetFigFont{10}{12.0}{\rmdefault}{\mddefault}{\updefault}$A$}}}}}
\put(2908,3190){\makebox(0,0)[b]{\smash{{{\SetFigFont{8}{9.6}{\rmdefault}{\mddefault}{\updefault}6}}}}}
\put(1558,3145){\makebox(0,0)[b]{\smash{{{\SetFigFont{8}{9.6}{\rmdefault}{\mddefault}{\updefault}3}}}}}
\put(388,3190){\makebox(0,0)[b]{\smash{{{\SetFigFont{8}{9.6}{\rmdefault}{\mddefault}{\updefault}1}}}}}
\put(1243,2605){\makebox(0,0)[b]{\smash{{{\SetFigFont{10}{12.0}{\rmdefault}{\mddefault}{\updefault}$L_2$}}}}}
\put(2863,2605){\makebox(0,0)[b]{\smash{{{\SetFigFont{10}{12.0}{\rmdefault}{\mddefault}{\updefault}$L_3$}}}}}
\put(1873,2695){\makebox(0,0)[b]{\smash{{{\SetFigFont{8}{9.6}{\rmdefault}{\mddefault}{\updefault}5}}}}}
\put(3088,760){\makebox(0,0)[b]{\smash{{{\SetFigFont{10}{12.0}{\rmdefault}{\mddefault}{\updefault}\child}}}}}
\put(1648,1615){\makebox(0,0)[b]{\smash{{{\SetFigFont{10}{12.0}{\rmdefault}{\mddefault}{\updefault}\following$^7$}}}}}
\put(208,760){\makebox(0,0)[b]{\smash{{{\SetFigFont{10}{12.0}{\rmdefault}{\mddefault}{\updefault}\child}}}}}
\put(1378,760){\makebox(0,0)[b]{\smash{{{\SetFigFont{10}{12.0}{\rmdefault}{\mddefault}{\updefault}\child}}}}}
\put(1648,265){\makebox(0,0)[b]{\smash{{{\SetFigFont{10}{12.0}{\rmdefault}{\mddefault}{\updefault}$B$}}}}}
\put(2233,265){\makebox(0,0)[b]{\smash{{{\SetFigFont{10}{12.0}{\rmdefault}{\mddefault}{\updefault}\following$^4$}}}}}
\put(2818,265){\makebox(0,0)[b]{\smash{{{\SetFigFont{10}{12.0}{\rmdefault}{\mddefault}{\updefault}$C$}}}}}
\put(478,265){\makebox(0,0)[b]{\smash{{{\SetFigFont{10}{12.0}{\rmdefault}{\mddefault}{\updefault}$A$}}}}}
\put(1018,265){\makebox(0,0)[b]{\smash{{{\SetFigFont{10}{12.0}{\rmdefault}{\mddefault}{\updefault}\following$^4$}}}}}
\put(1648,40){\makebox(0,0)[b]{\smash{{{\SetFigFont{10}{12.0}{\rmdefault}{\mddefault}{\updefault}(b)}}}}}
\put(1648,1300){\makebox(0,0)[b]{\smash{{{\SetFigFont{10}{12.0}{\rmdefault}{\mddefault}{\updefault}$L_2$}}}}}
\put(2818,1300){\makebox(0,0)[b]{\smash{{{\SetFigFont{10}{12.0}{\rmdefault}{\mddefault}{\updefault}$L_3$}}}}}
\put(478,1300){\makebox(0,0)[b]{\smash{{{\SetFigFont{10}{12.0}{\rmdefault}{\mddefault}{\updefault}$L_1$}}}}}
\put(1018,1030){\makebox(0,0)[b]{\smash{{{\SetFigFont{10}{12.0}{\rmdefault}{\mddefault}{\updefault}\following$^2$}}}}}
\put(2233,1030){\makebox(0,0)[b]{\smash{{{\SetFigFont{10}{12.0}{\rmdefault}{\mddefault}{\updefault}\following$^2$}}}}}
\put(343,2155){\makebox(0,0)[b]{\smash{{{\SetFigFont{10}{12.0}{\rmdefault}{\mddefault}{\updefault}$A$}}}}}
\put(2953,2155){\makebox(0,0)[b]{\smash{{{\SetFigFont{10}{12.0}{\rmdefault}{\mddefault}{\updefault}$C$}}}}}
\put(478,2605){\makebox(0,0)[b]{\smash{{{\SetFigFont{10}{12.0}{\rmdefault}{\mddefault}{\updefault}$L_1$}}}}}
\put(2098,2605){\makebox(0,0)[b]{\smash{{{\SetFigFont{10}{12.0}{\rmdefault}{\mddefault}{\updefault}$L_2$}}}}}
\put(253,2695){\makebox(0,0)[b]{\smash{{{\SetFigFont{8}{9.6}{\rmdefault}{\mddefault}{\updefault}2}}}}}
\put(1423,2695){\makebox(0,0)[b]{\smash{{{\SetFigFont{8}{9.6}{\rmdefault}{\mddefault}{\updefault}4}}}}}
\drawline(2818.000,1210.000)(2755.094,1245.230)(2691.128,1278.497)
	(2626.163,1309.770)(2560.262,1339.017)(2493.488,1366.212)
	(2425.904,1391.329)(2357.576,1414.342)(2288.569,1435.231)
	(2218.948,1453.975)(2148.781,1470.556)(2078.135,1484.959)
	(2007.076,1497.169)(1935.674,1507.176)(1863.997,1514.968)
	(1792.113,1520.540)(1720.091,1523.885)(1648.000,1525.000)
	(1575.909,1523.885)(1503.887,1520.540)(1432.003,1514.968)
	(1360.326,1507.176)(1288.924,1497.169)(1217.865,1484.959)
	(1147.219,1470.556)(1077.052,1453.975)(1007.431,1435.231)
	(938.424,1414.342)(870.096,1391.329)(802.512,1366.212)
	(735.738,1339.017)(669.837,1309.770)(604.872,1278.497)
	(540.906,1245.230)(478.000,1210.000)
\blacken\drawline(2745.157,1233.321)(2818.000,1210.000)(2759.730,1259.544)(2745.157,1233.321)
\end{picture}
}

%% file: child_following_qc.tex
\setlength{\unitlength}{0.00087489in}
\begingroup\makeatletter\ifx\SetFigFont\undefined%
\gdef\SetFigFont#1#2#3#4#5{%
  \reset@font\fontsize{#1}{#2pt}%
  \fontfamily{#3}\fontseries{#4}\fontshape{#5}%
  \selectfont}%
\fi\endgroup%
{\renewcommand{\dashlinestretch}{30}
\begin{picture}(1914,646)(0,-10)
\put(327,424){\circle*{40}}
\drawline(330,422)(15,17)
\drawline(318,417)(633,12)
\drawline(12,19)(642,19)
\put(1587,424){\circle*{40}}
\drawline(1590,422)(1275,17)
\drawline(1578,417)(1893,12)
\drawline(1272,19)(1902,19)
\put(957,604){\circle*{40}}
\drawline(957,604)(327,424)
\drawline(957,604)(1587,424)
\put(327,109){\makebox(0,0)[b]{\smash{{{\SetFigFont{10}{12.0}{\rmdefault}{\mddefault}{\updefault}$T$}}}}}
\put(1587,109){\makebox(0,0)[b]{\smash{{{\SetFigFont{10}{12.0}{\rmdefault}{\mddefault}{\updefault}$T$}}}}}
\end{picture}
}

%% file: auxiliary_nodes.tex
\setlength{\unitlength}{0.00087489in}
\begingroup\makeatletter\ifx\SetFigFont\undefined%
\gdef\SetFigFont#1#2#3#4#5{%
  \reset@font\fontsize{#1}{#2pt}%
  \fontfamily{#3}\fontseries{#4}\fontshape{#5}%
  \selectfont}%
\fi\endgroup%
{\renewcommand{\dashlinestretch}{30}
\begin{picture}(5996,1927)(0,-10)
\put(2503,1030){\circle*{40}}
\put(2503,625){\circle*{40}}
\put(2503,220){\circle*{40}}
\put(2188,625){\circle*{40}}
\put(2818,625){\circle*{40}}
\put(2413,805){\circle*{40}}
\put(2593,805){\circle*{40}}
\drawline(2503,1030)(2188,625)
\drawline(2503,1030)(2503,625)
\drawline(2503,1030)(2818,625)
\drawline(2503,625)(2503,220)
\drawline(2503,1030)(2413,805)
\drawline(2503,1030)(2593,805)
\put(2818,445){\makebox(0,0)[b]{\smash{{{\SetFigFont{10}{12.0}{\rmdefault}{\mddefault}{\updefault}$C$}}}}}
\put(2638,1030){\makebox(0,0)[b]{\smash{{{\SetFigFont{10}{12.0}{\rmdefault}{\mddefault}{\updefault}$L_3$}}}}}
\put(2458,1075){\makebox(0,0)[b]{\smash{{{\SetFigFont{8}{9.6}{\rmdefault}{\mddefault}{\updefault}6}}}}}
\put(2503,40){\makebox(0,0)[b]{\smash{{{\SetFigFont{10}{12.0}{\rmdefault}{\mddefault}{\updefault}$C$}}}}}
\put(2368,670){\makebox(0,0)[b]{\smash{{{\SetFigFont{8}{9.6}{\rmdefault}{\mddefault}{\updefault}$H$}}}}}
\put(2638,670){\makebox(0,0)[b]{\smash{{{\SetFigFont{8}{9.6}{\rmdefault}{\mddefault}{\updefault}$H$}}}}}
\put(2593,535){\makebox(0,0)[b]{\smash{{{\SetFigFont{8}{9.6}{\rmdefault}{\mddefault}{\updefault}7}}}}}
\put(2413,490){\makebox(0,0)[b]{\smash{{{\SetFigFont{10}{12.0}{\rmdefault}{\mddefault}{\updefault}$L_3$}}}}}
\put(343,1030){\circle*{40}}
\put(343,220){\circle*{40}}
\put(28,625){\circle*{40}}
\put(658,625){\circle*{40}}
\put(253,805){\circle*{40}}
\put(433,805){\circle*{40}}
\put(343,625){\circle*{40}}
\drawline(343,1030)(343,625)
\drawline(343,1030)(658,625)
\drawline(343,625)(343,220)
\drawline(343,1030)(28,625)
\drawline(343,1030)(253,805)
\drawline(343,1030)(433,805)
\put(28,445){\makebox(0,0)[b]{\smash{{{\SetFigFont{10}{12.0}{\rmdefault}{\mddefault}{\updefault}$A$}}}}}
\put(388,1075){\makebox(0,0)[b]{\smash{{{\SetFigFont{8}{9.6}{\rmdefault}{\mddefault}{\updefault}1}}}}}
\put(208,1030){\makebox(0,0)[b]{\smash{{{\SetFigFont{10}{12.0}{\rmdefault}{\mddefault}{\updefault}$L_1$}}}}}
\put(343,40){\makebox(0,0)[b]{\smash{{{\SetFigFont{10}{12.0}{\rmdefault}{\mddefault}{\updefault}$A$}}}}}
\put(478,670){\makebox(0,0)[b]{\smash{{{\SetFigFont{8}{9.6}{\rmdefault}{\mddefault}{\updefault}$H$}}}}}
\put(208,670){\makebox(0,0)[b]{\smash{{{\SetFigFont{8}{9.6}{\rmdefault}{\mddefault}{\updefault}$H$}}}}}
\put(253,535){\makebox(0,0)[b]{\smash{{{\SetFigFont{8}{9.6}{\rmdefault}{\mddefault}{\updefault}2}}}}}
\put(433,490){\makebox(0,0)[b]{\smash{{{\SetFigFont{10}{12.0}{\rmdefault}{\mddefault}{\updefault}$L_1$}}}}}
\put(1423,1435){\circle*{40}}
\put(1423,625){\circle*{40}}
\put(1423,1030){\circle*{40}}
\put(1108,625){\circle*{40}}
\put(793,625){\circle*{40}}
\put(1738,625){\circle*{40}}
\put(1738,220){\circle*{40}}
\put(1108,220){\circle*{40}}
\put(2053,625){\circle*{40}}
\put(1198,1210){\circle*{40}}
\put(1648,1210){\circle*{40}}
\put(1513,805){\circle*{40}}
\put(1333,805){\circle*{40}}
\put(1198,805){\circle*{40}}
\put(1648,805){\circle*{40}}
\drawline(1423,1435)(1423,1030)
\drawline(1423,1435)(343,1030)
\drawline(1423,1435)(2503,1030)
\drawline(1423,1030)(1423,625)
\drawline(1423,1030)(793,625)
\drawline(1423,1030)(1108,625)
\drawline(1423,1030)(2053,625)
\drawline(1423,1030)(1738,625)
\drawline(1108,625)(1108,220)
\drawline(1738,625)(1738,220)
\drawline(1423,1435)(1198,1210)
\drawline(1423,1435)(1648,1210)
\drawline(1423,1030)(1198,805)
\drawline(1423,1030)(1333,805)
\drawline(1423,1030)(1513,805)
\drawline(1423,1030)(1648,805)
\put(1423,445){\makebox(0,0)[b]{\smash{{{\SetFigFont{10}{12.0}{\rmdefault}{\mddefault}{\updefault}$B$}}}}}
\put(1738,40){\makebox(0,0)[b]{\smash{{{\SetFigFont{10}{12.0}{\rmdefault}{\mddefault}{\updefault}$B$}}}}}
\put(1108,40){\makebox(0,0)[b]{\smash{{{\SetFigFont{10}{12.0}{\rmdefault}{\mddefault}{\updefault}$B$}}}}}
\put(1558,1030){\makebox(0,0)[b]{\smash{{{\SetFigFont{10}{12.0}{\rmdefault}{\mddefault}{\updefault}$L_2$}}}}}
\put(1333,1030){\makebox(0,0)[b]{\smash{{{\SetFigFont{8}{9.6}{\rmdefault}{\mddefault}{\updefault}3}}}}}
\put(1108,1120){\makebox(0,0)[b]{\smash{{{\SetFigFont{8}{9.6}{\rmdefault}{\mddefault}{\updefault}$H$}}}}}
\put(1738,1120){\makebox(0,0)[b]{\smash{{{\SetFigFont{8}{9.6}{\rmdefault}{\mddefault}{\updefault}$H$}}}}}
\put(1558,670){\makebox(0,0)[b]{\smash{{{\SetFigFont{8}{9.6}{\rmdefault}{\mddefault}{\updefault}$H$}}}}}
\put(1783,670){\makebox(0,0)[b]{\smash{{{\SetFigFont{8}{9.6}{\rmdefault}{\mddefault}{\updefault}$H$}}}}}
\put(1333,670){\makebox(0,0)[b]{\smash{{{\SetFigFont{8}{9.6}{\rmdefault}{\mddefault}{\updefault}$H$}}}}}
\put(1063,670){\makebox(0,0)[b]{\smash{{{\SetFigFont{8}{9.6}{\rmdefault}{\mddefault}{\updefault}$H$}}}}}
\put(1198,535){\makebox(0,0)[b]{\smash{{{\SetFigFont{8}{9.6}{\rmdefault}{\mddefault}{\updefault}4}}}}}
\put(1018,490){\makebox(0,0)[b]{\smash{{{\SetFigFont{10}{12.0}{\rmdefault}{\mddefault}{\updefault}$L_2$}}}}}
\put(1648,535){\makebox(0,0)[b]{\smash{{{\SetFigFont{8}{9.6}{\rmdefault}{\mddefault}{\updefault}5}}}}}
\put(1828,490){\makebox(0,0)[b]{\smash{{{\SetFigFont{10}{12.0}{\rmdefault}{\mddefault}{\updefault}$L_2$}}}}}
\put(5653,1030){\circle*{40}}
\put(5653,625){\circle*{40}}
\put(5653,220){\circle*{40}}
\put(5338,625){\circle*{40}}
\put(5968,625){\circle*{40}}
\put(5563,805){\circle*{40}}
\put(5743,805){\circle*{40}}
\drawline(5653,1030)(5338,625)
\drawline(5653,1030)(5653,625)
\drawline(5653,1030)(5968,625)
\drawline(5653,625)(5653,220)
\drawline(5653,1030)(5563,805)
\drawline(5653,1030)(5743,805)
\put(5968,445){\makebox(0,0)[b]{\smash{{{\SetFigFont{10}{12.0}{\rmdefault}{\mddefault}{\updefault}$C$}}}}}
\put(5788,1030){\makebox(0,0)[b]{\smash{{{\SetFigFont{10}{12.0}{\rmdefault}{\mddefault}{\updefault}$L_3$}}}}}
\put(5608,1075){\makebox(0,0)[b]{\smash{{{\SetFigFont{8}{9.6}{\rmdefault}{\mddefault}{\updefault}6}}}}}
\put(5653,40){\makebox(0,0)[b]{\smash{{{\SetFigFont{10}{12.0}{\rmdefault}{\mddefault}{\updefault}$C$}}}}}
\put(5518,670){\makebox(0,0)[b]{\smash{{{\SetFigFont{8}{9.6}{\rmdefault}{\mddefault}{\updefault}$H$}}}}}
\put(5788,670){\makebox(0,0)[b]{\smash{{{\SetFigFont{8}{9.6}{\rmdefault}{\mddefault}{\updefault}$H$}}}}}
\put(5743,535){\makebox(0,0)[b]{\smash{{{\SetFigFont{8}{9.6}{\rmdefault}{\mddefault}{\updefault}7}}}}}
\put(5563,490){\makebox(0,0)[b]{\smash{{{\SetFigFont{10}{12.0}{\rmdefault}{\mddefault}{\updefault}$L_3$}}}}}
\put(3493,1030){\circle*{40}}
\put(3493,220){\circle*{40}}
\put(3178,625){\circle*{40}}
\put(3808,625){\circle*{40}}
\put(3403,805){\circle*{40}}
\put(3583,805){\circle*{40}}
\put(3493,625){\circle*{40}}
\drawline(3493,1030)(3493,625)
\drawline(3493,1030)(3808,625)
\drawline(3493,625)(3493,220)
\drawline(3493,1030)(3178,625)
\drawline(3493,1030)(3403,805)
\drawline(3493,1030)(3583,805)
\put(3178,445){\makebox(0,0)[b]{\smash{{{\SetFigFont{10}{12.0}{\rmdefault}{\mddefault}{\updefault}$A$}}}}}
\put(3538,1075){\makebox(0,0)[b]{\smash{{{\SetFigFont{8}{9.6}{\rmdefault}{\mddefault}{\updefault}1}}}}}
\put(3358,1030){\makebox(0,0)[b]{\smash{{{\SetFigFont{10}{12.0}{\rmdefault}{\mddefault}{\updefault}$L_1$}}}}}
\put(3493,40){\makebox(0,0)[b]{\smash{{{\SetFigFont{10}{12.0}{\rmdefault}{\mddefault}{\updefault}$A$}}}}}
\put(3628,670){\makebox(0,0)[b]{\smash{{{\SetFigFont{8}{9.6}{\rmdefault}{\mddefault}{\updefault}$H$}}}}}
\put(3358,670){\makebox(0,0)[b]{\smash{{{\SetFigFont{8}{9.6}{\rmdefault}{\mddefault}{\updefault}$H$}}}}}
\put(3403,535){\makebox(0,0)[b]{\smash{{{\SetFigFont{8}{9.6}{\rmdefault}{\mddefault}{\updefault}2}}}}}
\put(3583,490){\makebox(0,0)[b]{\smash{{{\SetFigFont{10}{12.0}{\rmdefault}{\mddefault}{\updefault}$L_1$}}}}}
\put(4573,1435){\circle*{40}}
\put(4573,625){\circle*{40}}
\put(4573,1030){\circle*{40}}
\put(4258,625){\circle*{40}}
\put(3943,625){\circle*{40}}
\put(4888,625){\circle*{40}}
\put(4888,220){\circle*{40}}
\put(4258,220){\circle*{40}}
\put(5203,625){\circle*{40}}
\put(4348,1210){\circle*{40}}
\put(4798,1210){\circle*{40}}
\put(4663,805){\circle*{40}}
\put(4483,805){\circle*{40}}
\put(4348,805){\circle*{40}}
\put(4798,805){\circle*{40}}
\drawline(4573,1435)(4573,1030)
\drawline(4573,1435)(3493,1030)
\drawline(4573,1435)(5653,1030)
\drawline(4573,1030)(4573,625)
\drawline(4573,1030)(3943,625)
\drawline(4573,1030)(4258,625)
\drawline(4573,1030)(5203,625)
\drawline(4573,1030)(4888,625)
\drawline(4258,625)(4258,220)
\drawline(4888,625)(4888,220)
\drawline(4573,1435)(4348,1210)
\drawline(4573,1435)(4798,1210)
\drawline(4573,1030)(4348,805)
\drawline(4573,1030)(4483,805)
\drawline(4573,1030)(4663,805)
\drawline(4573,1030)(4798,805)
\put(4573,445){\makebox(0,0)[b]{\smash{{{\SetFigFont{10}{12.0}{\rmdefault}{\mddefault}{\updefault}$B$}}}}}
\put(4888,40){\makebox(0,0)[b]{\smash{{{\SetFigFont{10}{12.0}{\rmdefault}{\mddefault}{\updefault}$B$}}}}}
\put(4258,40){\makebox(0,0)[b]{\smash{{{\SetFigFont{10}{12.0}{\rmdefault}{\mddefault}{\updefault}$B$}}}}}
\put(4708,1030){\makebox(0,0)[b]{\smash{{{\SetFigFont{10}{12.0}{\rmdefault}{\mddefault}{\updefault}$L_2$}}}}}
\put(4483,1030){\makebox(0,0)[b]{\smash{{{\SetFigFont{8}{9.6}{\rmdefault}{\mddefault}{\updefault}3}}}}}
\put(4258,1120){\makebox(0,0)[b]{\smash{{{\SetFigFont{8}{9.6}{\rmdefault}{\mddefault}{\updefault}$H$}}}}}
\put(4888,1120){\makebox(0,0)[b]{\smash{{{\SetFigFont{8}{9.6}{\rmdefault}{\mddefault}{\updefault}$H$}}}}}
\put(4708,670){\makebox(0,0)[b]{\smash{{{\SetFigFont{8}{9.6}{\rmdefault}{\mddefault}{\updefault}$H$}}}}}
\put(4933,670){\makebox(0,0)[b]{\smash{{{\SetFigFont{8}{9.6}{\rmdefault}{\mddefault}{\updefault}$H$}}}}}
\put(4483,670){\makebox(0,0)[b]{\smash{{{\SetFigFont{8}{9.6}{\rmdefault}{\mddefault}{\updefault}$H$}}}}}
\put(4213,670){\makebox(0,0)[b]{\smash{{{\SetFigFont{8}{9.6}{\rmdefault}{\mddefault}{\updefault}$H$}}}}}
\put(4348,535){\makebox(0,0)[b]{\smash{{{\SetFigFont{8}{9.6}{\rmdefault}{\mddefault}{\updefault}4}}}}}
\put(4168,490){\makebox(0,0)[b]{\smash{{{\SetFigFont{10}{12.0}{\rmdefault}{\mddefault}{\updefault}$L_2$}}}}}
\put(4798,535){\makebox(0,0)[b]{\smash{{{\SetFigFont{8}{9.6}{\rmdefault}{\mddefault}{\updefault}5}}}}}
\put(4978,490){\makebox(0,0)[b]{\smash{{{\SetFigFont{10}{12.0}{\rmdefault}{\mddefault}{\updefault}$L_2$}}}}}
\put(2998,1885){\circle*{40}}
\put(2998,1660){\circle*{40}}
\drawline(2998,1885)(2998,1660)
\put(2998,1525){\makebox(0,0)[b]{\smash{{{\SetFigFont{8}{9.6}{\rmdefault}{\mddefault}{\updefault}$H$}}}}}
\drawline(2998,1885)(4573,1435)
\drawline(1423,1435)(2998,1885)
\end{picture}
}

%% file: following_nextsibling.tex
\setlength{\unitlength}{0.00087489in}
\begingroup\makeatletter\ifx\SetFigFont\undefined%
\gdef\SetFigFont#1#2#3#4#5{%
  \reset@font\fontsize{#1}{#2pt}%
  \fontfamily{#3}\fontseries{#4}\fontshape{#5}%
  \selectfont}%
\fi\endgroup%
{\renewcommand{\dashlinestretch}{30}
\begin{picture}(3296,4047)(0,-10)
\drawline(2818.000,1215.000)(2755.094,1250.230)(2691.128,1283.497)
	(2626.163,1314.770)(2560.262,1344.017)(2493.488,1371.212)
	(2425.904,1396.329)(2357.576,1419.342)(2288.569,1440.231)
	(2218.948,1458.975)(2148.781,1475.556)(2078.135,1489.959)
	(2007.076,1502.169)(1935.674,1512.176)(1863.997,1519.968)
	(1792.113,1525.540)(1720.091,1528.885)(1648.000,1530.000)
	(1575.909,1528.885)(1503.887,1525.540)(1432.003,1519.968)
	(1360.326,1512.176)(1288.924,1502.169)(1217.865,1489.959)
	(1147.219,1475.556)(1077.052,1458.975)(1007.431,1440.231)
	(938.424,1419.342)(870.096,1396.329)(802.512,1371.212)
	(735.738,1344.017)(669.837,1314.770)(604.872,1283.497)
	(540.906,1250.230)(478.000,1215.000)
\blacken\drawline(2745.157,1238.321)(2818.000,1215.000)(2759.730,1264.544)(2745.157,1238.321)
\put(1648,3195){\circle*{40}}
\put(1648,3600){\circle*{40}}
\put(1648,4005){\circle*{40}}
\put(1018,3195){\circle*{40}}
\put(2278,3195){\circle*{40}}
\put(343,3600){\circle*{40}}
\put(343,3195){\circle*{40}}
\put(343,2790){\circle*{40}}
\put(2953,3600){\circle*{40}}
\put(2953,3195){\circle*{40}}
\put(2953,2790){\circle*{40}}
\put(28,3195){\circle*{40}}
\put(658,3195){\circle*{40}}
\put(2638,3195){\circle*{40}}
\put(3268,3195){\circle*{40}}
\put(1018,2790){\circle*{40}}
\put(2278,2790){\circle*{40}}
\put(703,2385){\circle*{40}}
\put(1018,2385){\circle*{40}}
\put(1333,2385){\circle*{40}}
\put(1963,2385){\circle*{40}}
\put(2278,2385){\circle*{40}}
\put(2593,2385){\circle*{40}}
\put(28,2790){\circle*{40}}
\put(658,2790){\circle*{40}}
\put(2638,2790){\circle*{40}}
\put(3268,2790){\circle*{40}}
\put(1648,1215){\circle*{40}}
\put(478,1215){\circle*{40}}
\put(2818,1215){\circle*{40}}
\put(1423,450){\circle*{40}}
\put(3043,450){\circle*{40}}
\put(253,450){\circle*{40}}
\put(1873,450){\circle*{40}}
\drawline(1648,4005)(1648,3600)
\drawline(1648,4005)(343,3600)
\drawline(1648,4005)(2953,3600)
\drawline(1648,3600)(1648,3195)
\drawline(1648,3600)(1018,3195)
\drawline(1648,3600)(2278,3195)
\drawline(343,3600)(658,3195)
\drawline(343,3195)(343,2790)
\drawline(343,3600)(28,3195)
\drawline(2953,3600)(2638,3195)
\drawline(2953,3600)(2953,3195)
\drawline(2953,3600)(3268,3195)
\drawline(2953,3195)(2953,2790)
\drawline(1652,3188)(1022,2783)
\drawline(1645,3190)(2275,2785)
\drawline(1018,2790)(1018,2385)
\drawline(1009,2797)(694,2392)
\drawline(1015,2788)(1330,2383)
\drawline(2275,2788)(2590,2383)
\drawline(2275,2792)(1960,2387)
\drawline(2278,2790)(2278,2385)
\drawline(2944,3202)(2629,2797)
\drawline(2959,3200)(3274,2795)
\drawline(340,3197)(25,2792)
\drawline(340,3193)(655,2788)
\drawline(343,3600)(343,3195)
\drawline(478,1215)(1648,1215)
\blacken\drawline(1573.000,1200.000)(1648.000,1215.000)(1573.000,1230.000)(1573.000,1200.000)
\drawline(1648,1215)(2818,1215)
\blacken\drawline(2743.000,1200.000)(2818.000,1215.000)(2743.000,1230.000)(2743.000,1200.000)
\drawline(2818,1215)(3043,450)
\blacken\drawline(3007.447,517.720)(3043.000,450.000)(3036.228,526.185)(3007.447,517.720)
\drawline(253,450)(1423,450)
\blacken\drawline(1348.000,435.000)(1423.000,450.000)(1348.000,465.000)(1348.000,435.000)
\drawline(1873,450)(3043,450)
\blacken\drawline(2968.000,435.000)(3043.000,450.000)(2968.000,465.000)(2968.000,435.000)
\drawline(251,451)(476,1216)
\blacken\drawline(469.228,1139.815)(476.000,1216.000)(440.447,1148.280)(469.228,1139.815)
\drawline(1648,1215)(1423,450)
\blacken\drawline(1429.772,526.185)(1423.000,450.000)(1458.553,517.720)(1429.772,526.185)
\drawline(1873,450)(1648,1215)
\blacken\drawline(1683.553,1147.280)(1648.000,1215.000)(1654.772,1138.815)(1683.553,1147.280)
\put(2278,2205){\makebox(0,0)[b]{{\SetFigFont{10}{12.0}{\rmdefault}{\mddefault}{\updefault}$L_2$}}}
\put(1018,2205){\makebox(0,0)[b]{{\SetFigFont{10}{12.0}{\rmdefault}{\mddefault}{\updefault}$L_2$}}}
\put(1648,2970){\makebox(0,0)[b]{{\SetFigFont{10}{12.0}{\rmdefault}{\mddefault}{\updefault}$L_2$}}}
\put(2953,2610){\makebox(0,0)[b]{{\SetFigFont{10}{12.0}{\rmdefault}{\mddefault}{\updefault}$L_3$}}}
\put(343,2610){\makebox(0,0)[b]{{\SetFigFont{10}{12.0}{\rmdefault}{\mddefault}{\updefault}$L_1$}}}
\put(1648,2025){\makebox(0,0)[b]{{\SetFigFont{10}{12.0}{\rmdefault}{\mddefault}{\updefault}(a)}}}
\put(1648,1620){\makebox(0,0)[b]{{\SetFigFont{10}{12.0}{\rmdefault}{\mddefault}{\updefault}\following$^{11}$}}}
\put(1648,45){\makebox(0,0)[b]{{\SetFigFont{10}{12.0}{\rmdefault}{\mddefault}{\updefault}(b)}}}
\put(1648,1305){\makebox(0,0)[b]{{\SetFigFont{10}{12.0}{\rmdefault}{\mddefault}{\updefault}$L_2$}}}
\put(2818,1305){\makebox(0,0)[b]{{\SetFigFont{10}{12.0}{\rmdefault}{\mddefault}{\updefault}$L_3$}}}
\put(478,1305){\makebox(0,0)[b]{{\SetFigFont{10}{12.0}{\rmdefault}{\mddefault}{\updefault}$L_1$}}}
\put(1018,1035){\makebox(0,0)[b]{{\SetFigFont{10}{12.0}{\rmdefault}{\mddefault}{\updefault}\following$^4$}}}
\put(2233,1035){\makebox(0,0)[b]{{\SetFigFont{10}{12.0}{\rmdefault}{\mddefault}{\updefault}\following$^4$}}}
\put(793,270){\makebox(0,0)[b]{{\SetFigFont{10}{12.0}{\rmdefault}{\mddefault}{\updefault}\following$^8$}}}
\put(2458,270){\makebox(0,0)[b]{{\SetFigFont{10}{12.0}{\rmdefault}{\mddefault}{\updefault}\following$^8$}}}
\put(1468,720){\makebox(0,0)[rb]{{\SetFigFont{10}{12.0}{\rmdefault}{\mddefault}{\updefault}\nextsibling}}}
\put(1828,720){\makebox(0,0)[lb]{{\SetFigFont{10}{12.0}{\rmdefault}{\mddefault}{\updefault}\nextsibling}}}
\put(343,810){\makebox(0,0)[rb]{{\SetFigFont{10}{12.0}{\rmdefault}{\mddefault}{\updefault}\nextsibling}}}
\put(2953,810){\makebox(0,0)[lb]{{\SetFigFont{10}{12.0}{\rmdefault}{\mddefault}{\updefault}\nextsibling}}}
\put(1558,3195){\makebox(0,0)[b]{{\SetFigFont{8}{9.6}{\rmdefault}{\mddefault}{\updefault}3}}}
\put(253,2790){\makebox(0,0)[b]{{\SetFigFont{8}{9.6}{\rmdefault}{\mddefault}{\updefault}2}}}
\put(928,2385){\makebox(0,0)[b]{{\SetFigFont{8}{9.6}{\rmdefault}{\mddefault}{\updefault}4}}}
\put(2188,2385){\makebox(0,0)[b]{{\SetFigFont{8}{9.6}{\rmdefault}{\mddefault}{\updefault}5}}}
\put(2863,2790){\makebox(0,0)[b]{{\SetFigFont{8}{9.6}{\rmdefault}{\mddefault}{\updefault}7}}}
\put(2863,3195){\makebox(0,0)[b]{{\SetFigFont{8}{9.6}{\rmdefault}{\mddefault}{\updefault}6}}}
\put(253,3195){\makebox(0,0)[b]{{\SetFigFont{8}{9.6}{\rmdefault}{\mddefault}{\updefault}1}}}
\put(433,3195){\makebox(0,0)[b]{{\SetFigFont{10}{12.0}{\rmdefault}{\mddefault}{\updefault}$L_1$}}}
\put(3043,3195){\makebox(0,0)[b]{{\SetFigFont{10}{12.0}{\rmdefault}{\mddefault}{\updefault}$L_3$}}}
\end{picture}
}

%% file: cq_apq.tex
\setlength{\unitlength}{0.00083333in}
\begingroup\makeatletter\ifx\SetFigFont\undefined%
\gdef\SetFigFont#1#2#3#4#5{%
  \reset@font\fontsize{#1}{#2pt}%
  \fontfamily{#3}\fontseries{#4}\fontshape{#5}%
  \selectfont}%
\fi\endgroup%
{\renewcommand{\dashlinestretch}{30}
\begin{picture}(8397,9105)(0,-10)
\thicklines
\drawline(1240,6870)(115,5895)
\blacken\thinlines
\drawline(161.853,5955.455)(115.000,5895.000)(181.501,5932.784)(161.853,5955.455)
\drawline(1240,6870)(2365,5895)
\blacken\drawline(2298.499,5932.784)(2365.000,5895.000)(2318.147,5955.455)(2298.499,5932.784)
\thicklines
\drawline(790,6120)(115,5895)
\blacken\thinlines
\drawline(181.408,5932.947)(115.000,5895.000)(190.895,5904.487)(181.408,5932.947)
\drawline(1690,6120)(2365,5895)
\blacken\drawline(2289.105,5904.487)(2365.000,5895.000)(2298.592,5932.947)(2289.105,5904.487)
\drawline(790,6120)(1690,6120)
\blacken\drawline(1615.000,6105.000)(1690.000,6120.000)(1615.000,6135.000)(1615.000,6105.000)
\put(1240,6945){\makebox(0,0)[b]{{\SetFigFont{10}{12.0}{\rmdefault}{\mddefault}{\updefault}$x$}}}
\put(115,5745){\makebox(0,0)[b]{{\SetFigFont{10}{12.0}{\rmdefault}{\mddefault}{\updefault}$y$}}}
\put(340,5820){\makebox(0,0)[lb]{{\SetFigFont{10}{12.0}{\rmdefault}{\mddefault}{\updefault}$\child^*$}}}
\put(1840,6420){\makebox(0,0)[lb]{{\SetFigFont{10}{12.0}{\rmdefault}{\mddefault}{\updefault}$\child^+$}}}
\put(640,6420){\makebox(0,0)[rb]{{\SetFigFont{10}{12.0}{\rmdefault}{\mddefault}{\updefault}$\child^+$}}}
\put(2365,5745){\makebox(0,0)[b]{{\SetFigFont{10}{12.0}{\rmdefault}{\mddefault}{\updefault}$z$}}}
\put(2140,5820){\makebox(0,0)[rb]{{\SetFigFont{10}{12.0}{\rmdefault}{\mddefault}{\updefault}$\child^*$}}}
\put(1240,6195){\makebox(0,0)[b]{{\SetFigFont{10}{12.0}{\rmdefault}{\mddefault}{\updefault}$\nextsibling^+$}}}
\put(1690,5970){\makebox(0,0)[b]{{\SetFigFont{10}{12.0}{\rmdefault}{\mddefault}{\updefault}$v$}}}
\put(790,5970){\makebox(0,0)[b]{{\SetFigFont{10}{12.0}{\rmdefault}{\mddefault}{\updefault}$u$}}}
\drawline(7165,5070)(6040,4095)
\blacken\drawline(6086.853,4155.455)(6040.000,4095.000)(6106.501,4132.784)(6086.853,4155.455)
\drawline(6715,4320)(7165,5070)
\blacken\drawline(7139.275,4997.971)(7165.000,5070.000)(7113.550,5013.405)(7139.275,4997.971)
\drawline(7615,4320)(6715,4320)
\blacken\drawline(6790.000,4335.000)(6715.000,4320.000)(6790.000,4305.000)(6790.000,4335.000)
\drawline(6715,4245)(7615,4245)
\blacken\drawline(7540.000,4230.000)(7615.000,4245.000)(7540.000,4260.000)(7540.000,4230.000)
\drawline(7167,5073)(8292,4098)
\blacken\drawline(8225.499,4135.784)(8292.000,4098.000)(8245.147,4158.455)(8225.499,4135.784)
\put(7165,5145){\makebox(0,0)[b]{{\SetFigFont{10}{12.0}{\rmdefault}{\mddefault}{\updefault}$x$}}}
\put(6040,3945){\makebox(0,0)[b]{{\SetFigFont{10}{12.0}{\rmdefault}{\mddefault}{\updefault}$y$}}}
\put(7765,4620){\makebox(0,0)[lb]{{\SetFigFont{10}{12.0}{\rmdefault}{\mddefault}{\updefault}$\child^+$}}}
\put(8290,3945){\makebox(0,0)[b]{{\SetFigFont{10}{12.0}{\rmdefault}{\mddefault}{\updefault}$z$}}}
\put(6640,4620){\makebox(0,0)[rb]{{\SetFigFont{10}{12.0}{\rmdefault}{\mddefault}{\updefault}$\child^+$}}}
\put(7165,4095){\makebox(0,0)[b]{{\SetFigFont{10}{12.0}{\rmdefault}{\mddefault}{\updefault}$\nextsibling^+$}}}
\put(7690,4245){\makebox(0,0)[b]{{\SetFigFont{10}{12.0}{\rmdefault}{\mddefault}{\updefault}$v$}}}
\put(6640,4245){\makebox(0,0)[b]{{\SetFigFont{10}{12.0}{\rmdefault}{\mddefault}{\updefault}$u$}}}
\put(7240,4395){\makebox(0,0)[b]{{\SetFigFont{10}{12.0}{\rmdefault}{\mddefault}{\updefault}$\child^*$}}}
\put(6940,4620){\makebox(0,0)[lb]{{\SetFigFont{10}{12.0}{\rmdefault}{\mddefault}{\updefault}$\child^*$}}}
\drawline(7165,6870)(6040,5895)
\blacken\drawline(6086.853,5955.455)(6040.000,5895.000)(6106.501,5932.784)(6086.853,5955.455)
\thicklines
\drawline(7165,6870)(7615,6120)
\blacken\thinlines
\drawline(7563.550,6176.595)(7615.000,6120.000)(7589.275,6192.029)(7563.550,6176.595)
\drawline(6715,6120)(7165,6870)
\blacken\drawline(7139.275,6797.971)(7165.000,6870.000)(7113.550,6813.405)(7139.275,6797.971)
\thicklines
\drawline(6715,6120)(7615,6120)
\blacken\thinlines
\drawline(7540.000,6105.000)(7615.000,6120.000)(7540.000,6135.000)(7540.000,6105.000)
\drawline(7615,6120)(8297,5918)
\blacken\drawline(8220.828,5924.917)(8297.000,5918.000)(8229.348,5953.682)(8220.828,5924.917)
\put(7165,6945){\makebox(0,0)[b]{{\SetFigFont{10}{12.0}{\rmdefault}{\mddefault}{\updefault}$x$}}}
\put(6040,5745){\makebox(0,0)[b]{{\SetFigFont{10}{12.0}{\rmdefault}{\mddefault}{\updefault}$y$}}}
\put(8290,5745){\makebox(0,0)[b]{{\SetFigFont{10}{12.0}{\rmdefault}{\mddefault}{\updefault}$z$}}}
\put(8065,5820){\makebox(0,0)[rb]{{\SetFigFont{10}{12.0}{\rmdefault}{\mddefault}{\updefault}$\child^*$}}}
\put(7165,5970){\makebox(0,0)[b]{{\SetFigFont{10}{12.0}{\rmdefault}{\mddefault}{\updefault}$\nextsibling^+$}}}
\put(6640,6120){\makebox(0,0)[b]{{\SetFigFont{10}{12.0}{\rmdefault}{\mddefault}{\updefault}$u$}}}
\put(6640,6420){\makebox(0,0)[rb]{{\SetFigFont{10}{12.0}{\rmdefault}{\mddefault}{\updefault}$\child^+$}}}
\put(7390,6570){\makebox(0,0)[lb]{{\SetFigFont{10}{12.0}{\rmdefault}{\mddefault}{\updefault}$\child^+$}}}
\put(6865,6270){\makebox(0,0)[lb]{{\SetFigFont{10}{12.0}{\rmdefault}{\mddefault}{\updefault}$\child^*$}}}
\put(7690,6195){\makebox(0,0)[b]{{\SetFigFont{10}{12.0}{\rmdefault}{\mddefault}{\updefault}$v$}}}
\drawline(4240,5070)(3115,4095)
\blacken\drawline(3161.853,4155.455)(3115.000,4095.000)(3181.501,4132.784)(3161.853,4155.455)
\drawline(4240,5070)(5365,4095)
\blacken\drawline(5298.499,4132.784)(5365.000,4095.000)(5318.147,4155.455)(5298.499,4132.784)
\thicklines
\drawline(3790,4320)(4240,5070)
\blacken\thinlines
\drawline(4214.275,4997.971)(4240.000,5070.000)(4188.550,5013.405)(4214.275,4997.971)
\thicklines
\drawline(4690,4320)(4240,5070)
\blacken\thinlines
\drawline(4291.450,5013.405)(4240.000,5070.000)(4265.725,4997.971)(4291.450,5013.405)
\drawline(3790,4320)(4690,4320)
\blacken\drawline(4615.000,4305.000)(4690.000,4320.000)(4615.000,4335.000)(4615.000,4305.000)
\put(4240,5145){\makebox(0,0)[b]{{\SetFigFont{10}{12.0}{\rmdefault}{\mddefault}{\updefault}$x$}}}
\put(3115,3945){\makebox(0,0)[b]{{\SetFigFont{10}{12.0}{\rmdefault}{\mddefault}{\updefault}$y$}}}
\put(4840,4620){\makebox(0,0)[lb]{{\SetFigFont{10}{12.0}{\rmdefault}{\mddefault}{\updefault}$\child^+$}}}
\put(5365,3945){\makebox(0,0)[b]{{\SetFigFont{10}{12.0}{\rmdefault}{\mddefault}{\updefault}$z$}}}
\put(4765,4320){\makebox(0,0)[b]{{\SetFigFont{10}{12.0}{\rmdefault}{\mddefault}{\updefault}$v$}}}
\put(4240,4170){\makebox(0,0)[b]{{\SetFigFont{10}{12.0}{\rmdefault}{\mddefault}{\updefault}$\nextsibling^+$}}}
\put(3715,4320){\makebox(0,0)[b]{{\SetFigFont{10}{12.0}{\rmdefault}{\mddefault}{\updefault}$u$}}}
\put(3715,4620){\makebox(0,0)[rb]{{\SetFigFont{10}{12.0}{\rmdefault}{\mddefault}{\updefault}$\child^+$}}}
\put(3865,4395){\makebox(0,0)[lb]{{\SetFigFont{10}{12.0}{\rmdefault}{\mddefault}{\updefault}$\child^*$}}}
\put(4540,4545){\makebox(0,0)[rb]{{\SetFigFont{10}{12.0}{\rmdefault}{\mddefault}{\updefault}$\child^*$}}}
\drawline(7165,3345)(6040,2370)
\blacken\drawline(6086.853,2430.455)(6040.000,2370.000)(6106.501,2407.784)(6086.853,2430.455)
\drawline(7615,2595)(7165,3345)
\blacken\drawline(7216.450,3288.405)(7165.000,3345.000)(7190.725,3272.971)(7216.450,3288.405)
\thicklines
\drawline(6715,2595)(7615,2595)
\blacken\thinlines
\drawline(7540.000,2580.000)(7615.000,2595.000)(7540.000,2610.000)(7540.000,2580.000)
\thicklines
\drawline(6715,2520)(7615,2520)
\blacken\thinlines
\drawline(7540.000,2505.000)(7615.000,2520.000)(7540.000,2535.000)(7540.000,2505.000)
\drawline(7167,3348)(8292,2373)
\blacken\drawline(8225.499,2410.784)(8292.000,2373.000)(8245.147,2433.455)(8225.499,2410.784)
\put(7165,3420){\makebox(0,0)[b]{{\SetFigFont{10}{12.0}{\rmdefault}{\mddefault}{\updefault}$x$}}}
\put(6040,2220){\makebox(0,0)[b]{{\SetFigFont{10}{12.0}{\rmdefault}{\mddefault}{\updefault}$y$}}}
\put(7765,2895){\makebox(0,0)[lb]{{\SetFigFont{10}{12.0}{\rmdefault}{\mddefault}{\updefault}$\child^+$}}}
\put(8290,2220){\makebox(0,0)[b]{{\SetFigFont{10}{12.0}{\rmdefault}{\mddefault}{\updefault}$z$}}}
\put(6640,2895){\makebox(0,0)[rb]{{\SetFigFont{10}{12.0}{\rmdefault}{\mddefault}{\updefault}$\child^+$}}}
\put(7165,2370){\makebox(0,0)[b]{{\SetFigFont{10}{12.0}{\rmdefault}{\mddefault}{\updefault}$\nextsibling^+$}}}
\put(7690,2520){\makebox(0,0)[b]{{\SetFigFont{10}{12.0}{\rmdefault}{\mddefault}{\updefault}$v$}}}
\put(6640,2520){\makebox(0,0)[b]{{\SetFigFont{10}{12.0}{\rmdefault}{\mddefault}{\updefault}$u$}}}
\put(7165,2670){\makebox(0,0)[b]{{\SetFigFont{10}{12.0}{\rmdefault}{\mddefault}{\updefault}$\child^*$}}}
\put(7390,2895){\makebox(0,0)[rb]{{\SetFigFont{10}{12.0}{\rmdefault}{\mddefault}{\updefault}$\child^*$}}}
\drawline(7165,8670)(6040,7695)
\blacken\drawline(6086.853,7755.455)(6040.000,7695.000)(6106.501,7732.784)(6086.853,7755.455)
\drawline(7240,8670)(6790,7920)
\blacken\drawline(6815.725,7992.029)(6790.000,7920.000)(6841.450,7976.595)(6815.725,7992.029)
\drawline(6715,7920)(7165,8670)
\blacken\drawline(7139.275,8597.971)(7165.000,8670.000)(7113.550,8613.405)(7139.275,8597.971)
\drawline(6715,7920)(7615,7920)
\blacken\drawline(7540.000,7905.000)(7615.000,7920.000)(7540.000,7935.000)(7540.000,7905.000)
\drawline(7615,7920)(8297,7718)
\blacken\drawline(8220.828,7724.917)(8297.000,7718.000)(8229.348,7753.682)(8220.828,7724.917)
\put(7165,8745){\makebox(0,0)[b]{{\SetFigFont{10}{12.0}{\rmdefault}{\mddefault}{\updefault}$x$}}}
\put(6040,7545){\makebox(0,0)[b]{{\SetFigFont{10}{12.0}{\rmdefault}{\mddefault}{\updefault}$y$}}}
\put(8290,7545){\makebox(0,0)[b]{{\SetFigFont{10}{12.0}{\rmdefault}{\mddefault}{\updefault}$z$}}}
\put(8065,7620){\makebox(0,0)[rb]{{\SetFigFont{10}{12.0}{\rmdefault}{\mddefault}{\updefault}$\child^*$}}}
\put(7165,7770){\makebox(0,0)[b]{{\SetFigFont{10}{12.0}{\rmdefault}{\mddefault}{\updefault}$\nextsibling^+$}}}
\put(6640,7920){\makebox(0,0)[b]{{\SetFigFont{10}{12.0}{\rmdefault}{\mddefault}{\updefault}$u$}}}
\put(6640,8220){\makebox(0,0)[rb]{{\SetFigFont{10}{12.0}{\rmdefault}{\mddefault}{\updefault}$\child^+$}}}
\put(6865,8070){\makebox(0,0)[lb]{{\SetFigFont{10}{12.0}{\rmdefault}{\mddefault}{\updefault}$\child^*$}}}
\put(7090,8295){\makebox(0,0)[lb]{{\SetFigFont{10}{12.0}{\rmdefault}{\mddefault}{\updefault}$\child^+$}}}
\put(7615,7995){\makebox(0,0)[b]{{\SetFigFont{10}{12.0}{\rmdefault}{\mddefault}{\updefault}$v$}}}
\drawline(3715.000,645.000)(3646.124,616.392)(3593.431,563.610)
	(3564.938,494.686)(3564.979,420.105)(3593.549,351.212)
	(3646.301,298.490)(3715.209,269.958)(3789.791,269.958)
	(3858.699,298.490)(3911.451,351.212)(3940.021,420.105)
	(3940.062,494.686)(3911.569,563.610)(3858.876,616.392)
	(3790.000,645.000)
\blacken\drawline(3864.872,629.372)(3790.000,645.000)(3853.101,601.777)(3864.872,629.372)
\drawline(4240,1470)(3115,495)
\blacken\drawline(3161.853,555.455)(3115.000,495.000)(3181.501,532.784)(3161.853,555.455)
\drawline(4690,720)(4240,1470)
\blacken\drawline(4291.450,1413.405)(4240.000,1470.000)(4265.725,1397.971)(4291.450,1413.405)
\drawline(3790,645)(4690,645)
\blacken\drawline(4615.000,630.000)(4690.000,645.000)(4615.000,660.000)(4615.000,630.000)
\drawline(4242,1473)(5367,498)
\blacken\drawline(5300.499,535.784)(5367.000,498.000)(5320.147,558.455)(5300.499,535.784)
\put(4240,1545){\makebox(0,0)[b]{{\SetFigFont{10}{12.0}{\rmdefault}{\mddefault}{\updefault}$x$}}}
\put(3115,345){\makebox(0,0)[b]{{\SetFigFont{10}{12.0}{\rmdefault}{\mddefault}{\updefault}$y$}}}
\put(4840,1020){\makebox(0,0)[lb]{{\SetFigFont{10}{12.0}{\rmdefault}{\mddefault}{\updefault}$\child^+$}}}
\put(5365,345){\makebox(0,0)[b]{{\SetFigFont{10}{12.0}{\rmdefault}{\mddefault}{\updefault}$z$}}}
\put(3715,1020){\makebox(0,0)[rb]{{\SetFigFont{10}{12.0}{\rmdefault}{\mddefault}{\updefault}$\child^+$}}}
\put(4765,645){\makebox(0,0)[b]{{\SetFigFont{10}{12.0}{\rmdefault}{\mddefault}{\updefault}$v$}}}
\put(3715,645){\makebox(0,0)[b]{{\SetFigFont{10}{12.0}{\rmdefault}{\mddefault}{\updefault}$u$}}}
\put(4465,1020){\makebox(0,0)[rb]{{\SetFigFont{10}{12.0}{\rmdefault}{\mddefault}{\updefault}$\child^*$}}}
\put(4240,720){\makebox(0,0)[b]{{\SetFigFont{10}{12.0}{\rmdefault}{\mddefault}{\updefault}$\nextsibling^+$}}}
\put(4015,345){\makebox(0,0)[lb]{{\SetFigFont{10}{12.0}{\rmdefault}{\mddefault}{\updefault}$\child^+$}}}
\drawline(7615.000,720.000)(7586.392,788.876)(7533.610,841.569)
	(7464.686,870.062)(7390.105,870.021)(7321.212,841.451)
	(7268.490,788.699)(7239.958,719.791)(7239.958,645.209)
	(7268.490,576.301)(7321.212,523.549)(7390.105,494.979)
	(7464.686,494.938)(7533.610,523.431)(7586.392,576.124)
	(7615.000,645.000)
\blacken\drawline(7599.753,570.050)(7615.000,645.000)(7572.099,581.679)(7599.753,570.050)
\drawline(7165,1470)(6040,495)
\blacken\drawline(6086.853,555.455)(6040.000,495.000)(6106.501,532.784)(6086.853,555.455)
\drawline(7615,720)(7165,1470)
\blacken\drawline(7216.450,1413.405)(7165.000,1470.000)(7190.725,1397.971)(7216.450,1413.405)
\drawline(7167,1473)(8292,498)
\blacken\drawline(8225.499,535.784)(8292.000,498.000)(8245.147,558.455)(8225.499,535.784)
\put(7165,1545){\makebox(0,0)[b]{{\SetFigFont{10}{12.0}{\rmdefault}{\mddefault}{\updefault}$x$}}}
\put(6040,345){\makebox(0,0)[b]{{\SetFigFont{10}{12.0}{\rmdefault}{\mddefault}{\updefault}$y$}}}
\put(7765,1020){\makebox(0,0)[lb]{{\SetFigFont{10}{12.0}{\rmdefault}{\mddefault}{\updefault}$\child^+$}}}
\put(8290,345){\makebox(0,0)[b]{{\SetFigFont{10}{12.0}{\rmdefault}{\mddefault}{\updefault}$z$}}}
\put(6640,1020){\makebox(0,0)[rb]{{\SetFigFont{10}{12.0}{\rmdefault}{\mddefault}{\updefault}$\child^+$}}}
\put(7690,645){\makebox(0,0)[b]{{\SetFigFont{10}{12.0}{\rmdefault}{\mddefault}{\updefault}$v$}}}
\put(7390,1020){\makebox(0,0)[rb]{{\SetFigFont{10}{12.0}{\rmdefault}{\mddefault}{\updefault}$\child^*$}}}
\put(7465,345){\makebox(0,0)[b]{{\SetFigFont{10}{12.0}{\rmdefault}{\mddefault}{\updefault}$\nextsibling^+$}}}
\drawline(4240,3345)(3565,2595)
\blacken\drawline(3604.023,2660.782)(3565.000,2595.000)(3626.322,2640.713)(3604.023,2660.782)
\drawline(4240,3345)(5365,2370)
\blacken\drawline(5298.499,2407.784)(5365.000,2370.000)(5318.147,2430.455)(5298.499,2407.784)
\drawline(4690,2595)(4240,3345)
\blacken\drawline(4291.450,3288.405)(4240.000,3345.000)(4265.725,3272.971)(4291.450,3288.405)
\drawline(3565,2595)(4690,2595)
\blacken\drawline(4615.000,2580.000)(4690.000,2595.000)(4615.000,2610.000)(4615.000,2580.000)
\drawline(3565,2595)(3565,2220)
\blacken\drawline(3550.000,2295.000)(3565.000,2220.000)(3580.000,2295.000)(3550.000,2295.000)
\put(4240,3420){\makebox(0,0)[b]{{\SetFigFont{10}{12.0}{\rmdefault}{\mddefault}{\updefault}$x$}}}
\put(4840,2895){\makebox(0,0)[lb]{{\SetFigFont{10}{12.0}{\rmdefault}{\mddefault}{\updefault}$\child^+$}}}
\put(5365,2220){\makebox(0,0)[b]{{\SetFigFont{10}{12.0}{\rmdefault}{\mddefault}{\updefault}$z$}}}
\put(3565,2070){\makebox(0,0)[b]{{\SetFigFont{10}{12.0}{\rmdefault}{\mddefault}{\updefault}$y$}}}
\put(3490,2370){\makebox(0,0)[rb]{{\SetFigFont{10}{12.0}{\rmdefault}{\mddefault}{\updefault}$\child^*$}}}
\put(3865,2970){\makebox(0,0)[rb]{{\SetFigFont{10}{12.0}{\rmdefault}{\mddefault}{\updefault}$\child^+$}}}
\put(4765,2445){\makebox(0,0)[b]{{\SetFigFont{10}{12.0}{\rmdefault}{\mddefault}{\updefault}$v$}}}
\put(4165,2445){\makebox(0,0)[b]{{\SetFigFont{10}{12.0}{\rmdefault}{\mddefault}{\updefault}$\nextsibling^+$}}}
\put(4465,2820){\makebox(0,0)[rb]{{\SetFigFont{10}{12.0}{\rmdefault}{\mddefault}{\updefault}$\child^*$}}}
\put(3490,2595){\makebox(0,0)[b]{{\SetFigFont{10}{12.0}{\rmdefault}{\mddefault}{\updefault}$u$}}}
\drawline(1240,8370)(790,7920)
\blacken\drawline(832.426,7983.640)(790.000,7920.000)(853.640,7962.426)(832.426,7983.640)
\drawline(1240,8370)(1690,7920)
\blacken\drawline(1626.360,7962.426)(1690.000,7920.000)(1647.574,7983.640)(1626.360,7962.426)
\drawline(790,7920)(1690,7920)
\blacken\drawline(1615.000,7905.000)(1690.000,7920.000)(1615.000,7935.000)(1615.000,7905.000)
\drawline(1240,3345)(565,2595)
\blacken\drawline(604.023,2660.782)(565.000,2595.000)(626.322,2640.713)(604.023,2660.782)
\thicklines
\drawline(565,2595)(1915,2595)
\blacken\thinlines
\drawline(1840.000,2580.000)(1915.000,2595.000)(1840.000,2610.000)(1840.000,2580.000)
\drawline(565,2595)(565,2220)
\blacken\drawline(550.000,2295.000)(565.000,2220.000)(580.000,2295.000)(550.000,2295.000)
\drawline(1915,2595)(1915,2220)
\blacken\drawline(1900.000,2295.000)(1915.000,2220.000)(1930.000,2295.000)(1900.000,2295.000)
\drawline(4240,6870)(3115,5895)
\blacken\drawline(3161.853,5955.455)(3115.000,5895.000)(3181.501,5932.784)(3161.853,5955.455)
\thicklines
\drawline(4240,6870)(5365,5895)
\blacken\thinlines
\drawline(5298.499,5932.784)(5365.000,5895.000)(5318.147,5955.455)(5298.499,5932.784)
\drawline(3790,6120)(4240,6870)
\blacken\drawline(4214.275,6797.971)(4240.000,6870.000)(4188.550,6813.405)(4214.275,6797.971)
\thicklines
\drawline(4690,6120)(5365,5895)
\blacken\thinlines
\drawline(5289.105,5904.487)(5365.000,5895.000)(5298.592,5932.947)(5289.105,5904.487)
\drawline(3790,6120)(4690,6120)
\blacken\drawline(4615.000,6105.000)(4690.000,6120.000)(4615.000,6135.000)(4615.000,6105.000)
\drawline(1240,5145)(565,4395)
\blacken\drawline(604.023,4460.782)(565.000,4395.000)(626.322,4440.713)(604.023,4460.782)
\thicklines
\drawline(1240,5145)(2365,4170)
\blacken\thinlines
\drawline(2298.499,4207.784)(2365.000,4170.000)(2318.147,4230.455)(2298.499,4207.784)
\thicklines
\drawline(1690,4395)(2365,4170)
\blacken\thinlines
\drawline(2289.105,4179.487)(2365.000,4170.000)(2298.592,4207.947)(2289.105,4179.487)
\drawline(565,4395)(1690,4395)
\blacken\drawline(1615.000,4380.000)(1690.000,4395.000)(1615.000,4410.000)(1615.000,4380.000)
\drawline(565,4395)(565,4020)
\blacken\drawline(550.000,4095.000)(565.000,4020.000)(580.000,4095.000)(550.000,4095.000)
\drawline(1240,1470)(565,720)
\blacken\drawline(604.023,785.782)(565.000,720.000)(626.322,765.713)(604.023,785.782)
\drawline(565,720)(1915,720)
\blacken\drawline(1840.000,705.000)(1915.000,720.000)(1840.000,735.000)(1840.000,705.000)
\drawline(565,720)(565,345)
\blacken\drawline(550.000,420.000)(565.000,345.000)(580.000,420.000)(550.000,420.000)
\drawline(1915,720)(1915,345)
\blacken\drawline(1900.000,420.000)(1915.000,345.000)(1930.000,420.000)(1900.000,420.000)
\thicklines
\drawline(1240,3345)(1915,2595)
\blacken\thinlines
\drawline(1853.678,2640.713)(1915.000,2595.000)(1875.977,2660.782)(1853.678,2640.713)
\put(1240,8445){\makebox(0,0)[b]{{\SetFigFont{10}{12.0}{\rmdefault}{\mddefault}{\updefault}$x$}}}
\put(940,8145){\makebox(0,0)[rb]{{\SetFigFont{10}{12.0}{\rmdefault}{\mddefault}{\updefault}$\child^+$}}}
\put(1540,8145){\makebox(0,0)[lb]{{\SetFigFont{10}{12.0}{\rmdefault}{\mddefault}{\updefault}$\child^+$}}}
\put(1765,7770){\makebox(0,0)[b]{{\SetFigFont{10}{12.0}{\rmdefault}{\mddefault}{\updefault}$z$}}}
\put(715,7770){\makebox(0,0)[b]{{\SetFigFont{10}{12.0}{\rmdefault}{\mddefault}{\updefault}$y$}}}
\put(1240,7695){\makebox(0,0)[b]{{\SetFigFont{10}{12.0}{\rmdefault}{\mddefault}{\updefault}$\following$}}}
\put(1240,3420){\makebox(0,0)[b]{{\SetFigFont{10}{12.0}{\rmdefault}{\mddefault}{\updefault}$x$}}}
\put(565,2070){\makebox(0,0)[b]{{\SetFigFont{10}{12.0}{\rmdefault}{\mddefault}{\updefault}$y$}}}
\put(490,2370){\makebox(0,0)[rb]{{\SetFigFont{10}{12.0}{\rmdefault}{\mddefault}{\updefault}$\child^*$}}}
\put(865,2970){\makebox(0,0)[rb]{{\SetFigFont{10}{12.0}{\rmdefault}{\mddefault}{\updefault}$\child^+$}}}
\put(1615,2970){\makebox(0,0)[lb]{{\SetFigFont{10}{12.0}{\rmdefault}{\mddefault}{\updefault}$\child^+$}}}
\put(1240,2670){\makebox(0,0)[b]{{\SetFigFont{10}{12.0}{\rmdefault}{\mddefault}{\updefault}$\nextsibling^+$}}}
\put(1990,2370){\makebox(0,0)[lb]{{\SetFigFont{10}{12.0}{\rmdefault}{\mddefault}{\updefault}$\child^*$}}}
\put(1915,2070){\makebox(0,0)[b]{{\SetFigFont{10}{12.0}{\rmdefault}{\mddefault}{\updefault}$z$}}}
\put(4240,6945){\makebox(0,0)[b]{{\SetFigFont{10}{12.0}{\rmdefault}{\mddefault}{\updefault}$x$}}}
\put(3115,5745){\makebox(0,0)[b]{{\SetFigFont{10}{12.0}{\rmdefault}{\mddefault}{\updefault}$y$}}}
\put(4840,6420){\makebox(0,0)[lb]{{\SetFigFont{10}{12.0}{\rmdefault}{\mddefault}{\updefault}$\child^+$}}}
\put(5365,5745){\makebox(0,0)[b]{{\SetFigFont{10}{12.0}{\rmdefault}{\mddefault}{\updefault}$z$}}}
\put(5140,5820){\makebox(0,0)[rb]{{\SetFigFont{10}{12.0}{\rmdefault}{\mddefault}{\updefault}$\child^*$}}}
\put(4765,6120){\makebox(0,0)[b]{{\SetFigFont{10}{12.0}{\rmdefault}{\mddefault}{\updefault}$v$}}}
\put(4240,5970){\makebox(0,0)[b]{{\SetFigFont{10}{12.0}{\rmdefault}{\mddefault}{\updefault}$\nextsibling^+$}}}
\put(3715,6120){\makebox(0,0)[b]{{\SetFigFont{10}{12.0}{\rmdefault}{\mddefault}{\updefault}$u$}}}
\put(3715,6420){\makebox(0,0)[rb]{{\SetFigFont{10}{12.0}{\rmdefault}{\mddefault}{\updefault}$\child^+$}}}
\put(3940,6270){\makebox(0,0)[lb]{{\SetFigFont{10}{12.0}{\rmdefault}{\mddefault}{\updefault}$\child^*$}}}
\put(1240,5520){\makebox(0,0)[b]{{\SetFigFont{10}{12.0}{\rmdefault}{\mddefault}{\updefault}$\Downarrow$}}}
\put(1240,3645){\makebox(0,0)[b]{{\SetFigFont{10}{12.0}{\rmdefault}{\mddefault}{\updefault}$\Downarrow$}}}
\put(1240,7320){\makebox(0,0)[b]{{\SetFigFont{10}{12.0}{\rmdefault}{\mddefault}{\updefault}$\Downarrow$}}}
\put(1240,1845){\makebox(0,0)[b]{{\SetFigFont{10}{12.0}{\rmdefault}{\mddefault}{\updefault}$\Downarrow$}}}
\put(2815,6270){\makebox(0,0)[b]{{\SetFigFont{10}{12.0}{\rmdefault}{\mddefault}{\updefault}$\Rightarrow$}}}
\put(5740,6270){\makebox(0,0)[b]{{\SetFigFont{10}{12.0}{\rmdefault}{\mddefault}{\updefault}$\Rightarrow$}}}
\put(4240,5520){\makebox(0,0)[b]{{\SetFigFont{10}{12.0}{\rmdefault}{\mddefault}{\updefault}$\Downarrow$}}}
\put(7165,7320){\makebox(0,0)[b]{{\SetFigFont{10}{12.0}{\rmdefault}{\mddefault}{\updefault}$\Uparrow$}}}
\put(7165,3720){\makebox(0,0)[b]{{\SetFigFont{10}{12.0}{\rmdefault}{\mddefault}{\updefault}(unsatisfiable)}}}
\put(2815,3420){\makebox(0,0)[b]{{\SetFigFont{10}{12.0}{\rmdefault}{\mddefault}{\updefault}\begin{rotate}{45}{$\Downarrow$}\end{rotate}}}}
\put(5665,3420){\makebox(0,0)[b]{{\SetFigFont{10}{12.0}{\rmdefault}{\mddefault}{\updefault}\begin{rotate}{45}{$\Downarrow$}\end{rotate}}}}
\put(1240,5220){\makebox(0,0)[b]{{\SetFigFont{10}{12.0}{\rmdefault}{\mddefault}{\updefault}$x$}}}
\put(1840,4695){\makebox(0,0)[lb]{{\SetFigFont{10}{12.0}{\rmdefault}{\mddefault}{\updefault}$\child^+$}}}
\put(2365,4020){\makebox(0,0)[b]{{\SetFigFont{10}{12.0}{\rmdefault}{\mddefault}{\updefault}$z$}}}
\put(2140,4095){\makebox(0,0)[rb]{{\SetFigFont{10}{12.0}{\rmdefault}{\mddefault}{\updefault}$\child^*$}}}
\put(1690,4245){\makebox(0,0)[b]{{\SetFigFont{10}{12.0}{\rmdefault}{\mddefault}{\updefault}$v$}}}
\put(1240,4470){\makebox(0,0)[b]{{\SetFigFont{10}{12.0}{\rmdefault}{\mddefault}{\updefault}$\nextsibling^+$}}}
\put(565,3870){\makebox(0,0)[b]{{\SetFigFont{10}{12.0}{\rmdefault}{\mddefault}{\updefault}$y$}}}
\put(490,4170){\makebox(0,0)[rb]{{\SetFigFont{10}{12.0}{\rmdefault}{\mddefault}{\updefault}$\child^*$}}}
\put(865,4770){\makebox(0,0)[rb]{{\SetFigFont{10}{12.0}{\rmdefault}{\mddefault}{\updefault}$\child^+$}}}
\put(490,2595){\makebox(0,0)[b]{{\SetFigFont{10}{12.0}{\rmdefault}{\mddefault}{\updefault}$u$}}}
\put(490,4395){\makebox(0,0)[b]{{\SetFigFont{10}{12.0}{\rmdefault}{\mddefault}{\updefault}$u$}}}
\put(1240,1545){\makebox(0,0)[b]{{\SetFigFont{10}{12.0}{\rmdefault}{\mddefault}{\updefault}$x$}}}
\put(565,195){\makebox(0,0)[b]{{\SetFigFont{10}{12.0}{\rmdefault}{\mddefault}{\updefault}$y$}}}
\put(490,495){\makebox(0,0)[rb]{{\SetFigFont{10}{12.0}{\rmdefault}{\mddefault}{\updefault}$\child^*$}}}
\put(865,1095){\makebox(0,0)[rb]{{\SetFigFont{10}{12.0}{\rmdefault}{\mddefault}{\updefault}$\child^+$}}}
\put(1240,795){\makebox(0,0)[b]{{\SetFigFont{10}{12.0}{\rmdefault}{\mddefault}{\updefault}$\nextsibling^+$}}}
\put(1990,495){\makebox(0,0)[lb]{{\SetFigFont{10}{12.0}{\rmdefault}{\mddefault}{\updefault}$\child^*$}}}
\put(1915,195){\makebox(0,0)[b]{{\SetFigFont{10}{12.0}{\rmdefault}{\mddefault}{\updefault}$z$}}}
\put(490,720){\makebox(0,0)[b]{{\SetFigFont{10}{12.0}{\rmdefault}{\mddefault}{\updefault}$u$}}}
\put(1990,720){\makebox(0,0)[b]{{\SetFigFont{10}{12.0}{\rmdefault}{\mddefault}{\updefault}$v$}}}
\put(1990,2595){\makebox(0,0)[b]{{\SetFigFont{10}{12.0}{\rmdefault}{\mddefault}{\updefault}$v$}}}
\put(5740,4470){\makebox(0,0)[b]{{\SetFigFont{10}{12.0}{\rmdefault}{\mddefault}{\updefault}$\Rightarrow$}}}
\put(4165,45){\makebox(0,0)[b]{{\SetFigFont{10}{12.0}{\rmdefault}{\mddefault}{\updefault}(unsatisfiable)}}}
\put(7165,45){\makebox(0,0)[b]{{\SetFigFont{10}{12.0}{\rmdefault}{\mddefault}{\updefault}(unsatisfiable)}}}
\put(7165,1845){\makebox(0,0)[b]{{\SetFigFont{10}{12.0}{\rmdefault}{\mddefault}{\updefault}$\Downarrow$}}}
\put(5740,1845){\makebox(0,0)[b]{{\SetFigFont{10}{12.0}{\rmdefault}{\mddefault}{\updefault}\begin{rotate}{-45}{$\Downarrow$}\end{rotate}}}}
\put(4165,1845){\makebox(0,0)[b]{{\SetFigFont{10}{12.0}{\rmdefault}{\mddefault}{\updefault}(unsatisfiable)}}}
\put(7165,8970){\makebox(0,0)[b]{{\SetFigFont{10}{12.0}{\rmdefault}{\mddefault}{\updefault}(unsatisfiable)}}}
\end{picture}
}

%% file: diamond.tex
\setlength{\unitlength}{0.00083333in}
\begingroup\makeatletter\ifx\SetFigFont\undefined%
\gdef\SetFigFont#1#2#3#4#5{%
  \reset@font\fontsize{#1}{#2pt}%
  \fontfamily{#3}\fontseries{#4}\fontshape{#5}%
  \selectfont}%
\fi\endgroup%
{\renewcommand{\dashlinestretch}{30}
\begin{picture}(3592,3918)(0,-10)
\put(2285,1700){\circle*{48}}
\put(2285,1400){\circle*{48}}
\put(2285,1100){\circle*{48}}
\put(2285,800){\circle*{48}}
\put(2285,3275){\circle*{48}}
\put(2285,2975){\circle*{48}}
\put(2285,2675){\circle*{48}}
\put(2285,2375){\circle*{48}}
\put(3560,1700){\circle*{48}}
\put(3560,1400){\circle*{48}}
\put(3560,1100){\circle*{48}}
\put(3560,800){\circle*{48}}
\put(3560,2975){\circle*{48}}
\put(3560,2675){\circle*{48}}
\put(3560,2375){\circle*{48}}
\put(3560,3275){\circle*{48}}
\put(2285,3575){\circle*{48}}
\put(3560,3575){\circle*{48}}
\put(2285,500){\circle*{48}}
\put(3560,500){\circle*{48}}
\drawline(462,3720)(12,3270)
\blacken\drawline(54.426,3333.640)(12.000,3270.000)(75.640,3312.426)(54.426,3333.640)
\drawline(462,3720)(912,3270)
\blacken\drawline(848.360,3312.426)(912.000,3270.000)(869.574,3333.640)(848.360,3312.426)
\drawline(912,3270)(462,2820)
\blacken\drawline(504.426,2883.640)(462.000,2820.000)(525.640,2862.426)(504.426,2883.640)
\drawline(12,3270)(462,2820)
\blacken\drawline(398.360,2862.426)(462.000,2820.000)(419.574,2883.640)(398.360,2862.426)
\drawline(462,2820)(12,2370)
\blacken\drawline(54.426,2433.640)(12.000,2370.000)(75.640,2412.426)(54.426,2433.640)
\drawline(462,2820)(912,2370)
\blacken\drawline(848.360,2412.426)(912.000,2370.000)(869.574,2433.640)(848.360,2412.426)
\drawline(912,2370)(462,1920)
\blacken\drawline(504.426,1983.640)(462.000,1920.000)(525.640,1962.426)(504.426,1983.640)
\drawline(12,2370)(462,1920)
\blacken\drawline(398.360,1962.426)(462.000,1920.000)(419.574,1983.640)(398.360,1962.426)
\drawline(462,1395)(12,945)
\blacken\drawline(54.426,1008.640)(12.000,945.000)(75.640,987.426)(54.426,1008.640)
\drawline(462,1395)(912,945)
\blacken\drawline(848.360,987.426)(912.000,945.000)(869.574,1008.640)(848.360,987.426)
\drawline(912,945)(462,495)
\blacken\drawline(504.426,558.640)(462.000,495.000)(525.640,537.426)(504.426,558.640)
\drawline(12,945)(462,495)
\blacken\drawline(398.360,537.426)(462.000,495.000)(419.574,558.640)(398.360,537.426)
\put(462,1545){\makebox(0,0)[b]{\smash{{{\SetFigFont{10}{12.0}{\rmdefault}{\mddefault}{\updefault}$\vdots$}}}}}
\put(2937,1995){\makebox(0,0)[b]{\smash{{{\SetFigFont{10}{12.0}{\rmdefault}{\mddefault}{\updefault}$\cdots$}}}}}
\put(2262,1995){\makebox(0,0)[lb]{\smash{{{\SetFigFont{10}{12.0}{\rmdefault}{\mddefault}{\updefault}$\vdots$}}}}}
\put(2187,1695){\makebox(0,0)[rb]{\smash{{{\SetFigFont{10}{12.0}{\rmdefault}{\mddefault}{\updefault}$Y_n$}}}}}
\put(2187,1395){\makebox(0,0)[rb]{\smash{{{\SetFigFont{10}{12.0}{\rmdefault}{\mddefault}{\updefault}$X_n$}}}}}
\put(2187,1095){\makebox(0,0)[rb]{\smash{{{\SetFigFont{10}{12.0}{\rmdefault}{\mddefault}{\updefault}$X_n'$}}}}}
\put(2187,795){\makebox(0,0)[rb]{\smash{{{\SetFigFont{10}{12.0}{\rmdefault}{\mddefault}{\updefault}$Y_{n+1}$}}}}}
\put(2187,3270){\makebox(0,0)[rb]{\smash{{{\SetFigFont{10}{12.0}{\rmdefault}{\mddefault}{\updefault}$Y_1$}}}}}
\put(2187,2970){\makebox(0,0)[rb]{\smash{{{\SetFigFont{10}{12.0}{\rmdefault}{\mddefault}{\updefault}$X_1$}}}}}
\put(2187,2670){\makebox(0,0)[rb]{\smash{{{\SetFigFont{10}{12.0}{\rmdefault}{\mddefault}{\updefault}$X_1'$}}}}}
\put(2187,2370){\makebox(0,0)[rb]{\smash{{{\SetFigFont{10}{12.0}{\rmdefault}{\mddefault}{\updefault}$Y_2$}}}}}
\put(3537,1995){\makebox(0,0)[lb]{\smash{{{\SetFigFont{10}{12.0}{\rmdefault}{\mddefault}{\updefault}$\vdots$}}}}}
\put(3462,1695){\makebox(0,0)[rb]{\smash{{{\SetFigFont{10}{12.0}{\rmdefault}{\mddefault}{\updefault}$Y_n$}}}}}
\put(3462,795){\makebox(0,0)[rb]{\smash{{{\SetFigFont{10}{12.0}{\rmdefault}{\mddefault}{\updefault}$Y_{n+1}$}}}}}
\put(3462,3270){\makebox(0,0)[rb]{\smash{{{\SetFigFont{10}{12.0}{\rmdefault}{\mddefault}{\updefault}$Y_1$}}}}}
\put(3462,2370){\makebox(0,0)[rb]{\smash{{{\SetFigFont{10}{12.0}{\rmdefault}{\mddefault}{\updefault}$Y_2$}}}}}
\put(3537,3045){\makebox(0,0)[lb]{\smash{{{\SetFigFont{10}{12.0}{\rmdefault}{\mddefault}{\updefault}$\vdots \; \big\} p(n)$}}}}}
\put(3462,1095){\makebox(0,0)[rb]{\smash{{{\SetFigFont{10}{12.0}{\rmdefault}{\mddefault}{\updefault}$X_n$}}}}}
\put(3462,1395){\makebox(0,0)[rb]{\smash{{{\SetFigFont{10}{12.0}{\rmdefault}{\mddefault}{\updefault}$X_n'$}}}}}
\put(3462,2670){\makebox(0,0)[rb]{\smash{{{\SetFigFont{10}{12.0}{\rmdefault}{\mddefault}{\updefault}$X_1$}}}}}
\put(3462,2970){\makebox(0,0)[rb]{\smash{{{\SetFigFont{10}{12.0}{\rmdefault}{\mddefault}{\updefault}$X_1'$}}}}}
\put(3537,2445){\makebox(0,0)[lb]{\smash{{{\SetFigFont{10}{12.0}{\rmdefault}{\mddefault}{\updefault}$\vdots \; \big\} p(n)$}}}}}
\put(3537,2745){\makebox(0,0)[lb]{\smash{{{\SetFigFont{10}{12.0}{\rmdefault}{\mddefault}{\updefault}$\vdots \; \big\} p(n)$}}}}}
\put(3537,1470){\makebox(0,0)[lb]{\smash{{{\SetFigFont{10}{12.0}{\rmdefault}{\mddefault}{\updefault}$\vdots \; \big\} p(n)$}}}}}
\put(3537,1170){\makebox(0,0)[lb]{\smash{{{\SetFigFont{10}{12.0}{\rmdefault}{\mddefault}{\updefault}$\vdots \; \big\} p(n)$}}}}}
\put(3537,870){\makebox(0,0)[lb]{\smash{{{\SetFigFont{10}{12.0}{\rmdefault}{\mddefault}{\updefault}$\vdots \; \big\} p(n)$}}}}}
\put(2262,3045){\makebox(0,0)[lb]{\smash{{{\SetFigFont{10}{12.0}{\rmdefault}{\mddefault}{\updefault}$\vdots \; \big\} p(n)$}}}}}
\put(2262,2745){\makebox(0,0)[lb]{\smash{{{\SetFigFont{10}{12.0}{\rmdefault}{\mddefault}{\updefault}$\vdots \; \big\} p(n)$}}}}}
\put(2262,2445){\makebox(0,0)[lb]{\smash{{{\SetFigFont{10}{12.0}{\rmdefault}{\mddefault}{\updefault}$\vdots \; \big\} p(n)$}}}}}
\put(2262,1470){\makebox(0,0)[lb]{\smash{{{\SetFigFont{10}{12.0}{\rmdefault}{\mddefault}{\updefault}$\vdots \; \big\} p(n)$}}}}}
\put(2262,1170){\makebox(0,0)[lb]{\smash{{{\SetFigFont{10}{12.0}{\rmdefault}{\mddefault}{\updefault}$\vdots \; \big\} p(n)$}}}}}
\put(2262,870){\makebox(0,0)[lb]{\smash{{{\SetFigFont{10}{12.0}{\rmdefault}{\mddefault}{\updefault}$\vdots \; \big\} p(n)$}}}}}
\put(2262,570){\makebox(0,0)[lb]{\smash{{{\SetFigFont{10}{12.0}{\rmdefault}{\mddefault}{\updefault}$\vdots \; \big\} p(n)$}}}}}
\put(3537,570){\makebox(0,0)[lb]{\smash{{{\SetFigFont{10}{12.0}{\rmdefault}{\mddefault}{\updefault}$\vdots \; \big\} p(n)$}}}}}
\put(3537,3345){\makebox(0,0)[lb]{\smash{{{\SetFigFont{10}{12.0}{\rmdefault}{\mddefault}{\updefault}$\vdots \; \big\} p(n)$}}}}}
\put(2262,3345){\makebox(0,0)[lb]{\smash{{{\SetFigFont{10}{12.0}{\rmdefault}{\mddefault}{\updefault}$\vdots \; \big\} p(n)$}}}}}
\put(462,45){\makebox(0,0)[b]{\smash{{{\SetFigFont{10}{12.0}{\rmdefault}{\mddefault}{\updefault}(a)}}}}}
\put(2937,45){\makebox(0,0)[b]{\smash{{{\SetFigFont{10}{12.0}{\rmdefault}{\mddefault}{\updefault}(b)}}}}}
\put(687,3495){\makebox(0,0)[lb]{\smash{{{\SetFigFont{9}{10.8}{\rmdefault}{\mddefault}{\updefault}$\child^+$}}}}}
\put(237,3495){\makebox(0,0)[rb]{\smash{{{\SetFigFont{9}{10.8}{\rmdefault}{\mddefault}{\updefault}$\child^+$}}}}}
\put(237,2895){\makebox(0,0)[rb]{\smash{{{\SetFigFont{9}{10.8}{\rmdefault}{\mddefault}{\updefault}$\child^+$}}}}}
\put(687,2895){\makebox(0,0)[lb]{\smash{{{\SetFigFont{9}{10.8}{\rmdefault}{\mddefault}{\updefault}$\child^+$}}}}}
\put(687,2595){\makebox(0,0)[lb]{\smash{{{\SetFigFont{9}{10.8}{\rmdefault}{\mddefault}{\updefault}$\child^+$}}}}}
\put(237,2595){\makebox(0,0)[rb]{\smash{{{\SetFigFont{9}{10.8}{\rmdefault}{\mddefault}{\updefault}$\child^+$}}}}}
\put(237,1995){\makebox(0,0)[rb]{\smash{{{\SetFigFont{9}{10.8}{\rmdefault}{\mddefault}{\updefault}$\child^+$}}}}}
\put(687,1995){\makebox(0,0)[lb]{\smash{{{\SetFigFont{9}{10.8}{\rmdefault}{\mddefault}{\updefault}$\child^+$}}}}}
\put(462,1770){\makebox(0,0)[b]{\smash{{{\SetFigFont{10}{12.0}{\rmdefault}{\mddefault}{\updefault}$Y_3$}}}}}
\put(12,2295){\makebox(0,0)[rb]{\smash{{{\SetFigFont{10}{12.0}{\rmdefault}{\mddefault}{\updefault}$X_2$}}}}}
\put(912,2295){\makebox(0,0)[lb]{\smash{{{\SetFigFont{10}{12.0}{\rmdefault}{\mddefault}{\updefault}$X_2'$}}}}}
\put(12,3195){\makebox(0,0)[rb]{\smash{{{\SetFigFont{10}{12.0}{\rmdefault}{\mddefault}{\updefault}$X_1$}}}}}
\put(912,3195){\makebox(0,0)[lb]{\smash{{{\SetFigFont{10}{12.0}{\rmdefault}{\mddefault}{\updefault}$X_1'$}}}}}
\put(462,3795){\makebox(0,0)[b]{\smash{{{\SetFigFont{10}{12.0}{\rmdefault}{\mddefault}{\updefault}$Y_1$}}}}}
\put(462,2595){\makebox(0,0)[b]{\smash{{{\SetFigFont{10}{12.0}{\rmdefault}{\mddefault}{\updefault}$Y_2$}}}}}
\put(612,1395){\makebox(0,0)[b]{\smash{{{\SetFigFont{10}{12.0}{\rmdefault}{\mddefault}{\updefault}$Y_n$}}}}}
\put(687,1170){\makebox(0,0)[lb]{\smash{{{\SetFigFont{9}{10.8}{\rmdefault}{\mddefault}{\updefault}$\child^+$}}}}}
\put(237,1170){\makebox(0,0)[rb]{\smash{{{\SetFigFont{9}{10.8}{\rmdefault}{\mddefault}{\updefault}$\child^+$}}}}}
\put(12,870){\makebox(0,0)[rb]{\smash{{{\SetFigFont{10}{12.0}{\rmdefault}{\mddefault}{\updefault}$X_n$}}}}}
\put(912,870){\makebox(0,0)[lb]{\smash{{{\SetFigFont{10}{12.0}{\rmdefault}{\mddefault}{\updefault}$X_n'$}}}}}
\put(462,345){\makebox(0,0)[b]{\smash{{{\SetFigFont{10}{12.0}{\rmdefault}{\mddefault}{\updefault}$Y_{n+1}$}}}}}
\put(237,570){\makebox(0,0)[rb]{\smash{{{\SetFigFont{9}{10.8}{\rmdefault}{\mddefault}{\updefault}$\child^+$}}}}}
\put(687,570){\makebox(0,0)[lb]{\smash{{{\SetFigFont{9}{10.8}{\rmdefault}{\mddefault}{\updefault}$\child^+$}}}}}
\end{picture}
}

%% file: stretch.tex
\setlength{\unitlength}{0.00083333in}
\begingroup\makeatletter\ifx\SetFigFont\undefined%
\gdef\SetFigFont#1#2#3#4#5{%
  \reset@font\fontsize{#1}{#2pt}%
  \fontfamily{#3}\fontseries{#4}\fontshape{#5}%
  \selectfont}%
\fi\endgroup%
{\renewcommand{\dashlinestretch}{30}
\begin{picture}(3936,2568)(0,-10)
\put(2006,1170){\circle*{48}}
\put(2006,570){\circle*{48}}
\put(2006,870){\circle*{48}}
\drawline(2006,1170)(2006,570)
\put(2606,1770){\circle*{48}}
\put(2606,1170){\circle*{48}}
\put(2606,1470){\circle*{48}}
\drawline(2606,1770)(2606,1170)
\put(3281,2370){\circle*{48}}
\put(3281,1770){\circle*{48}}
\put(3281,2070){\circle*{48}}
\drawline(3281,2370)(3281,1770)
\put(3806,2370){\circle*{48}}
\put(3806,1770){\circle*{48}}
\put(3806,2070){\circle*{48}}
\drawline(3806,2370)(3806,1770)
\put(3806,1770){\circle*{48}}
\put(3806,1170){\circle*{48}}
\put(3806,1470){\circle*{48}}
\drawline(3806,1770)(3806,1170)
\put(3806,1170){\circle*{48}}
\put(3806,570){\circle*{48}}
\put(3806,870){\circle*{48}}
\drawline(3806,1170)(3806,570)
\put(131,2370){\circle*{48}}
\put(131,1770){\circle*{48}}
\put(131,2070){\circle*{48}}
\drawline(131,2370)(131,1770)
\put(506,1770){\circle*{48}}
\put(506,1170){\circle*{48}}
\put(506,1470){\circle*{48}}
\put(881,1170){\circle*{48}}
\put(881,570){\circle*{48}}
\put(881,870){\circle*{48}}
\drawline(506,1770)(506,1170)
\drawline(881,1170)(881,570)
\drawline(131,2370)(506,1170)
\blacken\drawline(469.312,1237.112)(506.000,1170.000)(497.947,1246.060)(469.312,1237.112)
\drawline(502,1769)(877,569)
\blacken\drawline(840.312,636.112)(877.000,569.000)(868.947,645.060)(840.312,636.112)
\put(131,1545){\makebox(0,0)[b]{{\SetFigFont{10}{12.0}{\rmdefault}{\mddefault}{\updefault}$A$}}}
\put(881,1245){\makebox(0,0)[b]{{\SetFigFont{10}{12.0}{\rmdefault}{\mddefault}{\updefault}$B$}}}
\put(581,1470){\makebox(0,0)[lb]{{\SetFigFont{9}{10.8}{\rmdefault}{\mddefault}{\updefault}$\child^*$}}}
\put(206,2070){\makebox(0,0)[lb]{{\SetFigFont{9}{10.8}{\rmdefault}{\mddefault}{\updefault}$\child^*$}}}
\put(506,45){\makebox(0,0)[b]{{\SetFigFont{10}{12.0}{\rmdefault}{\mddefault}{\updefault}$(a)$}}}
\drawline(2006,1170)(2606,1170)
\blacken\drawline(2531.000,1155.000)(2606.000,1170.000)(2531.000,1185.000)(2531.000,1155.000)
\drawline(2606,1770)(3281,1770)
\blacken\drawline(3206.000,1755.000)(3281.000,1770.000)(3206.000,1785.000)(3206.000,1755.000)
\dottedline{45}(2006,570)(3806,570)
\blacken\drawline(3731.000,555.000)(3806.000,570.000)(3731.000,585.000)(3731.000,555.000)
\dottedline{45}(3281,2370)(3806,2370)
\blacken\drawline(3731.000,2355.000)(3806.000,2370.000)(3731.000,2385.000)(3731.000,2355.000)
\put(2081,1020){\makebox(0,0)[lb]{{\SetFigFont{9}{10.8}{\rmdefault}{\mddefault}{\updefault}$\child^*$}}}
\put(2681,1620){\makebox(0,0)[lb]{{\SetFigFont{9}{10.8}{\rmdefault}{\mddefault}{\updefault}$\child^*$}}}
\put(2006,345){\makebox(0,0)[b]{{\SetFigFont{10}{12.0}{\rmdefault}{\mddefault}{\updefault}$A$}}}
\put(3281,2445){\makebox(0,0)[b]{{\SetFigFont{10}{12.0}{\rmdefault}{\mddefault}{\updefault}$B$}}}
\put(3806,2445){\makebox(0,0)[b]{{\SetFigFont{10}{12.0}{\rmdefault}{\mddefault}{\updefault}$B$}}}
\put(3806,345){\makebox(0,0)[b]{{\SetFigFont{10}{12.0}{\rmdefault}{\mddefault}{\updefault}$A$}}}
\put(3738,548){\makebox(0,0)[lb]{{\SetFigFont{10}{12.0}{\rmdefault}{\mddefault}{\updefault}$\left. \begin{array}{l} \\[0.8ex] \\ \\ \\ \\ \\ \\ \\[3ex] \end{array} \right\} < |Q|$}}}
\put(2906,345){\makebox(0,0)[b]{{\SetFigFont{10}{12.0}{\rmdefault}{\mddefault}{\updefault}$\theta$}}}
\put(2906,45){\makebox(0,0)[b]{{\SetFigFont{10}{12.0}{\rmdefault}{\mddefault}{\updefault}$(b)$}}}
\end{picture}
}

%% file: compmerge.tex
\setlength{\unitlength}{0.00083333in}
\begingroup\makeatletter\ifx\SetFigFont\undefined%
\gdef\SetFigFont#1#2#3#4#5{%
  \reset@font\fontsize{#1}{#2pt}%
  \fontfamily{#3}\fontseries{#4}\fontshape{#5}%
  \selectfont}%
\fi\endgroup%
{\renewcommand{\dashlinestretch}{30}
\begin{picture}(6107,1680)(0,-10)
\put(877,1320){\blacken\ellipse{48}{48}}
\put(877,1320){\ellipse{48}{48}}
\put(877,720){\blacken\ellipse{48}{48}}
\put(877,720){\ellipse{48}{48}}
\put(877,1020){\blacken\ellipse{48}{48}}
\put(877,1020){\ellipse{48}{48}}
\path(877,1320)(877,720)
\put(1627,1020){\blacken\ellipse{48}{48}}
\put(1627,1020){\ellipse{48}{48}}
\put(1627,420){\blacken\ellipse{48}{48}}
\put(1627,420){\ellipse{48}{48}}
\put(1627,720){\blacken\ellipse{48}{48}}
\put(1627,720){\ellipse{48}{48}}
\path(1627,1020)(1627,420)
\put(2977,1320){\blacken\ellipse{48}{48}}
\put(2977,1320){\ellipse{48}{48}}
\put(2977,720){\blacken\ellipse{48}{48}}
\put(2977,720){\ellipse{48}{48}}
\put(2977,1020){\blacken\ellipse{48}{48}}
\put(2977,1020){\ellipse{48}{48}}
\path(2977,1320)(2977,720)
\put(4102,1620){\blacken\ellipse{48}{48}}
\put(4102,1620){\ellipse{48}{48}}
\put(4102,1020){\blacken\ellipse{48}{48}}
\put(4102,1020){\ellipse{48}{48}}
\put(4102,1320){\blacken\ellipse{48}{48}}
\put(4102,1320){\ellipse{48}{48}}
\path(4102,1620)(4102,1020)
\put(4102,1020){\blacken\ellipse{48}{48}}
\put(4102,1020){\ellipse{48}{48}}
\put(4102,420){\blacken\ellipse{48}{48}}
\put(4102,420){\ellipse{48}{48}}
\put(4102,720){\blacken\ellipse{48}{48}}
\put(4102,720){\ellipse{48}{48}}
\path(4102,1020)(4102,420)
\put(4627,1620){\blacken\ellipse{48}{48}}
\put(4627,1620){\ellipse{48}{48}}
\put(4627,1020){\blacken\ellipse{48}{48}}
\put(4627,1020){\ellipse{48}{48}}
\put(4627,1320){\blacken\ellipse{48}{48}}
\put(4627,1320){\ellipse{48}{48}}
\path(4627,1620)(4627,1020)
\put(4627,1320){\blacken\ellipse{48}{48}}
\put(4627,1320){\ellipse{48}{48}}
\put(4627,720){\blacken\ellipse{48}{48}}
\put(4627,720){\ellipse{48}{48}}
\put(4627,1020){\blacken\ellipse{48}{48}}
\put(4627,1020){\ellipse{48}{48}}
\path(4627,1320)(4627,720)
\path(4102,1320)(4627,1020)
\put(4402,45){\makebox(0,0)[b]{{\SetFigFont{10}{12.0}{\rmdefault}{\mddefault}{\updefault}(b)}}}
\put(4102,1545){\makebox(0,0)[rb]{{\SetFigFont{10}{12.0}{\rmdefault}{\mddefault}{\updefault}$x^{C_1}_1$}}}
\put(4102,1245){\makebox(0,0)[rb]{{\SetFigFont{10}{12.0}{\rmdefault}{\mddefault}{\updefault}$x^{C_1}_2$}}}
\put(4102,945){\makebox(0,0)[rb]{{\SetFigFont{10}{12.0}{\rmdefault}{\mddefault}{\updefault}$x^{C_1}_3$}}}
\put(4102,645){\makebox(0,0)[rb]{{\SetFigFont{10}{12.0}{\rmdefault}{\mddefault}{\updefault}$x^{C_4}_2$}}}
\put(4102,345){\makebox(0,0)[rb]{{\SetFigFont{10}{12.0}{\rmdefault}{\mddefault}{\updefault}$x^{C_4}_3$}}}
\put(4702,1545){\makebox(0,0)[lb]{{\SetFigFont{10}{12.0}{\rmdefault}{\mddefault}{\updefault}$x^{C_3}_1$}}}
\put(4702,1245){\makebox(0,0)[lb]{{\SetFigFont{10}{12.0}{\rmdefault}{\mddefault}{\updefault}$x^{C_2}_1$}}}
\put(4702,945){\makebox(0,0)[lb]{{\SetFigFont{10}{12.0}{\rmdefault}{\mddefault}{\updefault}$x^{C_2}_2$}}}
\put(4702,645){\makebox(0,0)[lb]{{\SetFigFont{10}{12.0}{\rmdefault}{\mddefault}{\updefault}$x^{C_2}_3$}}}
\put(4177,945){\makebox(0,0)[b]{{\SetFigFont{10}{12.0}{\rmdefault}{\mddefault}{\updefault}$A$}}}
\put(5902,1620){\blacken\ellipse{48}{48}}
\put(5902,1620){\ellipse{48}{48}}
\put(5902,1020){\blacken\ellipse{48}{48}}
\put(5902,1020){\ellipse{48}{48}}
\put(5902,1320){\blacken\ellipse{48}{48}}
\put(5902,1320){\ellipse{48}{48}}
\path(5902,1620)(5902,1020)
\put(5902,1020){\blacken\ellipse{48}{48}}
\put(5902,1020){\ellipse{48}{48}}
\put(5902,420){\blacken\ellipse{48}{48}}
\put(5902,420){\ellipse{48}{48}}
\put(5902,720){\blacken\ellipse{48}{48}}
\put(5902,720){\ellipse{48}{48}}
\path(5902,1020)(5902,420)
\put(5902,1545){\makebox(0,0)[rb]{{\SetFigFont{10}{12.0}{\rmdefault}{\mddefault}{\updefault}$x^{C_1}_1$}}}
\put(5902,1245){\makebox(0,0)[rb]{{\SetFigFont{10}{12.0}{\rmdefault}{\mddefault}{\updefault}$x^{C_1}_2$}}}
\put(5902,945){\makebox(0,0)[rb]{{\SetFigFont{10}{12.0}{\rmdefault}{\mddefault}{\updefault}$x^{C_1}_3$}}}
\put(5902,645){\makebox(0,0)[rb]{{\SetFigFont{10}{12.0}{\rmdefault}{\mddefault}{\updefault}$x^{C_4}_2$}}}
\put(5902,345){\makebox(0,0)[rb]{{\SetFigFont{10}{12.0}{\rmdefault}{\mddefault}{\updefault}$x^{C_4}_3$}}}
\put(5902,45){\makebox(0,0)[b]{{\SetFigFont{10}{12.0}{\rmdefault}{\mddefault}{\updefault}(c)}}}
\put(5977,945){\makebox(0,0)[b]{{\SetFigFont{10}{12.0}{\rmdefault}{\mddefault}{\updefault}$A$}}}
\put(2302,1320){\blacken\ellipse{48}{48}}
\put(2302,1320){\ellipse{48}{48}}
\put(2302,1020){\blacken\ellipse{48}{48}}
\put(2302,1020){\ellipse{48}{48}}
\path(877,1020)(1627,720)
\blacken\path(1551.793,733.927)(1627.000,720.000)(1562.935,761.781)(1551.793,733.927)
\path(1627,1020)(2302,1020)
\blacken\path(2227.000,1005.000)(2302.000,1020.000)(2227.000,1035.000)(2227.000,1005.000)
\path(2302,1320)(2302,1020)
\path(2302,1020)(2977,1020)
\blacken\path(2902.000,1005.000)(2977.000,1020.000)(2902.000,1035.000)(2902.000,1005.000)
\put(877,495){\makebox(0,0)[b]{{\SetFigFont{10}{12.0}{\rmdefault}{\mddefault}{\updefault}$A$}}}
\put(1702,870){\makebox(0,0)[lb]{{\SetFigFont{9}{10.8}{\rmdefault}{\mddefault}{\updefault}$\child^*$}}}
\put(3052,1245){\makebox(0,0)[lb]{{\SetFigFont{10}{12.0}{\rmdefault}{\mddefault}{\updefault}$x^{C_4}_1$}}}
\put(2452,870){\makebox(0,0)[lb]{{\SetFigFont{9}{10.8}{\rmdefault}{\mddefault}{\updefault}$\child^*$}}}
\put(877,1245){\makebox(0,0)[rb]{{\SetFigFont{10}{12.0}{\rmdefault}{\mddefault}{\updefault}$x^{C_1}_1$}}}
\put(877,945){\makebox(0,0)[rb]{{\SetFigFont{10}{12.0}{\rmdefault}{\mddefault}{\updefault}$x^{C_1}_2$}}}
\put(877,645){\makebox(0,0)[rb]{{\SetFigFont{10}{12.0}{\rmdefault}{\mddefault}{\updefault}$x^{C_1}_3$}}}
\put(3052,945){\makebox(0,0)[lb]{{\SetFigFont{10}{12.0}{\rmdefault}{\mddefault}{\updefault}$x^{C_4}_2$}}}
\put(3052,645){\makebox(0,0)[lb]{{\SetFigFont{10}{12.0}{\rmdefault}{\mddefault}{\updefault}$x^{C_4}_3$}}}
\put(2377,1320){\makebox(0,0)[rb]{{\SetFigFont{10}{12.0}{\rmdefault}{\mddefault}{\updefault}$x^{C_3}_1$}}}
\put(2302,720){\makebox(0,0)[b]{{\SetFigFont{10}{12.0}{\rmdefault}{\mddefault}{\updefault}$x^{C_3}_2$}}}
\put(1627,345){\makebox(0,0)[rb]{{\SetFigFont{10}{12.0}{\rmdefault}{\mddefault}{\updefault}$x^{C_2}_3$}}}
\put(2977,1395){\makebox(0,0)[b]{{\SetFigFont{10}{12.0}{\rmdefault}{\mddefault}{\updefault}$A$}}}
\put(1927,45){\makebox(0,0)[b]{{\SetFigFont{10}{12.0}{\rmdefault}{\mddefault}{\updefault}(a)}}}
\put(1627,645){\makebox(0,0)[rb]{{\SetFigFont{10}{12.0}{\rmdefault}{\mddefault}{\updefault}$x^{C_2}_2$}}}
\put(952,1020){\makebox(0,0)[lb]{{\SetFigFont{9}{10.8}{\rmdefault}{\mddefault}{\updefault}$\child^+$}}}
\put(1777,1095){\makebox(0,0)[rb]{{\SetFigFont{10}{12.0}{\rmdefault}{\mddefault}{\updefault}$x^{C_2}_1$}}}
\end{picture}
}

%% file: diamond_example.tex
\setlength{\unitlength}{0.00083333in}
\begingroup\makeatletter\ifx\SetFigFont\undefined%
\gdef\SetFigFont#1#2#3#4#5{%
  \reset@font\fontsize{#1}{#2pt}%
  \fontfamily{#3}\fontseries{#4}\fontshape{#5}%
  \selectfont}%
\fi\endgroup%
{\renewcommand{\dashlinestretch}{30}
\begin{picture}(5633,5160)(0,-10)
\drawline(1433,2295)(1433,1845)
\blacken\drawline(1418.000,1920.000)(1433.000,1845.000)(1448.000,1920.000)(1418.000,1920.000)
\drawline(533,2295)(533,1845)
\blacken\drawline(518.000,1920.000)(533.000,1845.000)(548.000,1920.000)(518.000,1920.000)
\drawline(533,1845)(983,1395)
\blacken\drawline(919.360,1437.426)(983.000,1395.000)(940.574,1458.640)(919.360,1437.426)
\drawline(1433,1845)(983,1395)
\blacken\drawline(1025.426,1458.640)(983.000,1395.000)(1046.640,1437.426)(1025.426,1458.640)
\drawline(1868,4950)(1418,4500)
\blacken\drawline(1460.426,4563.640)(1418.000,4500.000)(1481.640,4542.426)(1460.426,4563.640)
\drawline(1868,4950)(2318,4500)
\blacken\drawline(2254.360,4542.426)(2318.000,4500.000)(2275.574,4563.640)(2254.360,4542.426)
\drawline(2318,4500)(1868,4050)
\blacken\drawline(1910.426,4113.640)(1868.000,4050.000)(1931.640,4092.426)(1910.426,4113.640)
\drawline(1418,4500)(1868,4050)
\blacken\drawline(1804.360,4092.426)(1868.000,4050.000)(1825.574,4113.640)(1804.360,4092.426)
\drawline(1868,4050)(1418,3600)
\blacken\drawline(1460.426,3663.640)(1418.000,3600.000)(1481.640,3642.426)(1460.426,3663.640)
\drawline(1868,4050)(2318,3600)
\blacken\drawline(2254.360,3642.426)(2318.000,3600.000)(2275.574,3663.640)(2254.360,3642.426)
\drawline(2318,3600)(1868,3150)
\blacken\drawline(1910.426,3213.640)(1868.000,3150.000)(1931.640,3192.426)(1910.426,3213.640)
\drawline(1418,3600)(1868,3150)
\blacken\drawline(1804.360,3192.426)(1868.000,3150.000)(1825.574,3213.640)(1804.360,3192.426)
\put(2093,4725){\makebox(0,0)[lb]{{\SetFigFont{9}{10.8}{\rmdefault}{\mddefault}{\updefault}$\child^+$}}}
\put(1643,4725){\makebox(0,0)[rb]{{\SetFigFont{9}{10.8}{\rmdefault}{\mddefault}{\updefault}$\child^+$}}}
\put(1643,4125){\makebox(0,0)[rb]{{\SetFigFont{9}{10.8}{\rmdefault}{\mddefault}{\updefault}$\child^+$}}}
\put(2093,4125){\makebox(0,0)[lb]{{\SetFigFont{9}{10.8}{\rmdefault}{\mddefault}{\updefault}$\child^+$}}}
\put(2093,3825){\makebox(0,0)[lb]{{\SetFigFont{9}{10.8}{\rmdefault}{\mddefault}{\updefault}$\child^+$}}}
\put(1643,3825){\makebox(0,0)[rb]{{\SetFigFont{9}{10.8}{\rmdefault}{\mddefault}{\updefault}$\child^+$}}}
\put(1643,3225){\makebox(0,0)[rb]{{\SetFigFont{9}{10.8}{\rmdefault}{\mddefault}{\updefault}$\child^+$}}}
\put(2093,3225){\makebox(0,0)[lb]{{\SetFigFont{9}{10.8}{\rmdefault}{\mddefault}{\updefault}$\child^+$}}}
\put(1868,3000){\makebox(0,0)[b]{{\SetFigFont{10}{12.0}{\rmdefault}{\mddefault}{\updefault}$Y_3$}}}
\put(1418,3525){\makebox(0,0)[rb]{{\SetFigFont{10}{12.0}{\rmdefault}{\mddefault}{\updefault}$X_2$}}}
\put(2318,3525){\makebox(0,0)[lb]{{\SetFigFont{10}{12.0}{\rmdefault}{\mddefault}{\updefault}$X_2'$}}}
\put(1418,4425){\makebox(0,0)[rb]{{\SetFigFont{10}{12.0}{\rmdefault}{\mddefault}{\updefault}$X_1$}}}
\put(2318,4425){\makebox(0,0)[lb]{{\SetFigFont{10}{12.0}{\rmdefault}{\mddefault}{\updefault}$X_1'$}}}
\put(1868,5025){\makebox(0,0)[b]{{\SetFigFont{10}{12.0}{\rmdefault}{\mddefault}{\updefault}$Y_1$}}}
\put(1868,3825){\makebox(0,0)[b]{{\SetFigFont{10}{12.0}{\rmdefault}{\mddefault}{\updefault}$Y_2$}}}
\drawline(3983.000,2295.000)(3935.041,2237.108)(3895.526,2173.155)
	(3865.209,2104.363)(3844.669,2032.046)(3834.301,1957.588)
	(3834.301,1882.412)(3844.669,1807.954)(3865.209,1735.637)
	(3895.526,1666.845)(3935.041,1602.892)(3983.000,1545.000)
\blacken\drawline(3923.857,1593.499)(3983.000,1545.000)(3947.060,1612.515)(3923.857,1593.499)
\put(4306,4700){\circle*{48}}
\put(4306,4400){\circle*{48}}
\put(4306,4100){\circle*{48}}
\put(4306,3800){\circle*{48}}
\put(4306,3500){\circle*{48}}
\put(4306,3200){\circle*{48}}
\put(4306,2900){\circle*{48}}
\put(4306,2600){\circle*{48}}
\put(4306,2300){\circle*{48}}
\put(4306,2000){\circle*{48}}
\put(4306,1700){\circle*{48}}
\put(4306,1400){\circle*{48}}
\put(4306,1100){\circle*{48}}
\put(4306,800){\circle*{48}}
\put(4306,500){\circle*{48}}
\drawline(983,945)(1883,495)
\blacken\drawline(1809.210,515.125)(1883.000,495.000)(1822.626,541.957)(1809.210,515.125)
\drawline(2783,945)(1883,495)
\blacken\drawline(1943.374,541.957)(1883.000,495.000)(1956.790,515.125)(1943.374,541.957)
\drawline(983,1395)(983,945)
\blacken\drawline(968.000,1020.000)(983.000,945.000)(998.000,1020.000)(968.000,1020.000)
\drawline(2783,1395)(2783,945)
\blacken\drawline(2768.000,1020.000)(2783.000,945.000)(2798.000,1020.000)(2768.000,1020.000)
\drawline(2333,2295)(2333,1845)
\blacken\drawline(2318.000,1920.000)(2333.000,1845.000)(2348.000,1920.000)(2318.000,1920.000)
\drawline(2333,1845)(2783,1395)
\blacken\drawline(2719.360,1437.426)(2783.000,1395.000)(2740.574,1458.640)(2719.360,1437.426)
\drawline(4305,4680)(4305,480)
\put(1883,345){\makebox(0,0)[b]{{\SetFigFont{10}{12.0}{\rmdefault}{\mddefault}{\updefault}$Y_3$}}}
\put(1508,570){\makebox(0,0)[rb]{{\SetFigFont{9}{10.8}{\rmdefault}{\mddefault}{\updefault}$\child^+$}}}
\put(2258,570){\makebox(0,0)[lb]{{\SetFigFont{9}{10.8}{\rmdefault}{\mddefault}{\updefault}$\child^+$}}}
\put(983,870){\makebox(0,0)[rb]{{\SetFigFont{10}{12.0}{\rmdefault}{\mddefault}{\updefault}$X_2$}}}
\put(2783,870){\makebox(0,0)[lb]{{\SetFigFont{10}{12.0}{\rmdefault}{\mddefault}{\updefault}$X_2'$}}}
\put(908,1170){\makebox(0,0)[rb]{{\SetFigFont{9}{10.8}{\rmdefault}{\mddefault}{\updefault}$\child^+$}}}
\put(2858,1170){\makebox(0,0)[lb]{{\SetFigFont{9}{10.8}{\rmdefault}{\mddefault}{\updefault}$\child^+$}}}
\put(1058,1320){\makebox(0,0)[lb]{{\SetFigFont{10}{12.0}{\rmdefault}{\mddefault}{\updefault}$Y_2$}}}
\put(2708,1320){\makebox(0,0)[rb]{{\SetFigFont{10}{12.0}{\rmdefault}{\mddefault}{\updefault}$Y_2$}}}
\put(2558,1470){\makebox(0,0)[rb]{{\SetFigFont{9}{10.8}{\rmdefault}{\mddefault}{\updefault}$\child^+$}}}
\put(2333,1770){\makebox(0,0)[rb]{{\SetFigFont{10}{12.0}{\rmdefault}{\mddefault}{\updefault}$X_1$}}}
\put(2333,2370){\makebox(0,0)[b]{{\SetFigFont{10}{12.0}{\rmdefault}{\mddefault}{\updefault}$Y_1$}}}
\put(1208,1470){\makebox(0,0)[lb]{{\SetFigFont{9}{10.8}{\rmdefault}{\mddefault}{\updefault}$\child^+$}}}
\put(758,1470){\makebox(0,0)[rb]{{\SetFigFont{9}{10.8}{\rmdefault}{\mddefault}{\updefault}$\child^+$}}}
\put(1433,1770){\makebox(0,0)[lb]{{\SetFigFont{10}{12.0}{\rmdefault}{\mddefault}{\updefault}$X_1'$}}}
\put(533,1770){\makebox(0,0)[rb]{{\SetFigFont{10}{12.0}{\rmdefault}{\mddefault}{\updefault}$X_1$}}}
\put(533,2370){\makebox(0,0)[b]{{\SetFigFont{10}{12.0}{\rmdefault}{\mddefault}{\updefault}$Y_1$}}}
\put(1433,2370){\makebox(0,0)[b]{{\SetFigFont{10}{12.0}{\rmdefault}{\mddefault}{\updefault}$Y_1$}}}
\put(1358,2070){\makebox(0,0)[rb]{{\SetFigFont{9}{10.8}{\rmdefault}{\mddefault}{\updefault}$\child^+$}}}
\put(2408,2070){\makebox(0,0)[lb]{{\SetFigFont{9}{10.8}{\rmdefault}{\mddefault}{\updefault}$\child^+$}}}
\put(1883,2745){\makebox(0,0)[b]{{\SetFigFont{10}{12.0}{\rmdefault}{\mddefault}{\updefault}(a)}}}
\put(1883,45){\makebox(0,0)[b]{{\SetFigFont{10}{12.0}{\rmdefault}{\mddefault}{\updefault}(b)}}}
\put(4283,45){\makebox(0,0)[b]{{\SetFigFont{10}{12.0}{\rmdefault}{\mddefault}{\updefault}(c)}}}
\put(4208,4695){\makebox(0,0)[rb]{{\SetFigFont{10}{12.0}{\rmdefault}{\mddefault}{\updefault}$Y_1$}}}
\put(4208,4395){\makebox(0,0)[rb]{{\SetFigFont{10}{12.0}{\rmdefault}{\mddefault}{\updefault}$X_1$}}}
\put(4208,4095){\makebox(0,0)[rb]{{\SetFigFont{10}{12.0}{\rmdefault}{\mddefault}{\updefault}$Y_2$}}}
\put(4208,3795){\makebox(0,0)[rb]{{\SetFigFont{10}{12.0}{\rmdefault}{\mddefault}{\updefault}$X_2$}}}
\put(4208,3495){\makebox(0,0)[rb]{{\SetFigFont{10}{12.0}{\rmdefault}{\mddefault}{\updefault}$Y_3$}}}
\put(4208,3195){\makebox(0,0)[rb]{{\SetFigFont{10}{12.0}{\rmdefault}{\mddefault}{\updefault}$Y_1$}}}
\put(4208,2895){\makebox(0,0)[rb]{{\SetFigFont{10}{12.0}{\rmdefault}{\mddefault}{\updefault}$X_1$}}}
\put(4208,2595){\makebox(0,0)[rb]{{\SetFigFont{10}{12.0}{\rmdefault}{\mddefault}{\updefault}$Y_2$}}}
\put(4208,2295){\makebox(0,0)[rb]{{\SetFigFont{10}{12.0}{\rmdefault}{\mddefault}{\updefault}$X_2'$}}}
\put(4208,1995){\makebox(0,0)[rb]{{\SetFigFont{10}{12.0}{\rmdefault}{\mddefault}{\updefault}$Y_3$}}}
\put(4208,1695){\makebox(0,0)[rb]{{\SetFigFont{10}{12.0}{\rmdefault}{\mddefault}{\updefault}$Y_1$}}}
\put(4208,1395){\makebox(0,0)[rb]{{\SetFigFont{10}{12.0}{\rmdefault}{\mddefault}{\updefault}$X_1'$}}}
\put(4208,1095){\makebox(0,0)[rb]{{\SetFigFont{10}{12.0}{\rmdefault}{\mddefault}{\updefault}$Y_2$}}}
\put(4208,795){\makebox(0,0)[rb]{{\SetFigFont{10}{12.0}{\rmdefault}{\mddefault}{\updefault}$X_2$}}}
\put(4208,495){\makebox(0,0)[rb]{{\SetFigFont{10}{12.0}{\rmdefault}{\mddefault}{\updefault}$Y_3$}}}
\put(0,2070){\makebox(0,0)[lb]{{\SetFigFont{9}{10.8}{\rmdefault}{\mddefault}{\updefault}$\child^+$}}}
\put(5633,450){\makebox(0,0)[rb]{{\SetFigFont{10}{12.0}{\rmdefault}{\mddefault}{\updefault}$\left. \begin{array}{l} \\ \\ \\ \\ \\ \\[1.5ex] \end{array} \right\} LC(X_1' \land \neg X_2')$}}}
\put(5235,1956){\makebox(0,0)[rb]{{\SetFigFont{10}{12.0}{\rmdefault}{\mddefault}{\updefault}$\left. \begin{array}{l} \\ \\ \\ \\ \\ \\ \\ \\ \\ \\ \\ \\ \\[3ex] \end{array} \right\} LC(\neg X_1')$}}}
\end{picture}
}